\documentclass[notitlepage,aps,pra,superscriptaddress,showpacs,amsmath,amssymb,amsfonts,lengthcheck,twocolumn,longbibliography]{revtex4-2}
\usepackage[normalem]{ulem}   
\usepackage{xcolor}

\usepackage{dsfont}
\usepackage{graphicx} 
\usepackage{graphics}
\usepackage{mathtools}
\usepackage{dcolumn}    
\usepackage{makecell}
\usepackage{bm} 
\usepackage{graphicx}
\usepackage{amsmath}    
\usepackage{latexsym}
\usepackage{amsfonts}   
\usepackage{amssymb}
\usepackage{array}      
\usepackage{bbm}
\usepackage{xcolor}
\usepackage[colorlinks=true,linkcolor=blue,urlcolor=blue,citecolor=blue,pdfusetitle]{hyperref}

\newcommand{\Tr}{{\rm Tr}}

\newcommand{\bra}[1]{\langle #1 |}
\newcommand{\ket}[1]{| #1 \rangle}

\begin{document}
\title{Resource state generation for a multispin register in a hybrid matter-photon quantum information processor}
\author{Yu Liu}
\email{yu.liu@uni-ulm.de}
\affiliation{Institut f\"{u}r Theoretische Physik und IQST, Albert-Einstein-Allee 11, Universit\"{a}t Ulm, D-89081 Ulm, Germany}
\author{Martin B. Plenio}
\email{martin.plenio@uni-ulm.de}
\affiliation{Institut f\"{u}r Theoretische Physik und IQST, Albert-Einstein-Allee 11, Universit\"{a}t Ulm, D-89081 Ulm, Germany}
    \begin{abstract}
        Hybrid quantum architectures that integrate matter and photonic degrees of freedom present a promising pathway toward scalable, fault-tolerant quantum computing. This approach needs to combine well-established entangling operations between distant registers using photonic degrees of freedom with direct interactions between matter qubits within a solid-state register. The high-fidelity control of such a register, however, poses significant challenges. 
        In this work, we address these challenges with pulsed control sequences which modulate all inter-spin interactions to preserve the nearest-neighbor couplings while eliminating unwanted long-range interactions. We derive pulse sequences, including broadband and selective gates, using composite pulse and shaped pulse techniques as well as optimal control methods. This ensures a general pulse sequence in the presence of spin-position bias, and robustness against static offset detunings, and Rabi frequency fluctuations of the control fields. The control techniques developed here apply well beyond the present setting to a broad range of physical platforms.
        We demonstrate the efficacy of our methods for the resource state generation for fusion-based quantum computing in four- and six-spin systems encoded in the electronic ground states of nitrogen-vacancy centers or other molecular solid-state qubits. We also outline other elements of the proposed architecture, highlighting its potential for advancing quantum computing technology.	
    \end{abstract}
\maketitle
\section{Introduction}
Quantum computation is formulated in terms of either the gate-array model or the measurement-based model. In the gate-array model one- and two-qubit quantum gates, described as unitary operators, are used to build up a desired state transformation on a multi-qubit device \cite{Deutsch1985, vedral1998basics}. In contrast, in the measurement-based gate implementations \cite{GottesmannC1999, eisert2000optimal} Bell states are used to enable the synthesis of quantum gates based on local operations. More generally, a multipartite entangled resource state, such as the cluster state, can be prepared \cite{Raussendorf2001, Nielsen2004, Briegel2009}, and then generic quantum computation can be realized via single-qubit measurements and single-qubit rotations \cite{Nielsen2004, Briegel2009, Raussendorf2003, Nielsen2006}. 

The measurement-based approach is particularly attractive for optical quantum information processing, where a universal quantum gate set is difficult to implement directly as this would require large nonlinearities. Photon detection, however, is fast, readily available and in fact essential as photons have short lifetimes in integrated photonic structures and need to be measured before they are absorbed. A detailed measurement-based approach to optical quantum computing was first formulated in Ref.~\cite{knill2001scheme} and later refined into more resource-efficient designs within the language of the measurement-based approach based on cluster states \cite{Nielsen2004, Browne2005, nielsen2005fault}. However, the preparation of a large cluster state in advance is challenging due to environmental noise and the limitations of current control techniques. 

Recently, as an alternative, fusion-based quantum computation (FBQC)~\cite{Bombin2021, Bartolucci2023}, a highly modular architecture, has been formulated. Here, only small graph states (the so-called four- and six-ring states) are required as resource states, which are then connected by entangling measurements~\cite{Browne2005, Bose1999} between pairs of such resource states in a fusion network "on the fly" while simultaneously propagating the quantum computation. This all-optical approach requires highly multiplexed photonic down-conversion sources and fast optical switches to produce the required four- and six-ring resource states quasi-deterministically. Multiplexed sources, switches, and photon detectors need to be operated in a cryogenic environment to achieve high photon detection efficiency~\cite{Alexander2024}.

These daunting requirements have motivated the search for alternative, deterministic, photon sources, e.g., based on quantum dots \cite{Schwartz2016, Larsen2019, Istrati2020, Ferreira2024} where the internal spin degree of freedom can be used to control the generation of entangled photon states. The use of solid-state emitters with an internal degree of freedom suggests a further shift of paradigm towards hybrid schemes in which quantum information is held on matter degrees of freedom (stationary qubits) and connected via photonic degrees of freedom using entangling measurements \cite{Bose1999}. This represents an alternative approach that implements spin-optical quantum computing by generating source states in a spin-based system and performing fusion measurements through photonic methods. Such a hybrid architecture effectively integrates established matter-based and photonic-based techniques, establishing a potential platform to implement fault-tolerant quantum computing (see for example Ref.~\cite{deGliniasty2024} for a recent summary of such an approach).

While one might use entangling measurements on photons to build a cluster state and then execute measurement-based quantum computation, it seems tempting to transfer the concept of fusion-based quantum computing to a hybrid matter-photonic system. There are now two possible ways towards this goal. In the first approach, individual optically active spin-qubits are each placed in a separate optical resonator, which is then connected to a photonic network that allows for the implementation of entangling operations based on \cite{Bose1999} or reflection and detection of photons \cite{Nemoto2014}, or here we take an approach where each resonator contains a small number of optically active spin-qubits that are magnetically or otherwise coupled in order to create the desired resource states (e.g. four- and six-ring cluster states) locally on the matter qubits. The latter may offer additional opportunities, as matter qubits exhibit greater versatility and flexibility in their mutual interaction, which can be used to create desired quantum states and may therefore offer advantages. 

In this work, we are placing our emphasis on the matter-based quantum register in a fusion-based hybrid matter-photon quantum computer with the objective of demonstrating the potential for generating the desired resource state. The spin-based quantum information processor in a solid-state system employs the spin states of electrons or nuclei in solid materials to encode, manipulate, and process quantum information. However, the preparation of complex multi-partite quantum states in these systems remains a challenge. Especially in solid-state quantum registers based on magnetic interactions the nearest-neighbor (NN) couplings dominate but, nevertheless, beyond-NN interactions are still relevant when aiming to achieve the extremely high fidelities that are necessary for fault-tolerant quantum error correction. In conventional schemes capable of eliminating long-range spin interactions, the NN-ones tend to be suppressed as well. Moreover, Ref.~\cite{Zhang2011} proposes a general dynamical-decoupling-based protocol that applies appropriately $\pi$-pulse sequences on selected qubits to average out next-nearest-neighbor (NNN) couplings while preserving the desired NN ones, enabling high-fidelity preparation of one-dimensional cluster states. The evolution is divided into several short intervals, and spin-selective $\pi$ pulses are applied at suitable timing. Both $xy$ and $zz$ interactions can be refocused while NNN terms average over one dynamic cycle, leaving an effective Hamiltonian dominated by NN ones. The protocol, however, is formulated specifically for the one-dimensional spin chain and does not explicitly address how control sequences modify longer-range interactions, thereby limiting its direct applicability to more general architectures with extended couplings.

\begin{figure}[tb]
\centering
\includegraphics[width=0.48\textwidth]{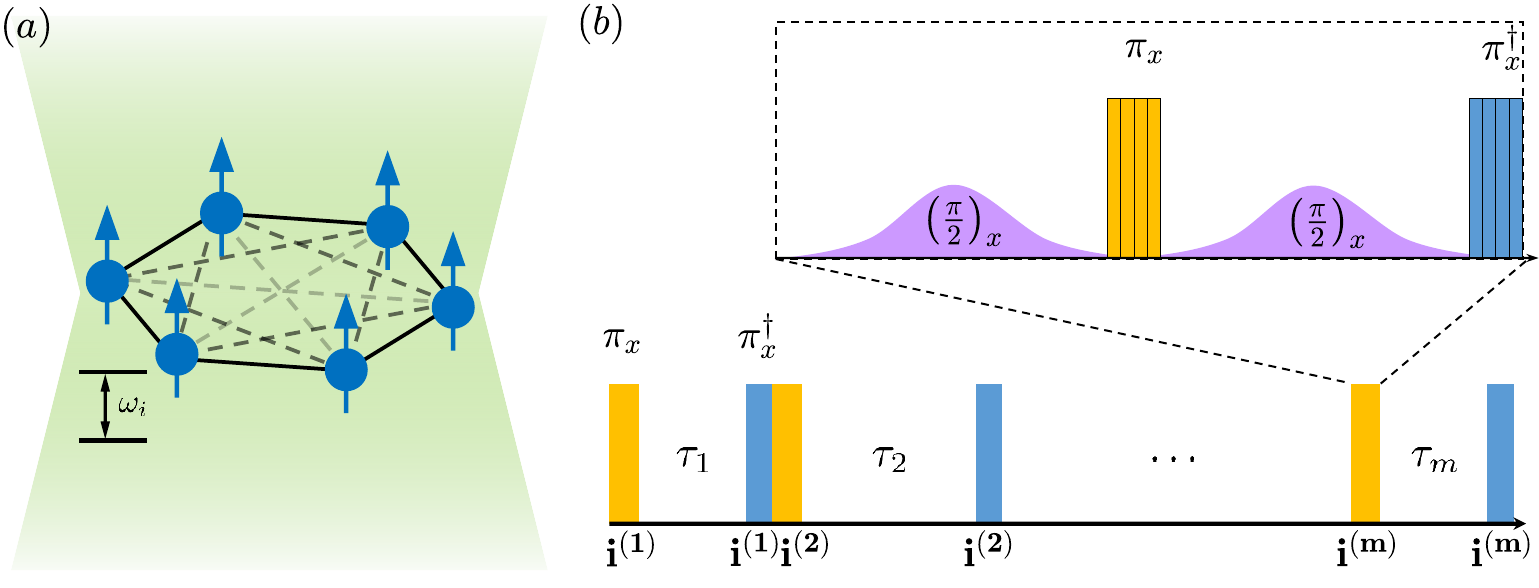}
\caption{(a) Scheme of the pulsed cluster preparation in a quantum multi-spin system, where all spins are driven by the global field (light green). The solid lines indicate the NN couplings that have to be preserved, while the dashed lines represent the unwanted long-range couplings. $\omega_i$ indicates the unique energy splittings of spin $i$. (b) Pulsed dynamical sequence to eliminate long-range couplings. The whole process is divided into $m$ segments, and each segment consists of a free evolution time $\tau_k$ sandwiched by two selective $\pi_x$ and $\pi_x^{\dagger}$ gates denoted as $R_\pi^x(\textbf{i}^{(k)})$ and $R_\pi^{-x}(\textbf{i}^{(k)})$. The index-vector $\textbf{i}^{(k)}$ indicates the spins flipped by the $\pi_x$ pulses in $k$-th segment. The foundation of the scheme lies in a selective single $\pi_x$ pulse, which selectively flips the resonant spin while leaving the remainder unchanged. As we discuss in Sec.~\ref{subsec_broad-band} and Sec.~\ref{subsec:selective_gate}, the selective $\pi_x$ and $\pi_x^\dagger$ pulses could be implemented by a sequence of two shaped $(\pi/2)_x$ pulses (purple curves) and two composite $\pi_x$ pulses (yellow and blue blocks), depicted within the dashed box. The generation of all pulses is attributed to the global driving field due to limitations in the spatial resolution of the control field.}		
\label{fig_scheme}
\end{figure}

In addition, conventional schemes typically assume an idealized scenario in which all spins are precisely positioned, usually forming a perfect one-dimensional spin chain or two-dimensional square lattice. In practice, however, deviations in the positions of the spins, for example, due to the implantation of color centers or those associated with nuclear spins, are common. The resulting variations in the mutual distances lead to inhomogeneities in the coupling strengths which, in turn, can significantly affect the effectiveness of most state preparation schemes.

Finally, in solid-state systems, strong interactions typically emerge when the spins are closely spaced, which poses significant challenges for performing local manipulations and measurements. Specifically, in electron spin or nuclear spin systems, sufficiently strong direct magnetic dipole interactions occur on the scale of tens down to a few nanometers. Consequently, it is crucial to investigate methods for quantum state preparation that are feasible with only a global drive. Note that the ability to manipulate spins without individual control is a fundamental and common necessity for quantum spin systems.

In our work, we present a new approach for generating cluster states using a carefully designed sequence of selective $\pi$ pulses. This sequence is designed to modify the inter-spin couplings in a non-uniform manner across different segments, ensuring the preservation of NN interactions while eliminating long-range interactions. Our scheme is illustrated in Fig.~\ref{fig_scheme}(a), where a quantum spin system is driven by a global control field (green light), and the solid and dashed lines between the spins denote the NN and long-range couplings. Fig.~\ref{fig_scheme}(b) shows the pulse sequence which is composed of $m$ segments. $R^{x}_{\pi}({\textbf{i}})$ is a selective $\pi$ pulse about the $x$-axis that flips all the spins indicated by an index-vector $\textbf{i}$, and the single selective $\pi$ pulse (the insert in Fig.~\ref{fig_scheme}(b)) is implemented by a composite pulse which only flips the resonant spin while leaving the others unaffected. To implement our protocol, we develop a general framework of a frequency-selective pulse using composite and shaped pulse techniques. These pulses, optimized using optimal control methods, are robust to possible imperfections in the global driving field. 

Our manuscript is organized as follows. Sec.~\ref{sec:framework} introduces the general framework of the cluster state preparation utilizing a pulsed control sequence. We then present several examples by giving a few multi-spin systems and discuss the extension to larger systems in Sec.~\ref{sec:examples}. Sec.~\ref{sec:implementations} demonstrates the feasibility of implementing the proposed scheme in nitrogen-vacancy (NV) center systems~\cite{Doherty2013, Wu2016}, with the development of selective pulses and the numerical simulation of preparation fidelity. A brief conclusion of our proposal and outlook towards the hybrid matter-photon quantum information processor are given in Sec.~\ref{sec:conclusion}. 

\section{Theoretical framework}
\label{sec:framework}
To generate a cluster state in a fully connected spin system, we first initialize all spins into a product state $\ket{\psi_0}=\otimes_i|+\rangle_i$ and then the cluster state $\ket{\psi}_{\mathcal{C}}$ is prepared by applying a nearest-neighbor controlled-$Z$ gate $S=\prod_{\langle i,j\rangle} S_{ij}$ where
\begin{align}
    S_{ij} =\ket{0}_i {}_i\bra{0}\otimes \mathbbm{1}_j+\ket{1}_i {}_i\bra{1}\otimes \sigma_j^z,
\end{align}
and $\langle i,j\rangle$ denotes NN spins. $S$ is a joint controlled-$Z$ gate and typically generated using an Ising-like $zz$ coupling Hamiltonian $H_{\text{C}}/\hbar=\sum_{\langle i,j\rangle} g(1-\sigma_i^z)(1-\sigma_j^z)$ with the required evolution time $t = \pi/4g$~\cite{Raussendorf2003}. $S$ can be decomposed into a product of three basic gates: $S = \prod_{\langle i,j\rangle}U^{zz}_{\pi}(i,j)R^z_{\pi/2}(i)R^z_{\pi/2}(j)$. Here and after, $R^{z}_{\alpha}(\textbf{i})$ indicates a joint $\alpha$ rotation along the $z$ axis of spins denoted by the index-vector $\textbf{i}$, while $U^{zz}_{\pi}(i,j)=\exp(-i\pi\sigma_{i}^z\sigma_{j}^z/4)$ acts between spin $i$ and $j$. Because all elements in $S$ commute, $S$ can be written as
\begin{align}
    S = \prod_i R_{z}^{\varphi_i}(i)\prod_{\langle i,j\rangle} U^{zz}_{\pi}(i,j),
\end{align}
where $\varphi_i = n_i\pi/2$ and $n_i$ indicates the number of NN pairs in which spin $i$ is involved.
We get a target unitary $U_{\text{t}}$ to prepare the cluster state after dropping the first local transforms 
\begin{align}
    U_{\text{t}} = \prod_{\langle i,j\rangle} e^{-i\frac{\pi}{4}\sigma_{i}^z\sigma_{j}^z}.
    \label{eq:Ut}
\end{align} 
This necessitates an engineered evolution process that ensures that the phases of the NN coupling terms are equal to $\pi$ while simultaneously eliminating the inherent long-range couplings. However, as the size of the spin system increases, the number of two-body interactions increases quadratically with the number of spins. This makes mitigation of long-range interactions more challenging. 
To explain our proposal to eliminate the long-range $zz$ interaction, we consider a fully connected $N$-spin Hamiltonian ($\hbar = 1$)
\begin{align}
    H_{zz} = \sum_{i\neq j} \frac{g_{ij}}{4}\sigma_{i}^z\sigma_{j}^z.
    \label{eq:ZZ_Hamiltonian}
\end{align}
For simplicity, we assume that the coupling terms $g_{ij}$ are arranged in descending order based on their corresponding strengths rather than the indices. 

The coupling strength $g_{ij}$ is typically determined by the relative position between spin $i$ and $j$. In practice, possible variations in spin positions lead to inhomogeneities in interaction strengths, causing differences in the accumulated phase (denoted as $\theta_{ij} = g_{ij}t$) for each interaction term during the evolution $U_{zz}(t)=\exp(-iH_{zz} t)$. The inhomogeneity results in unique phase shifts for each spin pairs, affecting the overall quantum state dynamics. To make the phases accumulated by all NN interactions equal to $\pi$ while eliminating long-range couplings, we introduce a pulsed dynamical decoupling sequence that selectively reverses specific interaction strengths.
As illustrated in Fig.~\ref{fig_scheme}(b), an elementary segment in the sequence consists of two multi-spin $\pi$ pulses $R_{\pi}^x(\mathbf{i}^{(k)})$ and $R_{\pi}^{-x}(\mathbf{i}^{(k)})$ before and after a period of free evolution period of a length $\tau_k$. This segment selectively inverts the coupling terms~\cite{Zhang2011}. Here 
\begin{align}
    R_{\pi}^{\phi}(\mathbf{i}) = \otimes_{j=1}^{n_{\text{i}}} \exp\left(-i\frac{\pi}{2}\sigma^{\phi}_{i_j}\right),
\end{align}
where index-vector $\mathbf{i} = (i_1,~i_2,~\cdots,~i_{n_{\text{i}}})$ indicating which spins are flipped are arranged in increasing order; and $\sigma_{i_j}^{\phi}=\cos(\phi)\sigma_{i_j}^x+\sin(\phi)\sigma_{i_j}^y$ denotes the rotation operator of the $i_j$-th spin. The evolution of the $k$-th segment is given by
\begin{align}
    \nonumber
    U\left(\mathbf{i}^{(k)},\tau_k\right) & = R_{\pi}^{-x}(\mathbf{i}^{(k)}) U_{zz}(\tau_k)R_{\pi}^{x}(\mathbf{i}^{(k)})\\
    & = \exp\left(-i\sum_{i\neq j} f_{ij}(k) \frac{g_{ij}\tau_k}{4}\sigma_i^z\sigma_j^z\right),
\end{align}
where the modulation factor $f_{ij}(k)=(-1)^{\mathcal{N}_{ij}(k)}$ and $\mathcal{N}_{ij}(k) = \left|\{i,j\}\cap\mathbf{i}^{(k)}\right|$ indicates the number of spins flipped by $R_{\pi}^x(\mathbf{i}^{(k)})$ in spin $\{i,j\}$.
Here $\tau_k$ denotes the duration of the $k-$th segment as $\pi$ pulses are assumed to be instantaneous in the analysis.
If only one of the spins $i$ and $j$ is affected by the $R_{\pi}^{x}(\mathbf{i}^{(k)})$ pulse, the effective coupling strength during this evolution is inverted,  $g_{ij}\rightarrow-g_{ij}$. If none or both of them are involved, the sign of the coupling strength $g_{ij}$ remains unchanged.

Therefore, we are able to tailor the accumulated phases by applying a pulse sequence. The sequence includes $m$ elementary segments with corresponding indices $\mathcal{S}=\{\mathbf{i}^{(1)},~\mathbf{i}^{(2)},~\cdots,~\mathbf{i}^{(m)}\}$ and corresponding duration $\vec{\tau}\ = (\tau_1,~\tau_2,~\cdots,~\tau_m)^T$.
The whole evolution unitary is
\begin{align}
    U(T_c) = \prod_{k=1}^{m} U\left(\mathbf{i}^{(k)},\tau_k\right)= \exp\left(-i\sum_{i\neq j}\frac{\theta_{ij}}{4}\sigma_i^z\sigma_j^z\right),
\end{align}
where the accumulated phases and total duration are
\begin{align}
    \theta_{ij} ={g_{ij}}\sum_{k=1}^mf_{ij}(k)\tau_k,\quad T_c &= \sum_{k=1}^m\tau_k.
\end{align}
These two equations naturally form a linear equation
\begin{align}
    \label{eq:F_tau_alpha}
    {M}\cdot\vec{\tau} = \vec\alpha,
\end{align}
with the factor matrix $M\in\mathbb{R}^{(n_g+1)\times m}$ and target phase vector $\vec{\alpha}\in\mathbb{R}^{(n_g+1)\times 1}$. The first $n_g+1$ elements of $\vec{\alpha}$ are $\theta_{ij}/g_{ij}$ and $\alpha(n_g+1) = T_c$. Here, $n_g =N(N+1)/2$ is the number of coupling terms. It is apparent that the form of the pulse sequence is determined by solving the linear equation. The solution of the equation is general and does not require a perfect lattice. After fabrication, all couplings are fixed, thereby enabling the determination of the corresponding pulse sequence.

In the next section, we will provide solutions in the small spin-ring systems employed in FQBC as examples to explain our procedure. Note that these decoupling pulses modify the interaction strengths while they do not change their forms. Hence, the order of applying the pulses can be arbitrary, as the unitary evolution operators of all segments mutually commute. In addition, each pulse can be applied at most once, as multiple segments with the same pulses can be moved together and merged.

\section{Resource state generation}
\label{sec:examples}
We proceed with discussing the specific forms of our pulsed cluster state preparation protocol in different systems.
First, we consider two small spin-ring systems that are fully solvable and are required for the implementation of FBQC. In these systems, we consider both the cases with and without uncertainties in the positions. We then investigate an ideal larger spin-ring system, utilizing its spatial symmetry, and demonstrate that our proposal could be extended to a larger quantum system with about 20 spins.  

\subsection{The spin-ring system} \label{subsec:solvable system}
\begin{figure}[tb]
    \centering
    \includegraphics[width=0.4\textwidth]{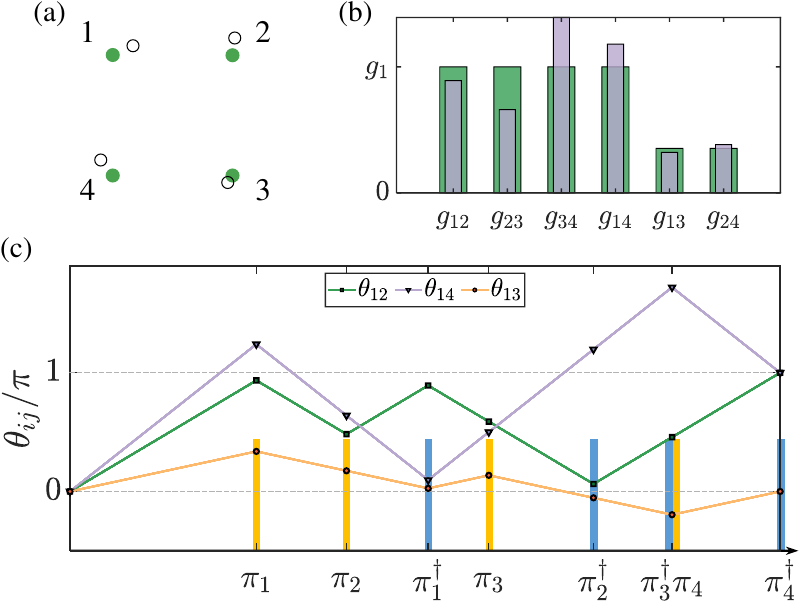}
    \caption{The preparation of a cluster state in a four-spin system, where the square lattice without and with position errors are represented by green dots and black circles in (a), respectively. The corresponding coupling strengths $g_{ij}$ are represented by green and purple bars in (b). $g_1$ is the NN coupling strength in the ideal case. (c) The evolution of the accumulated phase $\theta_{ij}$ with respect to the NN couplings $g_{12,14}$ and one NNN coupling $g_{13}$ are represented. In each segment, the derivatives of the phases with respect to time are determined by the corresponding modulation factors. The sequence of the selective $\pi$ pulses employed are $\mathcal{S}_4=$\{(0), (1), (1,~2), (2), (2,~3), (3), (4)\}, resulting in a pulse sequence [$\tau_1$-$\pi_1$-$\tau_2$-$\pi_2$-$\tau_3$-$\pi_1^\dagger$-$\tau_4$-$\pi_3$-$\tau_5$-$\pi_2^\dagger$-$\tau_6$-$\pi_3^\dagger\pi_4$-$\tau_7$-$\pi_4^\dagger$], the notation and meaning of $\tau_i$ and $\pi_i$ are introduced in main text. The parameters are identical to those presented in Sec.~\ref{subsec:numerical} within a solid-state system.}
    \label{fig_4_qubit}
\end{figure}

Our discussion initially focuses on two resource state generators associated with four- and six-spin-ring systems. These two systems are fully solvable, whereby the invertibility of the matrix $M$ and the relevance of the semi-positive solution can be guaranteed by selecting $n_g+1$ suitable pulses
\begin{align}
\vec{\tau} = M^{-1}\cdot \vec{\alpha} \geq 0.
\label{eq:tau}
\end{align}
In the four-spin system, as shown in Fig.~\ref{fig_4_qubit}, the spin index sequence is $\mathcal{S}_4 = $\{(0), (1), (1,~2), (2), (2,~3), (3), (4)\} where $(0)$ means that no pulses are applied in the first segment and $(i,~j)$ means that pulses are applied to spins $i$ and $j$. The corresponding time-duration solution is
\begin{align}
    \vec{\tau}=\frac{1}{4} \left(\begin{matrix}
	\alpha_{12}+\alpha_{23}+\alpha_{34}+\alpha_{14}\\
	T_c-\alpha_{12}-\alpha_{14}+\alpha_{24}\\
	\alpha_{12}+\alpha_{34}-\alpha_{13}-\alpha_{24}\\
	T_c-\alpha_{12}-\alpha_{23}+\alpha_{13}\\
	\alpha_{23}+\alpha_{14}-\alpha_{13}-\alpha_{24}\\
	T_c-\alpha_{23}-\alpha_{34}+\alpha_{24}\\
	T_c-\alpha_{34}-\alpha_{14}+\alpha_{13}
    \end{matrix}\right),
    \label{eq:tau_four}
\end{align}
where the first $n_g = 6$ elements of $\vec{\alpha}$ are defined as $\alpha_{ij}=\theta_{ij}/g_{ij}$.

This is a general solution of four-spin system, and the sequence $\mathcal{S}_4$ is chosen to minimize the number of single-spin pulses $N_{\pi}$ after canceling the pulses between adjacent segments, resulting in a pulse sequence [$\tau_1$-$\pi_1$-$\tau_2$-$\pi_2$-$\tau_3$-$\pi_1^\dagger$-$\tau_4$-$\pi_3$-$\tau_5$-$\pi_2^\dagger$-$\tau_6$-$\pi_3^\dagger\pi_4$-$\tau_7$-$\pi_4^\dagger$]. In the expression for the pulse sequence, the duration $\tau_i$ represents the $i$-th free evolution process while $\pi_i$ and $\pi_i^\dagger$ are the abbreviations of $R^x_{\pi}(i)$ and  $R^{-x}_{\pi}(i)$ in this section. A cluster state connecting NN pairs is generated by setting two phases of the long-range terms $\alpha_{13}=\alpha_{24}=0$. The evolution of the accumulated phases $\theta_{ij}$ and the pulse sequence are shown in Fig.~\ref{fig_4_qubit}(c), where the phases of NN couplings and the NNN coupling, respectively, are $\pi$ and $0$ in the end.

In the ideal case, all spins are located on the square lattice sites, denoted by the solid green dots in Fig.~\ref{fig_4_qubit}(a). Thus, the elements in vector $\vec{\alpha}$ associated with NN and NNN couplings are identical to $\pi/g_1$ and $0$. Here $g_1$ is the strength of the NN coupling. The minimal length of the sequence is $T_c=2\pi/g_1$ and the length of each segment is
\begin{align}
    \tau_1 = \frac{\pi}{g_1},~\tau_{2,4,6,7} = 0,~ \tau_{3,5} = \frac{\pi}{2g_1}.
\end{align}
Thereby, the corresponding index sequence is $\mathcal{S}_4$ = \{(0),~(1,~2),~(2,~3)\}, and only six pulses remain as the pulse sequence is [$\tau_1$-$\pi_1\pi_2$-$\tau_3$-$\pi_1^\dagger\pi_3$-$\tau_5$-$\pi_2^\dagger\pi_3^\dagger$] after the cancellation.

Similarly to the four-spin case, solving Eq.~\eqref{eq:tau}, a cluster state could be prepared by a sequence of two-body pulses in a six-spin system. We get a sequence of the pulse indices $\mathcal{S}_6=$\{(0), (1,~2), (2,~3), (3,~4), (4,~5), (5,~6), (1,~6), (1,~3), (1,~4), (4,~6), (2,~4), (2,~6), (3,~6), (3,~5), (1,~5), (2,~5)\}, which means that the required number $\pi$ pulses is $N_{\pi} =  32$. The six spins are labeled clockwise and the general pulse sequence is [$\tau_1$-$\pi_1\pi_2$-$\tau_2$-$\pi_1^\dagger\pi_3$-$\tau_3$-$\pi_2^\dagger\pi_4$-$\tau_4$-$\pi_3^\dagger\pi_5$-$\tau_5$-$\pi_4^\dagger\pi_6$-$\tau_6$-$\pi_5^\dagger\pi_1$-$\tau_7$-$\pi_6^\dagger\pi_1$-$\tau_8$-$\pi_3^\dagger\pi_4$-$\tau_9$-$\pi_1^\dagger\pi_6$-$\tau_{10}$-$\pi_6^\dagger\pi_2$-$\tau_{11}$-$\pi_4^\dagger\pi_6$-$\tau_{12}$-$\pi_2^\dagger\pi_3$-$\tau_{13}$-$\pi_6^\dagger\pi_5$-$\tau_{14}$-$\pi_3^\dagger\pi_1$-$\tau_{15}$-$\pi_1^\dagger\pi_2$-$\tau_{16}$-$\pi_2^\dagger\pi_5^\dagger$]. In the ideal case, the corresponding length of each segment is
\begin{align}
    \tau_1 = \frac{5\pi}{8g_1},~\tau_{2,3,5,6} =\frac{\pi}{g_1},~\tau_{4}=\frac{3\pi}{8g_1},\nonumber\\  \tau_{8,10,14\rightarrow16} = \frac{\pi}{8g_1},~\tau_{9,10,12,13} =0,
\end{align}
with the total duration of the sequence $T_c = 2\pi/g_1$. More details and sequences of quantum spin systems are provided in Appendix~\ref{app_sec_solvable}. It is essential to note that, for a multi-spin system, our proposal enables the generation of cluster states with diverse geometries and connectivities by appropriately choosing the target phase vector $\vec{\alpha}$. In particular, the same framework can be used either to effectively realize higher-dimensional cluster states, or to isolate a single desired coupling by suppressing all other interactions, as illustrated by the two representative examples in Appendix~\ref{app_subsec_example}.

\subsection{Larger spin-ring system}
As illustrated in the preceding section, the state generation proposals employ up to two-spin pulses in a small spin system. Inspired by this, we aim to explore the preparation of the cluster state in larger spin-ring systems. Here we assume that $N$ spins are equally spaced, and the periodic condition $\sigma_1^z=\sigma_{N+1}^z$ leads to a spatial symmetry which is the key in our proposal. First, we label the two-spin coupling with respect to the distance between them. Thus, all NN couplings are denoted as $\mathcal{C}^{(1)}$ and the NNN couplings are labeled as $\mathcal{C}^{(2)}$ and the $l$-th order of the collective coupling is
\begin{align}
    \mathcal{C}^{(l)} = \sum_{i=1}^{n_l} \frac{1}{4}\sigma_i^z\sigma_{i+l}^z,
\end{align} 
with $n_l$ is number of elements in $\mathcal{C}^{(l)}$.
Hamiltonian~(\ref{eq:ZZ_Hamiltonian}) can be rewrite as $H_{zz} = \sum_{l=1}^{L} g_l \mathcal{C}^{(l)}$ with $g_l$ is the $l$-th coupling strength and $L$ indicates the number of collective coupling.
The target unitary~(\ref{eq:Ut}) in a spin-ring system now is 
\begin{align}
    U_{\text{t}} = \exp{\left(-i\pi \mathcal{C}^{(1)}\right)},
\end{align}
which requires the preservation of the first collective coupling. In a system with an odd number of spins, the number of collective couplings is given as $L=(N-1)/2$. In the $l$-th collective coupling, the number of two-spin interactions is $n_l=N$. In contrast, in an even spin system, the number of collective couplings is $L=N/2$. Each coupling contains $n_l=N$ elements, except the number of elements in the $L$-th coupling is $n_L=N/2$.

Similarly, the $k$-th two-spin collective pulse $\mathcal{G}^{(k)}$, is defined to include all possible pulses applied on two spins with a specific distance $\mathcal{G}^{(k)}$=$\{R^x_{\pi}(1,1+k)$, $R^x_{\pi}(2,2+k)$, $\cdots\}$. The number of elements in $\mathcal{G}^{(k)}$ is $n_k$ that is identical to the number of coupling terms in $k$-th collective coupling $\mathcal{C}^{(k)}$.
All the two-spin pulses in $\mathcal{G}^{(k)}$ are applied before and after a unitary $U_{zz}$ sequentially, the evolution in $k$- segment is
\begin{align}
    \mathcal{G}^{(k)}[H_{zz},\tau_k] = \prod_{i=1}^{n_k} R^{-x}_{\pi}(i,i+k) U_{zz}(\tau_k) R^x_{\pi}(i,i+k).
\end{align}
Due to the symmetry of the system, the unitary operator can be expanded as  
\begin{align}
    \mathcal{G}^{(k)}[H_{zz},\tau_k] = \exp\left(-i\sum_{l=1}^{L}f'_{lk}g_l\mathcal{C}^{(l)}\tau_k\right)\, ,
\end{align}
where $f'_{lk}$ is the modulation factor of the $l$-th coupling $\mathcal{C}^{(l)}$ modified by $k$-th collective gate $\mathcal{G}^{(k)}$. 
The values of $f'_{lk}$ are listed in Table~\ref{table:f_collec}, and the last two columns are relevant to $f'_{lL}$ in the odd and even spin systems, respectively.
\begin{table}[t]
    \begin{center}
	\begin{tabular}{c|ccccccc}
		\hline
		$f'_{lk}$ & $\mathcal{G}^{(0)}$ & $\mathcal{G}^{(1)}$& $\mathcal{G}^{(2)}$& $\mathcal{G}^{(3)}$& $\cdots$& $\mathcal{G}^{(L)}_{\text{odd}}$& $\mathcal{G}^{(L)}_{\text{even}}$ \\
		\hline
		$\mathcal{C}^{(1)}$ & $1$ & $N-4$ & $N-8$& $N-8$& $\cdots$& $N-8$ & $N/2-4$ \\
		$\mathcal{C}^{(2)}$ & $1$ & $N-8$ & $N-4$& $N-8$& $\cdots$& $N-8$ & $N/2-4$ \\
		$\mathcal{C}^{(3)}$ & $1$ & $N-8$ & $N-8$& $N-4$& $\cdots$& $N-8$ & $N/2-4$ \\
		$\vdots$ & $\vdots$ & $\vdots$ & $\vdots$ & $\vdots$ & $\ddots$ & $\vdots$ & $\vdots$\\
		$\mathcal{C}^{(L)}$ & $1$ & $N-8$ & $N-8$& $N-8$& $\cdots$& $N-4$ & $N/2$\\
		\hline
	\end{tabular}
    \end{center}
\caption{The modulation factor of the $l$-th collective coupling $\mathcal{C}^{(l)}$ modified by the $k$-th two-spin collective pulse $\mathcal{G}^{(k)}$. The last two columns present $f'_{lL}$ in the system with odd and even spins, respectively. The $k$-th $(k>0)$ collective pulse modulates all the collective couplings in a unit manner, with the exception of the $k$-th coupling.
\label{table:f_collec}}
\end{table}

We take an example $C_{a,a+l} = \sigma_{a}^z\sigma_{a+l}^z$, an element in coupling $\mathcal{C}^{(l)}$, to explain the value of $f'_{lk}$. The involved spins $a$ and $(a+l)$ are only affected by four pulses $R_{\pi}^x(a-k,a)$, $R_{\pi}^x(a,a+k)$, $R_{\pi}^x(a+l-k,a+l)$ and $R_{\pi}^x(a+l,a+l+k)$ in collective pulse $\mathcal{G}^{(k)}$. Thus, if $l\neq k$, all four pulses inverse the coupling $C_{a,a+l}$ and the rest of $n_k-4$ elements in $\mathcal{G}^{(k)}$ do not affect the coupling, which gives the modulation factor of $n_k-8$. If $l=k$, only two unique pulses $R_{\pi}^x(a-k,a)$ and $R_{\pi}^x(a+k,a+2k)$ remain and inverse the coupling, resulting in a modulation factor $n_k-4$. 
One special case is the performance of the last collective pulse $\mathcal{G}^{(L)}$ in the even spin system $N=2L$. If $l\neq N/2$, $R_{\pi}^x(a-L,a)$, $R_{\pi}^x(a,a+L)$, $R_{\pi}^x(a+l-L,a+l)$ and $R_{\pi}^x(a+l,a+l+L)$ reduce to two pulses, $R_{\pi}^x(a-L,a)$, and $R_{\pi}^x(a+l-L,a+l)$ due to the symmetric property $a-L=a+L$. Thus, the modulation factor here is $N/2-4$. If $l=L$, neither of the two pulses inverses coupling $C_{a,a+l}$, giving a factor $N/2$.
	
As demonstrated in Table~\ref{table:f_collec}, a collective pulse $\mathcal{G}^{(k)}$ modulates all collective couplings in the same manner, with the exception of $\mathcal{C}^{(k)}$. Thus, we are able to preserve first-order interactions while nullifying others by choosing only the first two pulses $\mathcal{G}^{(0)}$ and $\mathcal{G}^{(1)}$ with the corresponding time intervals $(\tau_0,\tau_1)=(8-N,1)\tau_1$. The evolution is
\begin{align}
    \mathcal{G}^{(1)}[H_{zz},\tau_1]\mathcal{G}^{(0)}[H_{zz},\tau_0] = \exp\left(-i4g_1\tau_1\mathcal{C}^{(1)}\right).
\end{align}
We get the solution $(\tau_0,\tau_1) = (8-N,1)\pi/4g_1$ by setting $4g_1\tau_1 = \pi$. The total duration of the sequence is 
\begin{align}
    T_c = \tau_0+n_1\tau_1 = \frac{2\pi}{g_1}.
\end{align}
The corresponding preparation sequence of the indices is $\mathcal{S}_N$=\{(0), (1,~2), (2,~3), $\cdots$, ($N$,~1)\}, and the total number of single-spin $\pi$ pulses (including $\pi$ and $\pi^\dagger$) is $2N+2$. 
This two-spin cluster state preparation scheme is consistent with the sequence $\mathcal{S}_6$ we derived in Sec.~\ref{subsec:solvable system}.
It should be noted that this proposal is valid when $N\leq8$, to ensure that $8-N$ is a positive integer. The larger spin system will be discussed in the following text.  

Here, we extend our protocol to the generation of cluster states in a larger spin-ring system with $N>8$. 
The principle of the preparation requires multi-spin (up to $\mathcal{N}_g$ spins) pulses to increase the variety of coupling modulations, as opposed to relying solely on two-spin pulses.
Similarly, we define the $k$-th collective pulse $\mathcal{G}_{m_k}^{(k)}=\{R^{x}_\pi(\mathbf{i}_k^{(1)}),~R^{x}_\pi(\mathbf{i}_k^{(2)}),~\cdots,R^{x}_\pi(\mathbf{i}_k^{(N_k)})\}$ as a set of $m_k$-spin pulses possessing the $k$-th smallest internal distance $\mathcal{D}_k$. Here $\mathcal{D}=\sum_{m} \left|\vec{r}_{i_{m+1}}-\vec{r}_{i_m}\right|$ is the sum of the distance between every two nearby spins involved by a selective $\pi$ pulse $R^x_\pi({\mathbf{i}})$ with all indices in the vector $\textbf{i} = \{i_1,~i_2,~\cdots\}$ are listed in ascending order. 

Due to the symmetry of the quantum spin-ring system, the collective pulse modifies the free evolution in $k$-th segment as
\begin{align}
    \nonumber
    \mathcal{G}_{m_k}^{(k)}[H_{zz},\tau_k] &= \prod_{i=1}^{N_k} R^{-x}_\pi(\mathbf{i}_k^{(i)}) U_{zz}(\tau_k) R^{x}_\pi(\mathbf{i}_k^{(i)}) \\ 
    &= \exp\left(-i\sum_{l=1}^{L}\tilde{f}_{lk}g_l\mathcal{C}^{(l)}\tau_k\right).
\end{align}
The factor $\tilde{f}_{lk}$ quantifies the modulation achieved by the collective gate $\mathcal{G}_{m_k}^{(k)}$ to the $l$-th collective coupling. 
We apply the collective pulses sequentially with the corresponding sequence is ($k_1$, $k_2$,$ \cdots$, $k_{n_{\text{c}}}$), resulting in an overall unitary $U(T_c) = \prod_{j=1}^{n_{\text{c}}}\mathcal{G}_{m_{k_j}}^{(k_j)}[H_{zz},\tau_j]$ with the duration $T_c = \sum_{j=1}^{n_{\text{c}}}N_{k_j}\tau_j$. Here, $N_{k_j}$ denotes the number of $m_{k_j}$-spin pulses included in $\mathcal{G}_{m_{k_j}}^{(k_j)}$ and $\tau_j$ is the corresponding time interval.
Similar to Eq.~(\ref{eq:F_tau_alpha}), the linear equation is 
\begin{align}
    \tilde{M}\cdot \vec{\tau} = \vec{\alpha}.
    \label{eq:F_tau_alpha2}
\end{align}
Here, $\tilde{M}\in\mathbb{R}^{o\times n_{\text{c}}}$ represents the modulation factors corresponding to the $n_{\text{c}}$ collective pulses employed; $\vec{\tau}\in\mathbb{R}^{n_{\text{c}}\times 1}$ indicates the time interval within each segment and the target phases $\vec{\alpha}=0$ except the first element $\alpha(1) = \pi/g_1$.

While the utilization of collective gates to modify the coupling has notably simplified the computational complexity, the increase of the system size presents challenges. As the size of the system increases, the number of long-range couplings and multi-spin pulses increases quadratically and polynomially, respectively. Although serendipity may lead to a few solutions for quantum multi-spin systems, these solutions are not inherently optimal. Ideally, the number of involved single-spin pulses should be minimized, and the total evolution time should be kept to a minimum. 

\begin{figure}[t]
    \centering
    \includegraphics[width=0.48\textwidth]{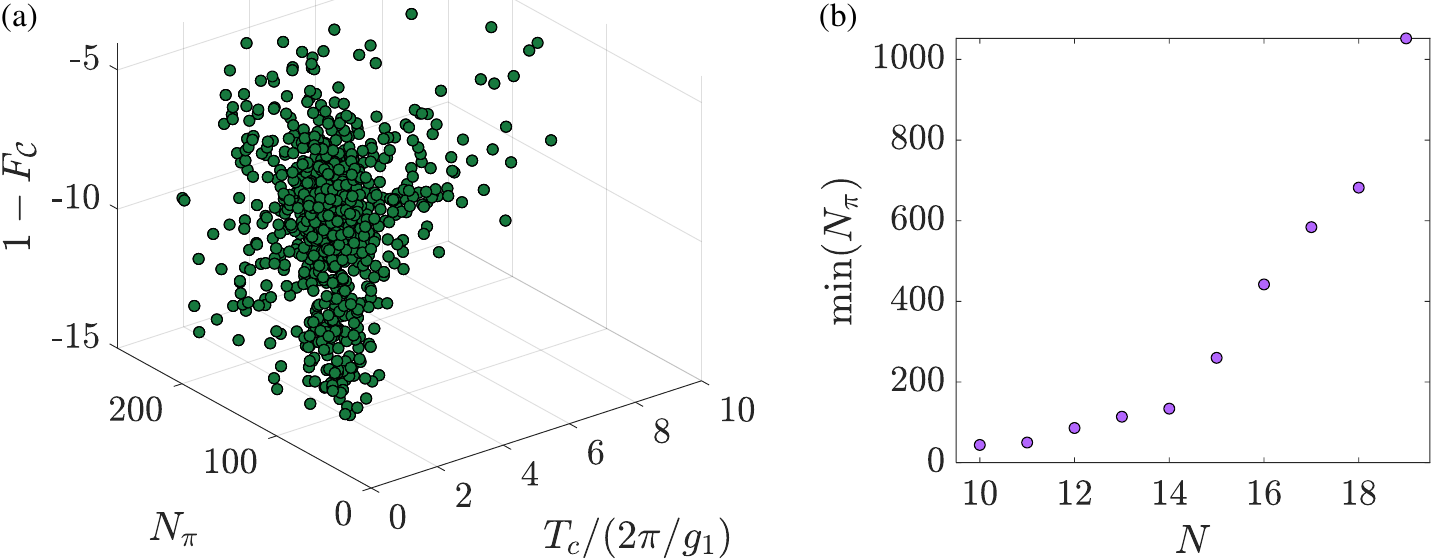}
    \caption{The preparation of cluster states in the larger spin-ring system. (a) The fidelity of the state preparation of the solutions obtained within a ten-spin system, along with the corresponding number of single-spin $\pi_x$ pulses and duration of the control field, $T_c$. (b) The minimum number of single-spin $\pi_x$ pulses required to prepare a cluster state in a multi-spin system, which increases with system size.}
    \label{fig_larger_system}
\end{figure}

The key challenge lies in developing an efficient search algorithm to address the scalability of our cluster state preparation proposal. 
In this context, we propose a feasible numerical solution: initially selecting $n_c=L-1$ elements from all collective pulses randomly, then finding the Least-Square-Solution (LSS) $\vec{\tau}_{\text{LSS}}$ of Eq.~(\ref{eq:F_tau_alpha2}), which minimizes the relative residual $\sigma=\|\tilde{M}\cdot\vec{\tau}_{\text{LSS}}-\vec{\alpha}\|_2/\|\vec{\alpha}\|_2$ ($\|\vec{x}\|_2 = \sqrt{\vec{x}^Tx}$). Finally, from all possible solutions, we identify the ones that satisfy the following criteria 
(i) positive time intervals; (ii) low residual, with a value no greater than a specified threshold $\sigma_0$; and (iii) limited duration, with a maximum value of a specified limit $T_0$.
The number of single-spin pulses can be further reduced by following two steps. Firstly, the segments whose time intervals are less than a certain value  $\tau_{\text{min}}$, are removed. Secondly, all the multi-spin pulses involved are arranged to cancel the pulses between two adjacent segments as much as possible.
In Fig.~\ref{fig_larger_system}(a), we present the solutions with high preparation fidelity ($F_{\mathcal{C}}>0.99999$) we get in ten-spin system, as well as the number of single-spin pulses and duration $T_c$. Furthermore, we show the minimal number of $N_{\pi}$ as a function of the size of the system size in Fig.~\ref{fig_larger_system}(b), with the parameters $\sigma_0=0.1,~T_0 = 10T_1,$ and $\tau_{\text{min}}=10^{-5}T_1$ with $T_1 = 2\pi/g_1$.  
   
\section{Potential experimental implementations}
\label{sec:implementations}
In this section, we present a potential implementation of our proposal for preparing a pulsed cluster state using an NV center system as an illustrative example. In principle, our proposal is well suited to entangling the quantum spin system using the inherent mutual $zz$ interactions as the resource. The NV center, an artificial defect in the diamond, represents a promising platform due to the desired $zz$ couplings arising from mutual direct magnetic dipole interactions. Moreover, nearby nuclear spins could be involved in the same interaction framework.

We begin this section by briefly introducing the NV center, outlining relevant practical challenges, and deriving the interaction Hamiltonian~(\ref{eq:ZZ_Hamiltonian}). Then, we develop the broadband and selective pulses necessary for our proposal, employing optimal control methods, as well as composite and shaped pulses. In the last, we conduct the numerical simulation of our proposal in the NV center system.

\subsection{NV centers in diamond}
\label{sec:NV}
NV centers in diamond are well-studied point defects with electronic spin-1 ground states that can be manipulated using microwave fields and initialized and read out optically. The electronic spin properties combined with the ability to couple to nearby nuclear spins make the NV center a promising candidate for scalable quantum networks and computing architectures~\cite{Doherty2013, Dolde2014, Bradley2019,Xie2023, Bartling2024, Joas2025}. However, several practical challenges must be addressed when developing a matter-photon quantum information processor based on NV centers. 

Firstly, achieving the desired Hamilton (\ref{eq:ZZ_Hamiltonian}) relies on the magnetic dipole-dipole interaction between NV centers, requiring nanometer-scale precision control of the positions of defects to achieve appropriate coupling strengths. In particular, the deterministic preparation of color centers in large numbers represents a significant challenge for the realization of scalable quantum computing. This constraint is somewhat relaxed in FBQC, where at most six spins are required in a single RSG. For scalable solid-state spin registers in diamond, a key requirement is the ability to position a small number of NV centers at nanometer-scale separations. A well-established method for this purpose is the nitrogen ion implantation process, where nitrogen ions are accelerated and implanted into diamond through a nano-hole to create dipolar-coupled NV centers. By optimizing the mask geometry and implantation parameters, interacting NV pairs with separations down to a few nanometers have been demonstrated~\cite{Jakobi2016}. Moreover, tailored ion beams and deterministic single-ion sources provide additional control over extraction voltage, beam focusing, and beam monochromaticity, further improving spatial accuracy and implantation yield for the creation of color centers~\cite{Tobalina2022, Jacob2016, Groot-Berning2019}. The number of coupled NV centers could be further increased by replacing the ion beam with ionized nitrogen-rich molecules. Consequently, the positions of NV centers are determined by the intrinsic structure of the molecule and implantation parameters~\cite{Haruyama2019, Joas2025}, offering a potential route toward the fabrication of regular spin arrays. A recent development in the field involves the fabrication of a system comprising four coupled NV centers, with a separation distance of approximately 9~nm~\cite{Kimura2022}. Moreover, coupled NV-center pairs exhibit to be entangled with a reported fidelity of 96\% in Ref.~\cite{Joas2025}, and the separation of the NV pair can be rapidly determined optically, as demonstrated in Ref.~\cite{Reinhardt2024}.

Secondly, the implementation of selective pulses is another issue to address. The ability to selectively address, manipulate and read out individual spins is fundamental to the development of matter-based quantum information processors. Constrained by the distance between NV centers, frequency-selective operations offer a viable alternative. Meanwhile, the coupling strengths can be extracted using the double electron-electron resonance technique~\cite{Grotz2011, Du2024}, which is essential for determining the preparation sequence.

Thirdly, the preparation proposal must be robust against noise, typically being consistent with the present methods to suppress the decoherence, such as the dynamical decoupling. The last challenge relates to increasing the low photonic connection efficiency between spatially separated registers, which is constrained by the low zero-phonon line (ZPL) emission efficiency and the low collection efficiency. Techniques such as lowering the temperature, placing the NV center in an optical resonant cavity, and fabricating diamond nanostructures can enhance the probability of detecting ZPL photons.

These challenges are specific to an NV center-based quantum information processor. In our work, focusing on state preparation, the main concern is realizing the selectivity to suppress the long-range interactions. Based on the designed broadband and selective pulses, the pulsed cluster proposal demonstrates robustness against uncertainties in spin positions, imperfections of the control fields and the dephasing noise.

\subsection{Hamiltonian}
We explore the preparation of a cluster state in a multi-spin system, where each spin is encoded in the electronic ground state of the NV center. Ideally, these NV centers are positioned close together, equidistant from one another, and aligned in the same plane with their orientations perpendicular to this plane.
Each NV center represents a spin $S=1$ system with three ground states that can be initialized, manipulated, and read out under ambient conditions. 

The Hamiltonian of the system is
\begin{align}
    H = \sum_{i}H_i^0 + \sum_{ij} H_{ij}^{\text{dd}}+H_c(t),
\end{align}
where $H_i^0$ is the Hamiltonian for the NV center at site $i$ under an external static magnetic field $B_z$ applied along $z$ to lift the degeneracy of the spin states $m_s = \pm1$
\begin{align}
    H_i^0 = (D_0 +\delta_i^D) (S_i^z)^2 - (\gamma_e B_z +\delta_i^z) S_i^z,
\end{align}
with $D_0=(2\pi)2.87$ GHz is the zero-field splitting, $\gamma_e$ is electronic spin gyromagnetic ratio. $\delta_i^D$ and $\delta_i^z$ present the energy disorder of site $i$, primarily attributed to lattice strain, the positioning of the NV centers and the inclusion of scattered paramagnetic impurities~\cite{Choi2017, Kucsko2018}.
A time-dependent global driving $H_c(t)=\sum_{i}\sqrt{2}\Omega(t)\cos[\omega_c t + \phi(t)]S_i^x$ is applied to the system, which generates all the control pulses required. And $H_{ij}^{\text{dd}} $ is the magnetic dipole-dipole interaction between spin $i$ and $j$. 

In our system, each spin-1/2 is encoded in the ground state of NV center as $\ket{0(1)} = \ket{m_s= +1(0)}$. 
Now moving to the interaction picture with respect to $H_0 = \sum_i \omega_c\sigma_i^z/2$, and applying rotating wave approximations (RWA), we obtain the effective Hamiltonian~\cite{Cai2013}
\begin{align}
    \nonumber
    H_I = \sum_i & \frac{\Delta_i}{2} \sigma_i^z + \sum_{ij} \frac{g_{ij}}{4}\sigma_i^z\sigma_j^z-\frac{g_{ij}}{2}\left(\sigma_i^+\sigma_j^-+\sigma_i^-\sigma_j^+\right) \\&+ \frac{\Omega(t)}{2}\sum_i \left[\cos\phi(t) \sigma_i^x +\sin\phi(t) \sigma_i^{y}\right],
    \label{eq:HI}
\end{align}
where the detuning $\Delta_i = \omega_i-\omega + \sum_{k\neq i} {g_{ik}}/{4}$ results from the disorder, $\Omega(t)$ and $\phi(t)$ are the amplitude and phase of the global field.

Although the presence of unwanted detunings $\Delta_i$ typically leads to deviation in evolution, it also offers an opportunity for individual spin control, provided that the spins acquire distinct energy splittings $\vec{\omega}=(\omega_1,~\omega_2,~\cdots,~\omega_N)$ which exhibit sufficiently large differences.
Selectivity is achieved by applying a resonant narrow-band $\pi$ pulse, which is composed of two shaped pulses and two broad-band $\pi$ pulses. Further details can be found in Sec.~\ref{subsec:selective_gate}. 
This ability to manipulate individual spins in nanoscale NV arrays paves the way for the realization of scalable quantum information processors~\cite{de_leon_2021}.

In the Hamiltonian, the $zz$ coupling term as delineated in Eq.~(\ref{eq:ZZ_Hamiltonian}) naturally emerges. Thus, a carefully designed evolution process becomes essential to mitigate the presence of undesired flip-flops and detuning terms. The disorder of spin resonant transition and optical transition of each NV center is intrinsic and can be modified by the application of external fields.~\cite{Zhang2017, Bodenstedt2018}. 
We establish a lower bound on the energy splitting difference for every two spins, set as $\Delta_{\text{min}}=(2\pi)4$ MHz. This is defined as $\forall (i,j), |\omega_i-\omega_j|=|\Delta_i-\Delta_j|\geq \Delta_{\text{min}}$. With neighboring spins at a distance of 30 nm, the resulting coupling strength i s $g_{ij} = (2\pi)1.924$ kHz. Thus, the flip-flop effect induced by $\sigma_i^+\sigma_j^-+\sigma_i^-\sigma_j^+$ is highly suppressed due to the large detuning difference $(g_{ij}\ll\Delta_{\text{min}})$.
In addition, the unique energy splitting enables the manipulation of individual spin (see the part of selective pulses).
Therefore, we can write the effective Hamiltonian in a free evolution process as~\cite{Dolde2014}
\begin{align}
    H_z = \sum_i \frac{\Delta_i}{2} \sigma_i^z + \sum_{ij}\frac{g_{ij}}{4}\sigma_i^z\sigma_j^z.
    \label{eq:Hz_NV}
\end{align}

To eliminate the detuning term, we design a basic pulse sequence that consists of a free evolution $U_z(t)=\exp(-iH_z t)$ and a following equal length free evolution sandwiched by two broadband $\pi$ pulses $R^x_{\pi}{(\mathbf{i}_N)}$ and $R^{-x}_{\pi}{(\mathbf{i}_N)}$. Here indices $\mathbf{i}_N = (1,~2,~\cdots,~N)$ indicate all the spins, which means that all spins will be flipped simultaneously. A unitary governed by $zz$ Hamiltonian (\ref{eq:ZZ_Hamiltonian}) is realized as 
\begin{align}
    \nonumber
    U_{zz}(2t) & = R^{-x}_{\pi}{(\mathbf{i}_N)} U(t)R^{x}_{\pi}{(\mathbf{i}_N)}U(t)\\&= \exp{\left(-i \sum_{i\neq j}\frac{g_{ij}t}{2}\sigma_i^z\sigma_j^z\right)}.
    \label{eq:Uzz_2t}
\end{align}
The design of the broadband pulse is explained in Sec.~\ref{subsec_broad-band}.
Note that this process is typically an application of pulsed dynamical decoupling in a multi-spin system~\cite{Bar-Gill2012}, thereby inheriting the property of robustness against the dephasing noise. 

In principle, the decoupling pulse could be replaced by more efficient pulses, such as Carr-Purcell-Meiboom-Gill (CPMG)~\cite{Carr1954, Meiboom1958} sequence or other periodic pulses~\cite{Ryan2010, Souza2011, Ali2013} to further extend the coherence time. Previous studies have reported coherence times of approximately 1~s~\cite{Abobeih2018}, which is a crucial prerequisite for the applicability of our proposal to solid-state systems.

In addition, during the whole process, we have to deal with the inherent imperfections of the control field $H_c(t)$~\cite{Rong2015}. The errors we consider are the offset error $\delta$ and the Rabi frequency error $\varepsilon$. The Hamiltonian with these two errors is 
\begin{align}
    H_c(\varepsilon,\delta,t) &=  \frac{\delta\Omega_0}{2}\sigma_z +\nonumber\\
    &\frac{\Omega(t)(1+\varepsilon)}{2} \left[\cos\phi(t) \sigma_x +\sin\phi(t) \sigma_y\right],
\end{align}
where $\Omega_0 = \max{|\Omega(t)|}$ is the peak value of Rabi frequency $\Omega(t)$.
In the following section, we develop general frameworks for generating broadband and narrowband pulses, which are crucial to the experimental implementation, utilizing the global field $H_c(t)$ and the quantum optimal control method. This approach integrates composite pulse and shaped pulse techniques to achieve the desired pulse characteristics, while also demonstrating robustness against the errors of the control field.
In the simulation of the construction of the pulses, the weak interaction $zz$ terms are ignored since their coupling strengths are much weaker than the Rabi frequency ($\sim$MHz) of the driving field. For simplicity, we denote the single-spin pulses as $\alpha_{\phi} = \exp(-i\alpha\sigma_{\phi}/2)$ and $\alpha_z =\exp(-i\alpha\sigma_{z}/2)$.

\subsection{Broadband pulses} 
\label{subsec_broad-band}
We first design the broadband pulses $\pi_x$ and $(\pi/2)_x$ (i.e $\pi_0$ and $(\pi/2)_0$) employed to flip and initialize all spins simultaneously. To this end, we develop a composite pulse that is robust against the detuning error.
The composite pulse is an easily implemented technique to generate a robust quantum pulse using a series of optimally chosen pulses to cancel out imperfections~\cite{Torosov2021, Levitt1986, Tycko1983, Jones2009, Jones2013}. The pulses inside the composite pulse are realized by a constant Rabi frequency (i.e.,~rectangular pulses) with the Hamiltonian $H = {(\delta\Omega_{\text{b}}/2)}\sigma_z + [{\Omega_{\text{b}}(1+\epsilon)/2}]\sigma_{\phi}$. The composite pulse is designed to be robust against the detuning error with a broad high fidelity $(1-F<10^{-5})$ range $|\delta|<\delta_0$ that covers all spins
\begin{align}
    2\delta_0\Omega_{\text{b}}> \omega_{\max}-\omega_{\min},
\end{align}
with $\omega_{\max(\min)} = \max(\min)\{\vec{\omega}\}$.
The frequency of the driving field is chosen to be $\omega_c=(\omega_{\max}+\omega_{\min})/2$.
We construct a $\pi_x$ gate with a symmetric pulse sequence $U(\vec{\phi}_{n_{\phi}})=\pi_{\phi_1}\pi_{\phi_2}\cdots\pi_{\phi_n}\cdots\pi_{\phi_2}\pi_{\phi_1}$, where the phases $\vec{\phi}_{n_{\phi}}=(\phi_1\cdots\phi_n)$ are the parameters to be optimized. 
To develop a broadband $\pi_x$ gate, we define a collective cost function $G(\vec{\phi}_{n_{\phi}}) = \sum_{i}a_i{g(\delta_i,\vec{\phi}_{n_{\phi}}})$ with $a_i$ is weight of the contribution $g$ at working point $\delta=\delta_i$, and a single component $g(\delta_i,\vec{\phi}_{n_{\phi}})$ is
\begin{align}
    g(\delta_i,\vec{\phi}_{n_{\phi}}) = 1-F_i+ D_i(\delta).
\end{align}
with the fidelity is $F_i=\operatorname{Tr}[\pi_x U^{\dagger}(\delta_i,\vec{\phi}_{n_{\phi}})]/2$.
The derivative is
\begin{align}
    D_i(\delta) = \sum_{k=1}^{k_i}\left\|c^{-k}\left.\frac{\partial^{k} U(\delta_i,\vec{\phi}_{n_{\phi}})}{\partial\delta^{k}}\right|_{\delta = \delta_i}\right\|_F,
\end{align}  
where $k_i$ is the derivative order with respect to $\delta$ at each working point $\delta_i$, the coefficient $c^{-k}$ is the weight of each order of derivative and we usually set $c=10$; $\left\|X\right\|_F=\sqrt{\sum_{ij}|X_{ij}|^2}$ is the Frobenius norm.
By choosing the parameters and minimizing the cost function, a series of broadband composite pulses are generated~(see Fig.~\ref{fig_app_broadband}).
In addition, a broadband $(\pi/2)_x$ is developed by using a different symmetric pulse sequence $U(\alpha,\vec{\phi}_{n_{\phi}})=\alpha_{\phi_1}\pi_{\phi_2}\pi_{\phi_3}\cdots\pi_{\phi_{n_{\phi}}}\cdots\pi_{\phi_3}\pi_{\phi_2}\alpha_{\phi_1}$~\cite{Gevorgyan2021}, where $\alpha$ and $\vec{\phi}_{n_{\phi}}$ are the parameters to optimize. 

\begin{figure}[t]
    \centering
    \includegraphics[width=0.48\textwidth]{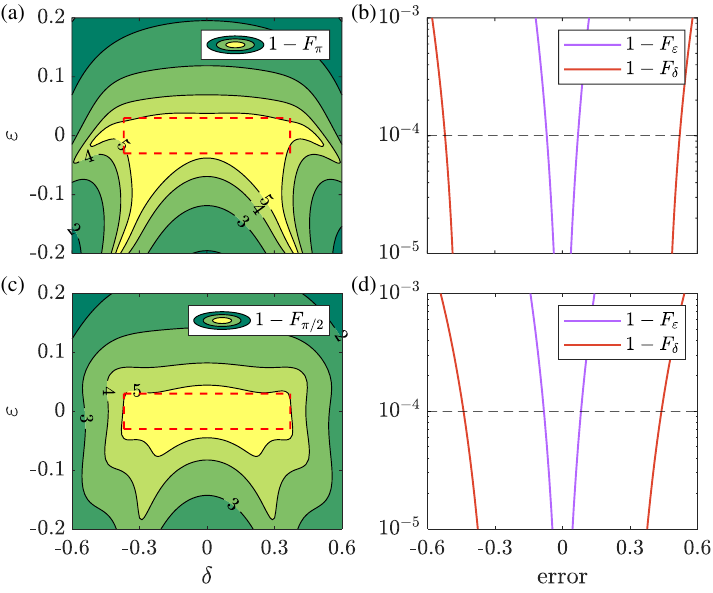}
    \caption{Simulated infidelities between the optimal composite pulses and the target pulses. (a) Infidelity of the composite pulse $\pi_x$ versus static detuning and control amplitude error with $n_{\phi}=10$. (b) The vertical and horizontal cut along $\delta=0$ and $\varepsilon=0$, respectively. (c) and (d) are the same with (a) and (b) but for an optimal $(\pi/2)_x$ gate and the corresponding parameter $n_{\phi}=8$. The integers $m$ in (a,~c) indicate the infidelity $10^{-m}$. Additionally, two regions of high fidelity ($F>1-10^{-5}$) are identified by red dashed boxes, bounded by $|\delta|\leq0.35$ and $|\varepsilon|\leq0.03$. }
    \label{fig_broadband_pulse}
\end{figure}

\begin{figure*}[t]
    \centering
    \includegraphics[width=0.8\textwidth]{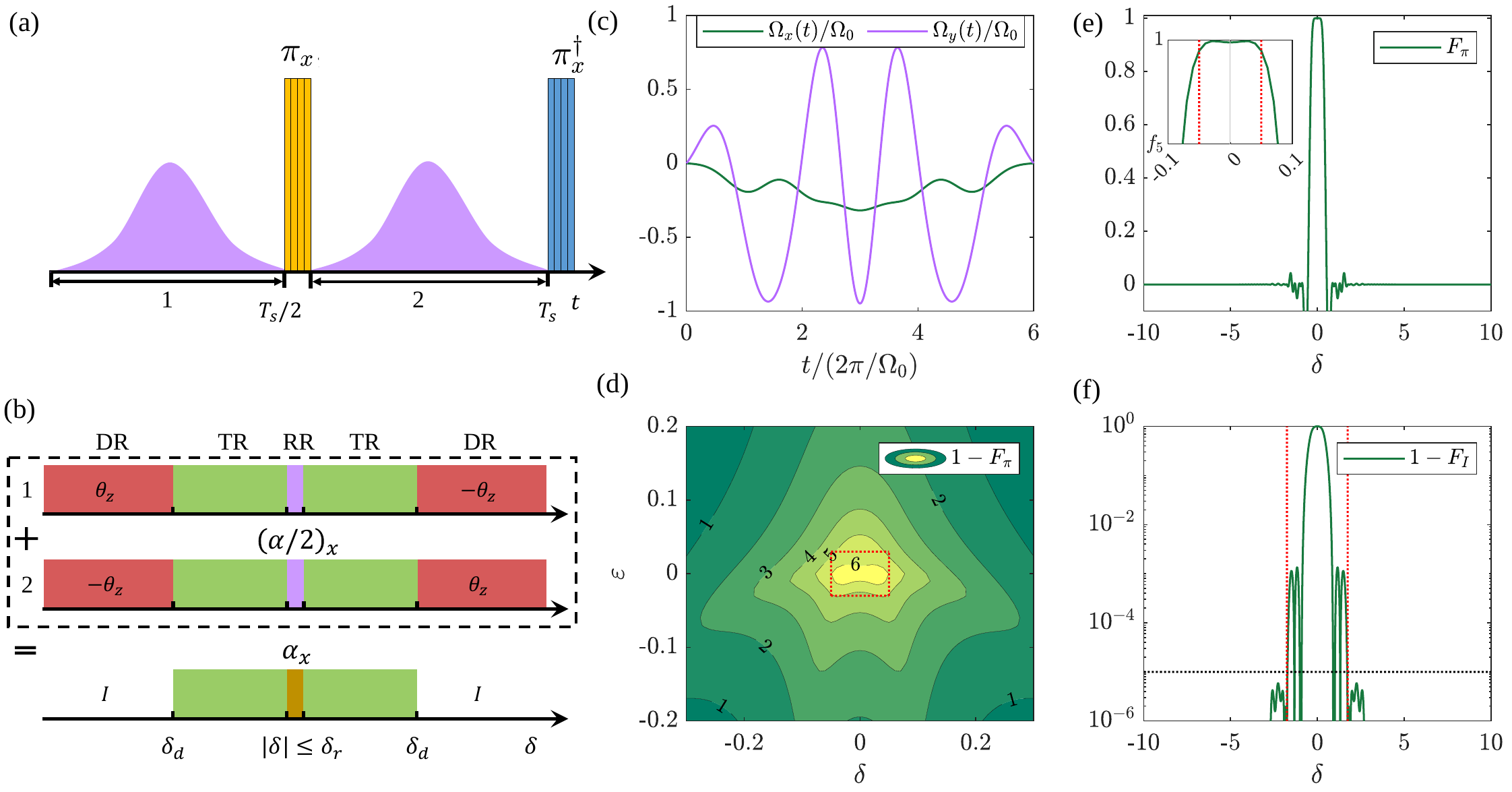}
    \caption{The scheme of the composite selective pulse $\alpha_x$. (a) Time-domain of the selective pulse $\alpha_x$. The pulse is composed of two sequential shaped $(\alpha/2)_x$ (purple curves) and two broadband $\pi_x$ and $\pi_x^\dagger$ pulses (yellow and blue blocks) applied at $t=T_s/2, T_s$ respectively. (b) Frequency domain of the pulse $\alpha_x$. The first two diagrams represent the performances of two pulses in the time domain with the corresponding unitary $(\alpha/2)_x$ and $\pi_x^\dagger(\alpha/2)_x\pi_x$, respectively. In the far detuning region (DR), defined as $|\delta|\geq\delta_d$, the large detuning term $\delta\Omega_0\sigma_z/2$ dominates and results in two nearly pure $z$ pulses $\pm\theta_z$ with $\theta = \delta\Omega_0 T_s$. As shown in the third diagram, the two $z$ pulses ultimately cancel, yielding an identity operator Meanwhile, in the resonant region (RR), $|\delta|\leq\delta_r$, both of the two unitary operators remain $(\alpha/2)_x$ (purple blocks), resulting in a desired $\alpha_x$ pulse. The transition region (TR) is given by $|\delta|\in(\delta_s, \delta_d)$, where the evolution (green blocks) is usually quite different from the desired gate. (c) Gaussian-shaped pulses $\Omega_{x(y)}$ to implement a single $(\alpha/2)_x$ pulse. (d) Contour plots of the simulated infidelity of the selective $\pi_x$ pulse versus $\delta$ and $\varepsilon$. The high fidelity area is highlighted by the red box bounded by $|\delta|=0.05$ and $|\varepsilon|=0.03$. (e) The fidelity between the composite pulse and ideal $\pi_x$ gate as a function of detuning, and the insert shows the fidelity in RR and $f_5=0.99999$. The resonant region is indicated by two red dotted lines $\delta = \pm 0.05$. (f) The infidelity between the composite pulse and identity gate as a function of detuning, and the DRs are defined as  $|\delta|>\delta_d=1.75$ here.}
    \label{fig_select_regime}
\end{figure*}

We present numerical simulations illustrating the infidelities of the optimal composite broadband pulses $\pi_x$ and $(\pi/2)_x$ in Figs.~\ref{fig_broadband_pulse}(a,c) respectively. 
These simulation results demonstrate the robustness of composite pulses against both detuning and amplitude errors. The red dotted boxes in the figures highlight regions of high fidelity, where the relative biases of detuning and control amplitude errors are tolerable, with values of $\delta_0=0.35$ and $\epsilon_0=0.03$, respectively. For a large Rabi frequency of $\Omega_{\text{b}}=(2\pi)40$ MHz, at most 7 spins could be manipulated with high fidelity simultaneously.
Figs.~\ref{fig_broadband_pulse}(b,d) display vertical and horizontal slices of the infidelities along $\delta=0$ and $\varepsilon = 0$, respectively.   
Note that by adding a global phase $\Phi$ to the optimal phase $\vec{\phi}_{n_{\phi}}$, arbitrary broadband gates $\pi_\Phi$ and $(\pi/2)_\Phi$ are obtained. These various broadband pulses $\pi$ and $\pi/2$ enable the implementation of pulse sequences such as the XY-series pulse~\cite{Maudsley1986, Gullion1990, Casanova2015}.

\subsection{Selective pulses} 
\label{subsec:selective_gate}
Now we introduce the approach to implement selective pulses that are the building blocks of our proposal. 
These tailored pulses ensure precise control over the state of the targeted spin while minimizing disruption to the states of other spins.
Within our framework, the unique energy splitting exhibited by each spin allows the manipulation of spin $j$ at the spatial site to be equated with the application of a resonant narrowband pulse that closely matches the intrinsic frequency $\omega_j$.
In the absence of spin interactions, the system Hamiltonian (\ref{eq:HI}) simplifies to the single-spin case 
\begin{align}
    H_{\text{s}}(t) = \frac{\delta \Omega_0}{2} \sigma_z + \frac{\Omega_x(t)}{2}\sigma_x + \frac{\Omega_y(t)}{2}\sigma_y,
\end{align}
resulting in an unitary $U_{\text{s}}=\mathcal{T}e^{-i\int H_{\text{s}}(\tau) d\tau }$.  	

The third diagram depicted in Fig.~\ref{fig_select_regime}(b) illustrates the application of a desired selective $\alpha_x$ (i.e.,~$U_{\text{s}}=\alpha_x$) to the spin in the resonant region (RR), defined as $|\delta| \leq \delta_r$.
This RR exhibits a finite width, which serves to maintain the accuracy of a selective pulse in the presence of an offset error. On the other side, the states of spins within the far detuning regions (DR), defined as $|\delta|\geq\delta_d$, remain unperturbed. Thus, to implement a selective $\alpha_x$ to a specific spin, it is necessary to ensure that the driving is nearly resonant to the target spin, and that the detunings of the rest spins are strictly within the DR.
This condition is naturally satisfied by setting the minimum disorder value greater than the TR width ($\Delta_{\min}\geq\delta_d\Omega_0$).

Fig.~\ref{fig_select_regime}(a) illustrates the scheme of the composite selective pulse $\alpha_x$ that is composed of two distinct processes. The first is a single $(\alpha/2)_x$ pulse, while the second is a $(\alpha/2)_x$ pulse sandwiched between two broadband $\pi_x$ and $\pi_x^{\dagger}$ pulses.
Fig.~\ref{fig_select_regime}(b) demonstrates the details of the selective pulse in the frequency domain, which is divided into five ranges by $\delta_r$ and $\delta_d$.
In RR, the evolution of the entire process is $\pi_x^\dagger\left({\alpha}/{2}\right)_x\pi_x\left({\alpha}/{2}\right)_x = \alpha_x$,
which is the desired gate. However, within DR, the offset term $\delta\Omega_0\sigma_z/2$ dominates the Hamiltonian, resulting in a nearly pure $z$ rotation. Thus the evolution in DR is $\pi_x^\dagger\theta_z\pi_x\theta_z=\mathbbm{1}$ with $\theta = \delta\Omega_0 T_s/2$. 
Lastly, the evolution operator of the shaped pulse in the transition region (TR), defined as $|\delta|\in(\delta_s, \delta_d)$, is of no consequence.

At this point, the objective is to design the pulses to implement the $\alpha_x$ and identities within RR and DR, respectively. The values of $\delta_d$ and $\delta_r$ characterize the selectivity in the frequency domain and robustness against the detuning, respectively.
Therefore, it is recommended to obtain a relatively low value of $\delta_d$ to achieve high selectivity, while a larger value of $\delta_d$ to ensure robustness against detuning in the optimization process. We introduce a shaped pulse control field to achieve these objectives, enabling additional robustness against the Rabi frequency fluctuation error $\varepsilon$. 
Two components of the shaped pulse are both the sum of half-sine functions that are modulated by a Gaussian envelope~\cite{Steffen2007}
\begin{align}
    \label{eq_amplitude}
    \tilde{\Omega}_{x(y)}(t) = \sum_{k=1}^{N_{a(b)}} a(b)_k \sin\left[(2k-1)\omega_s t\right] e^{-\frac{(t-T_s/4)^2}{2\sigma^2}}, 
\end{align}
where $\omega_s = 2\pi/T_s$, $\vec{a}$, $\vec{b}$ and $\sigma$ are the parameters to generate a smooth pulse.
Gaussian-shaped functions in the time domain usually result in selectivity in the frequency domain, which has been widely used~\cite{Vandersypen2005, Haase2018}.
Employing sine components guarantees that the Rabi frequency $\tilde{\Omega}_{x(y)}$ begins and ends with zero values.
To limit the magnitude of the Rabi frequency to the value $\Omega_0$, both components are adjusted relative to a normalization factor, $\Omega_{x(y)}(t) = r\tilde{\Omega}_{x(y)}(t)$, where $r = {\Omega_0}/{\max\left(\sqrt{\tilde{\Omega}_x^2(t)+\tilde{\Omega}_y^2(t)}\right)}$. 
In the presence of errors $\delta$ and $\varepsilon$, we define a cost function
\begin{align}
    \label{eq_cost_selective}
    G = \bar{I}_{I} + \bar{I}_{\alpha}+ c \bar{D}(\varepsilon), 	
\end{align}
with the quantities $\bar{I}_{\alpha}$ and $\bar{I}_{I}$ represent the average infidelities associated with the desired $\alpha_x$ and identity gate within RR and DR, respectively. 
The third term is the average derivative with respect to $\varepsilon$ in RR and $c$ is the weight.

By minimizing the cost function, the obtained optimal pulse shapes are determined by the parameters $\vec{a}$ and $\vec{b}$. The shaped pulses and the resulting infidelity $1-F_{\pi}$ associated with $\pi_x$ near RR are shown in Fig.~\ref{fig_select_regime}(c) and (d). We highlight the high-fidelity area with the red dotted box that is bounded by $|\delta|\leq\delta_r=0.05$ and $|\varepsilon|\leq0.03$. The horizontal cut of $F_{\pi}$ along $\varepsilon=0$ is shown in Fig.~\ref{fig_select_regime}(e). Furthermore, we show the infidelity associated with the identity operator in Fig.~\ref{fig_select_regime}(f). The corresponding boundary of the DR, denoted by the red dotted lines, is located at $\delta_d=1.75$. 
The narrow peaks of fidelity $F_\pi$ and infidelity $I_{I}$ in the vicinity of $\delta=0$ illustrate the frequency selectivity of the optimal pulse, which allows the realization of the cluster state preparation proposal. More details and parameters can be found in Appendix~\ref{app_sec_optimization}. In addition, an optimal selective $(\pi/2)_x$ pulse is presented in Fig.~\ref{fig_piover2_selective}.
It is worth noting that we use composite pulses to generate selective $\pi$ pulse coupling modulation, to demonstrate more intuitively the importance of selectivity.
However, a simpler modulation way with only the Gaussian-shaped pulse (\ref{eq_amplitude}) to implement $\pi_x$ pulse within RR, regardless of $z$ rotations within DR, is achievable, given that all these local $z$ rotations commute with the evolution of each segment.

\subsection{Numerical simulation of the preparation protocol}
\label{subsec:numerical}
We complete this section with the numerical simulations for both the four- and six-spin systems using the broadband and selective pulses we developed. The systems we consider here are both spin-ring systems, in which the position of each spin slightly shifts from the desired position. Thus, the position-dependent coupling strengths are determined, giving us the details of the time intervals $\vec{\tau}$. Using the pulse sequence $\mathcal{S}_4$ and $\mathcal{S}_6$ we developed, we get the total evolution process in the presence of errors are 
\begin{align}
    U_{\mathcal{C}}(\vec{\tau},\delta,\varepsilon) = \prod_{k} R_{\pi}^x (\textbf{i}^{(k)}) e^{-i H_{zz} \tau_k} R_{\pi}^{-x} (\textbf{i}^{(k)}).
\end{align}
In our simulation, the selective $\pi$ pulse $R_{\pi}^x (\textbf{i}^{(k)})$ is applied sequentially as $R_{\pi}^x (\textbf{i}^{(k)}) = R_{\pi}^x(i^{(k)}_{n_k})\cdots R_{\pi}^x(i^{(k)}_2)R_{\pi}^x(i^{(k)}_1)$ to each spin involved in $\textbf{i}^{(k)}$. We choose the maximum value of the selective pulses as $\Omega_0=\Delta_{\min}/\delta_d=(2\pi)$ 2.28 MHz resulting in $T_s = 6.2132~\mu$s. The Rabi frequency in the broadband pulse is $\Omega_{\text{b}}/2\pi = 30(40)$ MHz for the four(six)-spin system.
The achievable fidelity of the preparation $F_{\mathcal{C}}=|\bra{\psi_0}U^\dagger_{\mathcal{C}} U_{\text{t}}\ket{\psi_0}|^2$ depends on the fidelity of each selective pulse $R_{\pi}^x (\textbf{i}^{(k)})$, usually constrained by the presence of $zz$ crosstalk. In this context, we propose an optimization procedure to mitigate the impact of $zz$ coupling through adjustments in the time intervals $\vec{\tau}_o$.
The optimal procedure is demonstrated under the non-error conditions
\begin{align}
    1-F_{\text{op}}=\min_{\vec{\tau}}\frac{1}{2^N} \operatorname{Tr}\left\{U_t U_{\mathcal{C}}^{\dagger}(\vec{\tau},\delta=0,\varepsilon=0)\right\},
    \label{eq_op_last}
\end{align}
with $F_{\text{op}}$ is the optimal cluster state preparation fidelity.
It is pertinent to note that the fidelity of preparation in the spin-lattice system should be comparable to that in the spin-ring system, given that they both require identical pulse sequences albeit with differing intervals. Moreover, the fidelity of cluster state preparation can be further enhanced by applying a series of proper local unitary operations on each spin. 
\begin{figure}[t]
    \centering
    \includegraphics[width=0.49\textwidth]{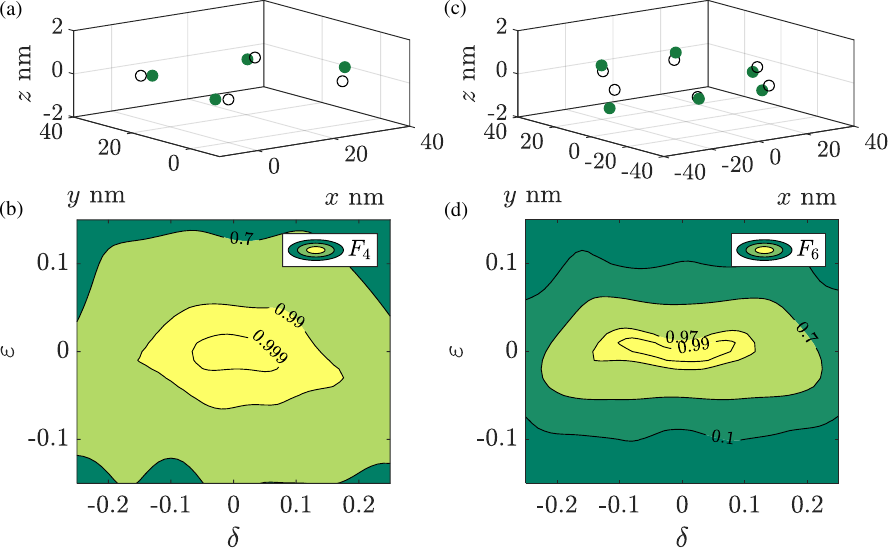}
    \caption{Simulated fidelities of cluster state preparation in four and six-spin systems. (a,c) The green solid circles represent the positions of spin, while the black blank circles denote the ideal positions. (b,d) Contour plots of the state preparation fidelities $F_{\text{op}}$ versus detuning error $\delta$ and Rabi frequency error $\varepsilon$. The values of the fidelities in the four- and six-spin systems are 0.9997 and 0.9937 in non-error cases, respectively.}
    \label{fig_simulation}
\end{figure}

Here, we show the simulated optimal fidelities within four and six spin-ring systems in Fig.~\ref{fig_simulation}. In Fig.~\ref{fig_simulation}(a,c), the green dots and black circles denote the positions of individual spins and their corresponding ideal positions, respectively. By choosing the total time of the cluster state preparation  $T_{c} = 4\pi/g_1=1.0395$ ms, the optimal time intervals $\vec{\tau}_o$ are determined by demonstrating the optimization process (\ref{eq_op_last}), the corresponding contour plots of preparation fidelity is shown in Fig.~\ref{fig_simulation}b(d), in which a high fidelity area over 0.999(0.99) exists in four(six)-spin system. The values of fidelity without error are 0.9997 and 0.9937. More details of the simulation are given in Appendix~\ref{app_sec_optimization}. Moreover, in Appendix~\ref{app_sec_erorr}, we perform numerical evaluations of the impact of possible error sources during the preparation process. The results indicate that the preparation scheme maintains high fidelity across the range of experimentally feasible parameters. 

\subsection{Towards the hybrid quantum computing}
We conclude this section by discussing the potential approaches for developing a hybrid FBQC using spin-ring systems as resource state generators (RSGs). Once the resource states are prepared, two main steps are required to complete the entangling fusion measurement. First, the target spins within each RSG must be entangled with photonic qubits. Then, the photons from two RSGs are collected and interfered at a beam splitter to carry out the required fusion measurements. 

Several schemes have been proposed and implemented with the objective of entangling photons and spins~\cite{Johnson2017, Aharonovich2011, Beukers2024, Ruf2021a}. These schemes are based on the principle that certain degrees of freedom inherent to photons exhibit responses conditional to spin states and transitions. Typically, the spin systems are embedded in separate optical cavities to enhance spin-photon interference and improve photon-collecting efficiency. In our approach, the selective pulses that we have developed are used to populate the state of the target spin to the far-detuned third $\ket{a}=\ket{m_s=-1}$ level. This enables selective entanglement of the target spin with photons, owing to the significantly different optical transition of the target spin. Additionally, we assume that the energy splittings between $\ket{1}$ and $\ket{a}$ of each spin in an RSG exhibit sufficiently large differences, which is essential to ensure the effectiveness of the selective pulses.

One entanglement method starts by applying a selective $\pi$ pulse to flip the target spin from $\ket{1}=\ket{m_s=0}$ to $\ket{a}$. Subsequently, the state $\ket{a}\ket{v}$ couples with $\ket{1}\ket{f}$ (i.e.,~the vacuum Rabi oscillation) when the cavity mode is resonant with the transition between $\ket{1}$ and $\ket{a}$. Here, $\ket{v}$ represents the vacuum state of the cavity mode, and $\ket{f}$ denotes the single-photon Fock state in the cavity~\cite{Bose1999, Browne2003, Barrett2005}. An alternative entanglement protocol involves photon emission from an excited state. For example, we could use a short laser pulse to excite $\ket{1}$ to the excited state $\ket{E_{x/y}}$ presenting the same spin projection. The spontaneous emission back to $\ket{1}$ then directly entangles the spin with the number of photons~\cite{Bernien2013, Hensen2015, Pompili2021}. 

Once the spin-photon entangled states have been prepared, time must be allowed for the photons to leak from the cavities. The required entangling measurements are then performed using linear quantum optical methods~\cite{Kok2007, Reiserer2015}. A common technique is "path erasure", where photons are directed to interfere on a beam splitter to remove the which-path information \cite{Bose1999}. The detection of a single photon in this setup probabilistically projects the spins in the two RSGs into an entangled state. The success of the path erasure depends on the photons being indistinguishable in all their degrees of freedom, including the frequency, spatial wave packet, polarization, and arrival time. To ensure that photons from different cavities share identical frequencies, the ability to tune optical transitions with local external electromagnetic fields is essential~\cite{Bernien2013, Hensen2015}.

In the spin-photon entanglement scheme based on the conditional amplitude reflection~\cite{Nemoto2014, Bhaskar2020}, satisfying the indistinguishability condition may be less challenging, because the photons are generated from the same coherent source. In this scheme, the photon is reflected by a high-cooperativity cavity when the transition of the target spin is resonant with the cavity. This mechanism entangles the spin state and photon numbers. The required high cooperativity not only facilitates entanglement generation but also demonstrates robustness against detuning. The cooperativity $C\sim1$~\cite{Riedel2017, Yurgens2024} achieved thus far by coupling a single NV center with an optical cavity is significantly lower than the requisite level of high cooperativity. However, it is possible to achieve high entanglement fidelity by sacrificing entanglement probability using the reflected or transmitted photons as resources. Ref.~\cite{Omlor2025} presents an entanglement protocol based on reflection from two remote NV–cavity nodes. A single photon splits in a Mach–Zehnder interferometer and is reflected from both nodes; the conditional detection in one output port then heralds an entangled state between the two NV centers. By appropriately choosing the cavity decay rate and mirror asymmetry, the scheme ensures the high-fidelity heralded entanglement with moderate cooperativity. Ref.~\cite{Koshino2012} proposes another entangling scheme based on the transmitted photon through a symmetric cavity, operating within the bad cavity regime and with a homogeneously broadened matter spin. Moreover, the frequency-bin encoding scheme provides an alternative method for spin and photon entanglement without employing indistinguishable photons. In this approach, the target spins in two RSGs are resonant with the corresponding cavity, and the photonic qubit is encoded in two distinct frequencies, typically enabling a potential platform for the photonic quantum information process~\cite{Lukens2017, Clementi2023}.

The atom-like properties of the NV centers enable the application of photonic connection schemes originally developed on the basis of a cavity with single atoms and optical photons. However, in our configuration, a basic register consists of a spin-ring system placed in a cavity. Therefore, existing connection schemes need to be modified to achieve selective connectivity associated with the target spin. One advantage of this configuration is that it requires fewer resources to construct a quantum network, as it demands fewer optical connections.

Finally, additional optimizations or improvements may be necessary to enhance the probability of establishing a successful connection. These may include implementing a repeat-until-success protocol with timing encoding bin, improving photon detection efficiency, and addressing photon loss. The resource state generation methodology and photonic connection techniques suggest the possibility of developing a hybrid quantum computing technology that leverages both matter spins and light.

\section{Conclusion and outlook}	
\label{sec:conclusion}
In conclusion, we establish a scheme to generate cluster states within the quantum spin system, where spins are spatially localized and separated. The principle of our proposal is to design an optimal pulse control sequence that exhibits non-uniform modulation among two-spin couplings in each segment, eventually preserving the NN couplings while eliminating all long-range ones. Our proposal applies to systems of varying configurations, including one-dimensional and even high-dimensional systems. We theoretically derive the corresponding pulse sequences for few-spin systems, accounting for the possible positional errors of each spin, and explore the feasibility of extending this approach to larger systems. The present preparation proposal is a versatile tool and can be extended to more general systems with similar forms of interaction, such as other single defects and molecular spin systems etc. To illustrate our proposal in a solid-state system, we develop the broadband and selective gates required in the pulse sequence using composite pulse and pulse-shaped techniques with experimentally feasible parameters. All pulses are designed to be robust against the static offset detuning and Rabi frequency fluctuations to mitigate inaccuracies caused by these control imperfections. 

Our preparation scheme facilitates the realization of various quantum information scenarios. The first point is the construction of the hybrid matter-photon quantum computing architecture. The preparation of four- and six-ring cluster states in NV center systems represents a crucial first step towards FBQC, serving as RSGs~\cite{Bartolucci2023}. The second step involves implementing an entangling fusion measurement between two spins located in distinct resource states. This is achieved by placing each in an optical cavity and employing probabilistic linear optical operations~\cite{Nemoto2014}. Furthermore, the proposed scheme offers a candidate quantum control method for the direct interaction-based spin systems~\cite{Tscherbul2023}, with the potential to advance quantum information processing in spin arrays~\cite{Le2023}.

The second potential application is that the capacity to exert selective control over the system represents a possibility for building a quantum register within the solid-state spin system. Firstly, the obtained solutions of the linear equation~(\ref{eq:F_tau_alpha}) allow the preparation of different entangled states with the various target phase vectors $\Vec{\alpha}$. This approach is expected to facilitate the generation of the entangled states, such as the weighted cluster state, Greenberger–Horne–Zeilinger states, and even provides the possibility of simulating high-dimensional entangled states within a lower-dimensional system~\cite{Rajabi2019}. Secondly, the selective pulses we developed enable three basic single-spin gates to be implemented as $\theta_x =(\pi/2)_{\pi/2} \pi_{(\theta-\pi)/2}(\pi/2)_{\pi/2}$, $\theta_y=(\pi/2)_{\pi} \pi_{\theta/2}(\pi/2)_{\pi}$ and $\theta_z=\pi_{\pi+\theta/2}\pi_0$~\cite{Torosov2014}. 
And a selective controlled-$Z$ gate could be implemented by setting one element, $\alpha_{ij}$, equal to $\pi$. Consequently, a set of universal quantum gates is implemented~\cite{Barenco1995}. Lastly, the sizes of the system could be extended by involving the proximal nuclear spins (e.g., N, ${}^{13}$C)~\cite{Dutt2007, Neumann2010, Cai2013, Wu2019}.

We would like to stress that our approach is versatile and can be translated directly to other systems such as molecular color centers in suitable organic molecules. As an example, the optically addressable chromium ion (Cr$^{4+}$)-based molecular qubit exhibits a spin-1 ground state, which can be manipulated by microwave and tuned by strong field (aryl) ligands in a high-symmetry configuration~\cite{Bayliss2020}. The established synthetic chemical method allows for precise control over the positions of molecular qubits, thereby ensuring the emergence of dipole interactions~\cite{Gaita2019}. Furthermore, the coherence time of the molecular qubit can be significantly extended by the symmetry control over the host crystal surroundings~\cite{Bayliss2022}. Due to its excellent spin properties and rich degrees of control, it offers an alternative platform for implementing our cluster state preparation scheme and, more importantly, is a crucial platform for realizing spin-based quantum information processing~\cite{de_leon_2021}. Moreover, in the cluster state preparation scheme proposed in other quantum systems, such as trapped ions, external controls are employed to engineer the interactions and thereby adjust their accumulated phases~\cite{Ivanov2008, Wunderlich2009}. Our approach, leveraging pulsed coupling modulation, may introduce an additional and auxiliary means that reduces the required control resources.

In the last place, some other improvements to our proposal for preparation are worthy of further discussion and investigation. The first of these improvements is developing an advanced and efficient algorithm to generate cluster states within larger systems, thus expanding the scope and scalability of our preparation proposal. In the NV center system, this requires developing more selective narrow-band pulses to lower the detuning difference demanded. The important steps remain the preparation of a large NV center system with the same orientations and the appropriate manipulations, including the possible modulation of the energy splittings of electronic spins using an electric field through the electrode or applying a magnetic gradient~\cite{Zhang2017, Bodenstedt2018}. The second is to use composite gates to implement a robust two-qubit $zz$ gate, in which random fluctuations in the coupling strength are corrected~\cite{Ichikawa2013, Ivanov2015}. Although this scheme is developed for a two-qubit system, it suggests that residual errors arising from uncertainties in the coupling characterization can be mitigated by suitable composite-pulse techniques.

In addition, although our state preparation scheme achieves high fidelity by suppressing long-range $zz$ couplings during free evolution, the $zz$ crosstalk during the implementation of selective pulses still constrains our fidelity. Therefore, we set $d$ = 30 nm to maintain small coupling strengths. Several approaches have been proposed to suppress $zz$ crosstalk. A widely used one is to introduce additional hardware (a tunable coupler~\cite{Niskanen2007, Sung2021, Li2020}) or extra qubits (shunted flux qubits~\cite{Zhao2020, Ku2020}), to eliminate undesired $zz$ coupling terms, albeit at the expense of increased system complexity. Alternative approaches, such as adjusting the control fields~\cite{Liang2024} applied to spins or suppressing the impact of $zz$ crosstalk through pulse and scheduling co-optimization~\cite{xie2022} or pulsed dynamical decouplings~\cite{Li2020, Tripathi2022}, however, require the capability of individual control of spins. Thus, in solid-state spin systems, advancing high-fidelity selective pulses under the limitations of global control and non-tunable coupling is imperative. 

\section{Acknowledgments}
We are grateful for the comments on the manuscript by Daniel Dulog and Jonas Breustedt. Discussions with Jan Haase at the early stages of this work are acknowledged. Y. L. thanks Qingyun Cao for useful discussions on the experimental aspects of the NV center. This work is supported by the BMBF under the funding program ‘quantum technologies—from basic research to market’ in the project Spinning (project No.~13N16215) and project CoGeQ (project No.~13N16101).  This work was supported by EU-project C-QuENS (Grant No.~101135359) and EU-Project SPINUS (Grant No.~101135699). The authors acknowledge support by the state of Baden-W\"{u}rttemberg through bwHPC and the German Research Foundation (DFG) through grant no INST 40/575-1 FUGG (JUSTUS 2 cluster). 

\appendix
\section{Additional information on solvable systems.}
\label{app_sec_solvable}
\subsection{Solvable systems}
\label{subsec_app_solvable}
In this section, we are going to present more details about the solutions in the four-, five- and six-spin systems. In a four-qubit system, the index sequence of spin is $\mathcal{S}_4$= \{(0), (1), (1,~2), (2), (2,~3), (3), (4)\}, with the corresponding matrix $M_4$ is
\begin{align}
	M_4 = 
	\begin{pmatrix}
		1 & -1 &  1 & -1 & -1 &  1 &  1 \\
		1 &  1 & -1 & -1 &  1 & -1 &  1 \\
		1 &  1 &  1 &  1 & -1 & -1 &  1 \\			
		1 & -1 & -1 &  1 &  1 &  1 & -1 \\			
		1 & -1 & -1 &  1 & -1 & -1 &  1 \\			
		1 &  1 & -1 & -1 & -1 &  1 & -1 \\						
		1 &  1 &  1 &  1 &  1 &  1 &  1 \\
	\end{pmatrix}
\end{align}
and the corresponding inverse matrix is 
\begin{align}
	M_4^{-1} = \frac{1}{4}
	\begin{pmatrix}
		1 &  1 &  1 &  1 &  0 &  0 &  0 \\
		-1 &  0 &  0 & -1 &  0 &  1 &  1 \\
		1 &  0 &  1 &  0 & -1 & -1 &  0 \\
		-1 & -1 &  0 &  0 &  1 &  0 &  1 \\
		0 &  1 &  0 &  1 & -1 & -1 &  0 \\
		0 & -1 & -1 &  0 &  0 &  1 &  1 \\
		0 &  0 & -1 & -1 &  1 &  0 &  1 \\
	\end{pmatrix}.
\end{align}
Here $\vec{\alpha}$ $ = \left(\alpha_{12}, \alpha_{23}, \alpha_{34}, \alpha_{14}, \alpha_{13}, \alpha_{24}, T_c\right)^T$ is the target vector.

In a five-spin system, we identify two distinct sequences of spin indices. The first sequence is $\mathcal{S}_5$ = \{(0), (1), (1,~2), (2), (2,~3), (3), (3,~4), (4), (4,~5), (5), (1,~5)\}, and the corresponding inverse matrix is
\begin{align}
	\nonumber   
	&M_5^{-1}=\nonumber\\ 
        &\frac{1}{4}\left( 
	\begin{array}{ccccccccccc}
		1 & 1 & 1 & 1 & 1 & 0 & 0 & 0 & 0 & 0 &-1\\
		-1 & 0 & 0 & 0 &-1 & 0 & 0 & 0 & 0 & 1 & 1\\
		-1 &-1 & 0 & 0 & 0 & 1 & 0 & 0 & 0 & 0 & 1\\
		0 &-1 &-1 & 0 & 0 & 0 & 1 & 0 & 0 & 0 & 1\\
		0 & 0 &-1 &-1 & 0 & 0 & 0 & 1 & 0 & 0 & 1\\
		0 & 0 & 0 &-1 &-1 & 0 & 0 & 0 & 1 & 0 & 1\\
		1 & 0 & 0 & 0 & 0 &-1 & 0 & 1 & 0 &-1 & 0\\
		0 & 1 & 0 & 0 & 0 &-1 &-1 & 0 & 1 & 0 & 0\\
		0 & 0 & 1 & 0 & 0 & 0 &-1 &-1 & 0 & 1 & 0\\
		0 & 0 & 0 & 1 & 0 & 1 & 0 &-1 &-1 & 0 & 0\\
		0 & 0 & 0 & 0 & 1 & 0 & 1 & 0 &-1 &-1 & 0\\
	\end{array}\right).
\end{align}
The second sequence is \{(0), (1,~2), (2,~3), (3,~4), (4,~5), (1,~5), (1,~4), (2,~4), (2,~5), (3,~5), (1,~3)\}, and the corresponding inverse matrix is
\begin{align}
	\nonumber   
	&M_5^{-1}=\nonumber\\ &\frac{1}{12}\left( 
	\begin{array}{ccccccccccc}
		1 &  1 &  1 &  1 &  1 &  1 &  1 &  1 &  1 &  1 & 2 \\
		-1 & -1 &  2 &  2 & -1 & -1 & -1 &  2 & -1 & -1 & 1 \\
		-1 & -1 & -1 &  2 &  2 & -1 & -1 & -1 &  2 & -1 & 1 \\
		2 & -1 & -1 & -1 &  2 & -1 & -1 & -1 & -1 &  2 & 1 \\
		2 &  2 & -1 & -1 & -1 &  2 & -1 & -1 & -1 & -1 & 1 \\
		-1 &  2 &  2 & -1 & -1 & -1 &  2 & -1 & -1 & -1 & 1 \\
		-1 & -1 & -1 &  2 & -1 & -1 &  2 & -1 & -1 &  2 & 1 \\
		-1 & -1 & -1 & -1 &  2 &  2 & -1 &  2 & -1 & -1 & 1 \\
		2 & -1 & -1 & -1 & -1 & -1 &  2 & -1 &  2 & -1 & 1 \\
		-1 &  2 & -1 & -1 & -1 & -1 & -1 &  2 & -1 &  2 & 1 \\
		-1 & -1 &  2 & -1 & -1 &  2 & -1 & -1 &  2 & -1 & 1 \\
	\end{array}\right).
\end{align}
The solution of the time intervals are directly get as $\vec{\tau}=M_5^{-1}\cdot \vec{\alpha}$, with target vector is
\begin{align}
	\vec{\alpha} = \left(\alpha_{12}, \alpha_{23}, \alpha_{34}, \alpha_{45}, \alpha_{15}, \alpha_{13}, \alpha_{24}, \alpha_{35}, \alpha_{14}, \alpha_{25}, T_c\right)^T.
\end{align}
The numbers of $\pi$ pulses employed in these two sequences are 12 and 22, respectively.

In the six-spin system, the sequence is $\mathcal{S}_6=$\{(0), (1,~2), (2,~3), (3,~4), (4,~5), (5,~6), (1,~6), (1,~3), (1,~4), (4,~6), (2,~4), (2,~6), (3,~6), (3,~5), (1,~5), (2,~5)\}, which means that the required number $\pi$ pulses is $N_{\pi} =  32$. The target vector is
\begin{align}
    \vec{\alpha} &= (\alpha_{12}, \alpha_{23}, \alpha_{34}, \alpha_{45}, \alpha_{56}, \alpha_{16}, \alpha_{13}, \alpha_{24}, \alpha_{35}, \alpha_{46},\nonumber\\
    &\alpha_{15}, \alpha_{26}, \alpha_{14}, \alpha_{25}, \alpha_{36}, T_c)^T.
\end{align} 
The corresponding inverse matrix is
\begin{widetext}
    \begin{align}
		M_6^{-1}=\frac{1}{16}\left(
		\begin{array}{cccccccccccccccc}
			1  &  1  &  1  &  1  &  1  &  1  &  1  &  1  &  1  &  1  &  1  &  1  &  1  &  1  &  1  &  1  \\
			1  & -1  &  1  &  1  &  1  & -1  & -1  & -1  &  1  &  1  & -1  & -1  & -1  & -1  &  1  &  1  \\
			-1  &  1  & -1  &  1  &  1  &  1  & -1  & -1  & -1  &  1  &  1  & -1  &  1  & -1  & -1  &  1  \\
			1  & -1  &  1  & -1  &  1  &  1  & -1  & -1  & -1  & -1  &  1  &  1  & -1  &  1  & -1  &  1  \\
			1  &  1  & -1  &  1  & -1  &  1  &  1  & -1  & -1  & -1  & -1  &  1  & -1  & -1  &  1  &  1  \\
			1  &  1  &  1  & -1  &  1  & -1  &  1  &  1  & -1  & -1  & -1  & -1  &  1  & -1  & -1  &  1  \\
			-1  &  1  &  1  &  1  & -1  &  1  & -1  &  1  &  1  & -1  & -1  & -1  & -1  &  1  & -1  &  1  \\
			-1  & -1  & -1  &  1  &  1  & -1  &  1  &  1  & -1  &  1  & -1  &  1  & -1  &  1  & -1  &  1  \\
			-1  &  1  & -1  & -1  &  1  & -1  & -1  & -1  &  1  & -1  & -1  &  1  &  1  &  1  &  1  &  1  \\
			1  &  1  & -1  & -1  & -1  & -1  &  1  & -1  &  1  &  1  &  1  & -1  & -1  &  1  & -1  &  1  \\
			-1  & -1  & -1  & -1  &  1  &  1  &  1  &  1  &  1  & -1  &  1  & -1  & -1  & -1  &  1  &  1  \\
			-1  & -1  &  1  &  1  & -1  & -1  &  1  & -1  &  1  & -1  &  1  &  1  &  1  & -1  & -1  &  1  \\
			1  & -1  & -1  &  1  & -1  & -1  & -1  &  1  & -1  & -1  &  1  & -1  &  1  &  1  &  1  &  1  \\
			1  & -1  & -1  & -1  & -1  &  1  & -1  &  1  &  1  &  1  & -1  &  1  &  1  & -1  & -1  &  1  \\
			-1  &  1  &  1  & -1  & -1  & -1  & -1  &  1  & -1  &  1  &  1  &  1  & -1  & -1  &  1  &  1  \\
			-1  & -1  &  1  & -1  & -1  &  1  &  1  & -1  & -1  &  1  & -1  & -1  &  1  &  1  &  1  &  1  
		\end{array}
		\right).
    \end{align}
\end{widetext}
\subsection{Two examples in four-spin system}
\label{app_subsec_example}
Here, we introduce two interesting examples with different target vectors $\Vec{\alpha}$ of the four-spin system, demonstrating the broad applicability of our proposal. The first is preparing a cluster state within a small 3-D system, where four spins are located on the vertices of a regular tetrahedron. Now we set all phases $\theta_{ij}=\pi$, resulting in $\alpha_{12}=\alpha_{23}=\alpha_{34}=\alpha_{14}=\pi/g_1$ and $\alpha_{13} = \alpha_{24}=\pi/(\sqrt{8}g_1)$ assuming the coupling strength $g\propto r^{-3}$. The duration $T_c = (2-1/\sqrt{8})\pi/g_1$ and the length of each segment is
\begin{align}
	\tau_1 = \frac{\pi}{g_1},~\tau_{2,4,6,7} = 0,~ \tau_{3,5} = \left(1-\frac{1}{\sqrt{8}}\right)\frac{\pi}{2g_1}.
\end{align}

The second example is to demonstrate an identity evolution, i.e., eliminating all the coupling terms. This is achieved by setting $\theta_{12} = 4\pi$ and the rest phases equal to 0, which resulting a unitary $U(T_c)=\exp\left(-i\pi\sigma_1^z\sigma_2^z\right)=-\mathbbm{1}$. The corresponding time duration $T_c = 4\pi/g_1$ and
\begin{align}
	\tau_{1,3,6,7} = \frac{\pi}{g_1},~\tau_{2,4,5} = 0.
\end{align}

\section{Details of states preparation in NV center systems}	
\label{app_subsec_NV}
First, we initialize all NV centers to the ground state $\ket{1}=\ket{m_s=0}$, and a product state $\otimes_i\ket{+}_i$ could be prepared by applying a broadband pulse $(\pi/2)_y$. 
In the four-spin system, the detuning of each NV center we set is $(2\pi)[-6,~-2,~2,~6]$ MHz.
As shown in Fig.~4 in the main text, a wide high-fidelity range $2\delta_0\Omega_{\text{b}}=(2\pi)21$ MHz with $\Omega_{\text{b}}=(2\pi)30$ MHz allows one to rotate all NV centers simultaneously.
The corresponding positions of the four NV centers are
\begin{align}
	\vec{r}_1 &= [-1.3077,~2.7694,~-0.0631]\nonumber\\
	\vec{r}_2 &= [29.5664,~-1.3499,~0.7147]\nonumber\\	
	\vec{r}_3 &= [30.3426,~33.0349,~-0.2050]\nonumber\\
	\vec{r}_4 &= [3.5784,~30.7254,~-0.1241]~\text{nm},
\end{align}
denoted by the green dots in Fig.~\ref{fig_simulation}(a) in the main text, which results in the coupling strengths
\begin{align}
    [g_{12},\,g_{23},\,g_{34},\,g_{14}]_1 & = (2\pi)[1.7143,~1.2728,~2.6796,~2.2727]\nonumber,\\
    [g_{13},\,_{24}]_2 & = (2\pi)[0.6186,~0.7370]~\text{kHz}.
\end{align}
The corresponding ideal NN and NNN coupling strengths are $g=(2\pi)[1.9241,~0.6802]$. Choosing a total time of evolution $T_c = 4\pi/g_1$, and the corresponding optimal time intervals are 
\begin{align}
	\vec{\tau}_o &= [267.6897,~135.7387,~111.6563,~93.3421,\\
        &149.3451,~119.6501,~163.0152]~\mu \text{s}.
\end{align}
In the six-spin system, the detunings are $(2\pi)$[$-10$, $-6$, $-2$, 2, 6, 10] MHz. And the Rabi frequency is $\Omega_{\text{b}} = (2\pi)40$ MHz. The spin positions are
\begin{align}
	\vec{r}_1 &= [29.2127,~1.4384,~-0.2414]\nonumber\\
	\vec{r}_2 &= [15.8884,~26.3060,~0.3192]\nonumber\\	
	\vec{r}_3 &= [-16.1471,~25.2258,~0.3129]\nonumber\\
	\vec{r}_4 &= [-31.0689,~1.3703,~-0.8649]\nonumber\\	
	\vec{r}_5 &= [-15.8095,~-27.6923,~-0.0301]\nonumber\\
	\vec{r}_6 &= [12.0557,~-26.0830,~-0.1649]~\text{nm},
\end{align}
which are indicated by the green dots in Fig.~\ref{fig_simulation}(c) in the main text. These spins are slightly different from the ideal position represented by the black circles. The corresponding coupling strengths are
\begin{align}
	&[g_{12},~g_{23},~g_{34},~g_{45},~g_{56},~g_{16}]_1 \nonumber\\
        &= (2\pi)[2.3094,~1.5774,~2.3136,~1.4645,~2.3888,~1.5229]\nonumber\\	
	&[g_{13},~g_{24},~g_{35},~g_{46},~g_{26},~g_{15}]_2 \nonumber\\
        &= (2\pi)[0.3864,~0.3449,~0.3505,~0.3885,~0.3583,~0.3369]\nonumber\\
	&[g_{14},~g_{25},~g_{36}]_3 = (2\pi)[0.2370,~0.2116,~0.2588]~ \text{kHz},
\end{align} 
and the ideal coupling strengths are $(2\pi)$ [1.9241, 0.3703, 0.2405], respectively. By choosing $T_c = 4\pi/g_1$, the optimal intervals are
\begin{align}
	\vec{\tau}_o = [&175.8969,~92.8740,~118.1285,~87.8812,\nonumber\\
                        &122.2728,~88.8799,~117.1140,~31.2665,\nonumber\\
                        &30.8702,~34.4913,~31.2770,~36.4844,\nonumber\\
                        &38.8677,~35.8695,~35.4856,~35.2711]~\mu\text{s}.       
\end{align}

\section{Details of optimization}
\label{app_sec_optimization}
Here we present details of the optimization process employed in the development of broadband and selective pulses. All optimizations are demonstrated in the Matlab Optimization Toolbox \cite{Optimization2023} to minimize the corresponding cost functions.

\subsection{Derivative of the optimal control field}
Achieving a robust and optimal control field requires aligning $U(T,\vec{\lambda})$ closely with the target gate while minimizing its variations with respect to errors $\vec{\lambda}$.
To get the derivatives $\partial U/\partial\vec{\lambda}$, we first expand the whole process with the Dyson series with a general Hamiltonian
\begin{align}
	H(t) = H_0(t) + H_p(t),
\end{align}
where $H_p(t)$ represents a perturbation error. Thus evolution is
\begin{align}
	U(T)=U_0&(T)\sum_{n=0}^{\infty}\int_{0}^{T}dt_1 \int_{0}^{t_1}dt_2 \int_{0}^{t_2}dt_3 \cdots\nonumber\int_{0}^{t_{n-1}}dt_n \nonumber\\
    &\times(-i)^n\tilde{H}_p(t_1)\tilde{H}_p(t_2)\tilde{H}_p(t_3)\cdots \tilde{H}_p(t_n),
	\label{eq:U(T)}
\end{align} 
with $U_0(T) = \mathcal{T}\exp\left[-i\int_{0}^{T}d\tau H_0(\tau)\right]$ and $\tilde{H}_p(t) = U_0(t)^{-1} H_p(t) U_0(t)$.

In a single error case, i.e.,~$\tilde{H}_p(t) = \varepsilon H_{\varepsilon}(t)$, the $k$-th order derivative is
\begin{align}
	\frac{d^kU(T)}{d \varepsilon^k} = U_0(T)\sum_{n=k}^{\infty}(-i)^n \frac{ n!}{(n-k)!}\varepsilon^{n-k}\nonumber\\
        \times\mathcal{D}\left(\tilde{H}_{\varepsilon}^{(1)},\tilde{H}_{\varepsilon}^{(2)},\cdots,\tilde{H}_{\varepsilon}^{(n)}\right),
\end{align}
with $\tilde{H}_{\varepsilon}^{(i)} = \tilde{H}_{\varepsilon}(t_i)$ and $\mathcal{D}\left(\tilde{H}_{\varepsilon}^{(1)},\tilde{H}_{\varepsilon}^{(2)},\cdots,\tilde{H}_{\varepsilon}^{(n)}\right)$ denotes the multiple integrals in Eq.~(\ref{eq:U(T)}) times $\varepsilon^{-n}$, which could be numerically calculated as a exponential of a matrix~\cite{VanLoan1978,Najfeld1995,Carbonell2008,Goodwin2015,Haas2019}:
\begin{align}
	\nonumber
	L(t)=-i \left(\begin{array}{cccccc}
		H_0(t) & A_1(t) & 0 &\cdots & 0 & 0\\
		0 & H_0(t) & A_2(t) &\cdots & 0 & 0\\
		0   &   0  & H_0(t) &\cdots & 0 & 0\\
		\vdots&\vdots&\vdots&\ddots &\vdots&\vdots\\
		0 & 0 & 0 &\cdots & H_0(t) & A_n(t)\\
		0 & 0 & 0 & \cdots & 0 & H_0(t)\\
	\end{array}\right),
\end{align}
resulting in
\begin{align}
	\mathcal{T} &\exp\left[\int_{0}^{T}d\tau L(\tau)\right]=\nonumber\\
	&\left(\begin{array}{cccc}
		U_0(T) & \mathcal{D}\left(\tilde{A}_1^{(1)}\right) & \cdots & \mathcal{D}\left(\tilde{A}_1^{(1)},\cdots\tilde{A}_n^{(n)}\right)\\		0 & U_0(T) & \cdots & \mathcal{D}\left(\tilde{A}_2^{(1)},\cdots \tilde{A}_n^{(n-1)}\right)\\
		0 & 0 & \cdots & \mathcal{D}\left(\tilde{A}_3^{(1)},\cdots \tilde{A}_n^{(n-2)}\right)\\		
		\vdots & \vdots & \ddots & \vdots\\		
		0 & 0 & \cdots & \mathcal{D}\left(\tilde{A}_n^{(1)}\right)\\
		0 & 0 & \cdots & U_0(T)\nonumber\\
	\end{array}\right),
\end{align}
with $\tilde{A}^{(i)}_k = U_0^{-1}(t_i)A_k(t_i)U_0(t_i)$. Here all the diagonal blocks are $U_0(T)$.
By choosing all $A_k(t) = H_{\varepsilon}(t)$, we directly get all the order of derivatives. In the zero point $\varepsilon = 0$, the derivative is
\begin{align}
	\frac{d^kU(T)}{d \varepsilon^k} = U_0(T)(-i)^k k! \mathcal{D}\left(\tilde{H}_p^{(1)},\tilde{H}_p^{(2)},\cdots,\tilde{H}_p^{(n)}\right).
\end{align}
\begin{figure}[tb]
	\centering
	\includegraphics[width=0.49\textwidth]{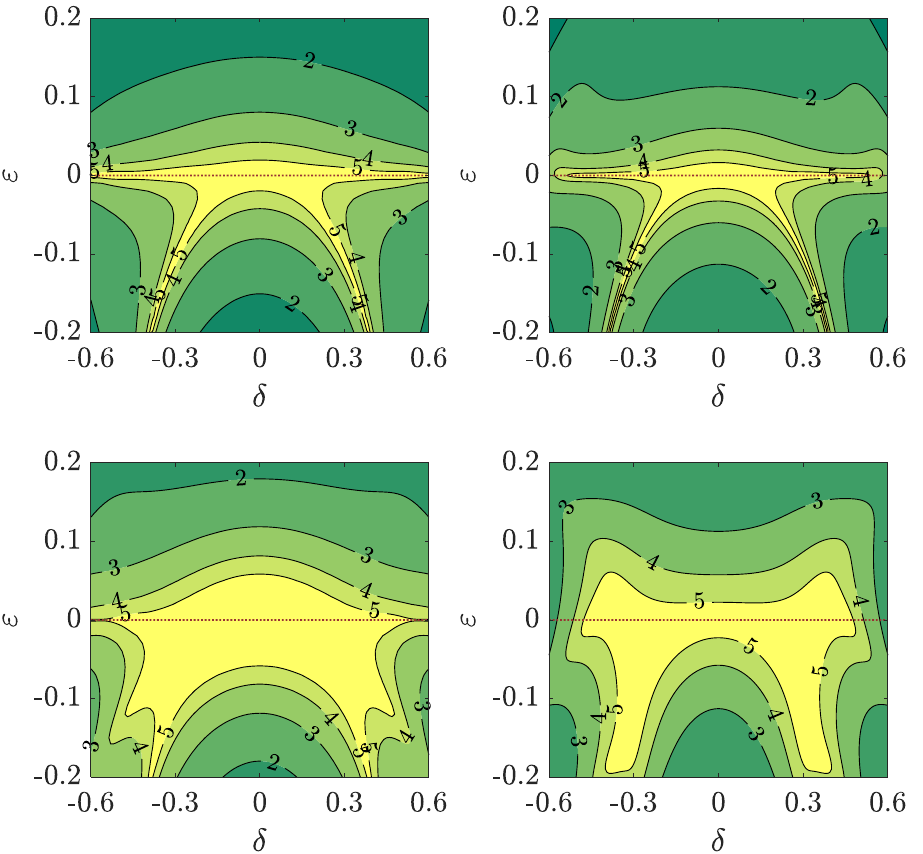}
	\caption{Simulated infidelities of the optimal broadband pulses $\pi_x$. In the upper panels, we show two optimal pulses with wide high fidelity ranges $|\delta|\leq0.6$ with the corresponding $n_{\phi}$=8,~10. And two pulses in the bottom panels are robust against both $\delta$ and $\varepsilon$, with $n_{\phi}$=11,~10.} 		
	\label{fig_app_broadband}
\end{figure}
In the two-error case, the perturbation Hamiltonian reads
\begin{align}
	H_p(t) = \varepsilon H_{\varepsilon}(t) + \delta H_{\delta}(t).
\end{align} 
Expanding to the second order, the evolution operator is
\begin{align}
    U(T) =&U_0(T) \bigg\{\mathbbm{1} -i\int_{0}^{T} dt_1 \left[\varepsilon \tilde{H}_{\varepsilon}(t_1) + \delta \tilde{H}_{\delta}(t_1) \right]\nonumber\\
    &- \int_{0}^{T} dt_1\int_{0}^{t_1} dt_2 \Big[\delta^2 \tilde{H}_{\delta}^{(1)}\tilde{H}_{\delta}^{(2)}+\varepsilon^2 \tilde{H}_{\varepsilon}^{(1)}\tilde{H}_{\varepsilon}^{(2)}\nonumber\\
    &+\delta\varepsilon\left(\tilde{H}_{\delta}^{(1)} \tilde{H}_{\varepsilon}^{(2)}+\tilde{H}_{\varepsilon}^{(1)}\tilde{H}_{\delta}^{(2)}\right)\Big]\bigg\}+ \cdots,
\end{align}
which gives a multiple derivative at $\delta=0$ and $\varepsilon=0$
\begin{align}
	\frac{\partial}{\partial\varepsilon}\frac{\partial}{\partial\delta}U(T) = -U_0(T) \left[\mathcal{D}\left(\tilde{H}^{(1)}_\delta,\tilde{H}^{(2)}_\varepsilon\right)+\mathcal{D}\left(\tilde{H}^{(1)}_\varepsilon,\tilde{H}^{(2)}_\delta\right)\right].
	\nonumber
\end{align}
In our simulation, the optimization is demonstrated on multi-working points $\delta_i$. The required corresponding derivatives on these points are calculated by choosing a new Hamiltonian $\tilde{H}_0(t) = H_0(t) + \delta_i H_{\delta}(t).$ 
All the exponential functions of the time-dependent matrices are approximated numerically by the production of constant $N_T$ slices of the total duration as
\begin{align}
	\mathcal{T}\exp\left[\int_{0}^{T}d\tau B(\tau)\right] = \prod_{k=1}^{N_T}\exp[B(t_j)\Delta t],
\end{align}
where $t_j =j\Delta t = jT/N_T$.

\subsection{Details of Optimal broadband and selective pulses}
\label{subsec:optimal}
In this section, we present further details on the development of broadband and selective pulses.
To illustrate the practicality of our broadband pulse optimization scheme, we present four additional optimal broadband pulses $\pi_x$ in Fig.~\ref{fig_app_broadband}. Unlike the top two pulses, which are only robust against the detuning error, the two pulses at the bottom are robust against both errors. Furthermore, the parameters for the broadband pulse $\pi_x$ in Fig.~\ref{fig_broadband_pulse} are $\vec{\phi}=$[0.4531, 1.1692, $-$0.2510, 0.8733, 0.9023, 0.7061, 0.6578, $-$0.3322, 1.3085]$\pi$ and for $(\pi/2)_x$ are $(\alpha,~\vec{\phi})=$[0.5355, 0.4645, 0.1322, 0.3789, 1.3156, 0.8444, 0.8222, 1.1370, $-$0.1073]$\pi$.

\begin{figure}[t]
	\centering
	\includegraphics[width=0.49\textwidth]{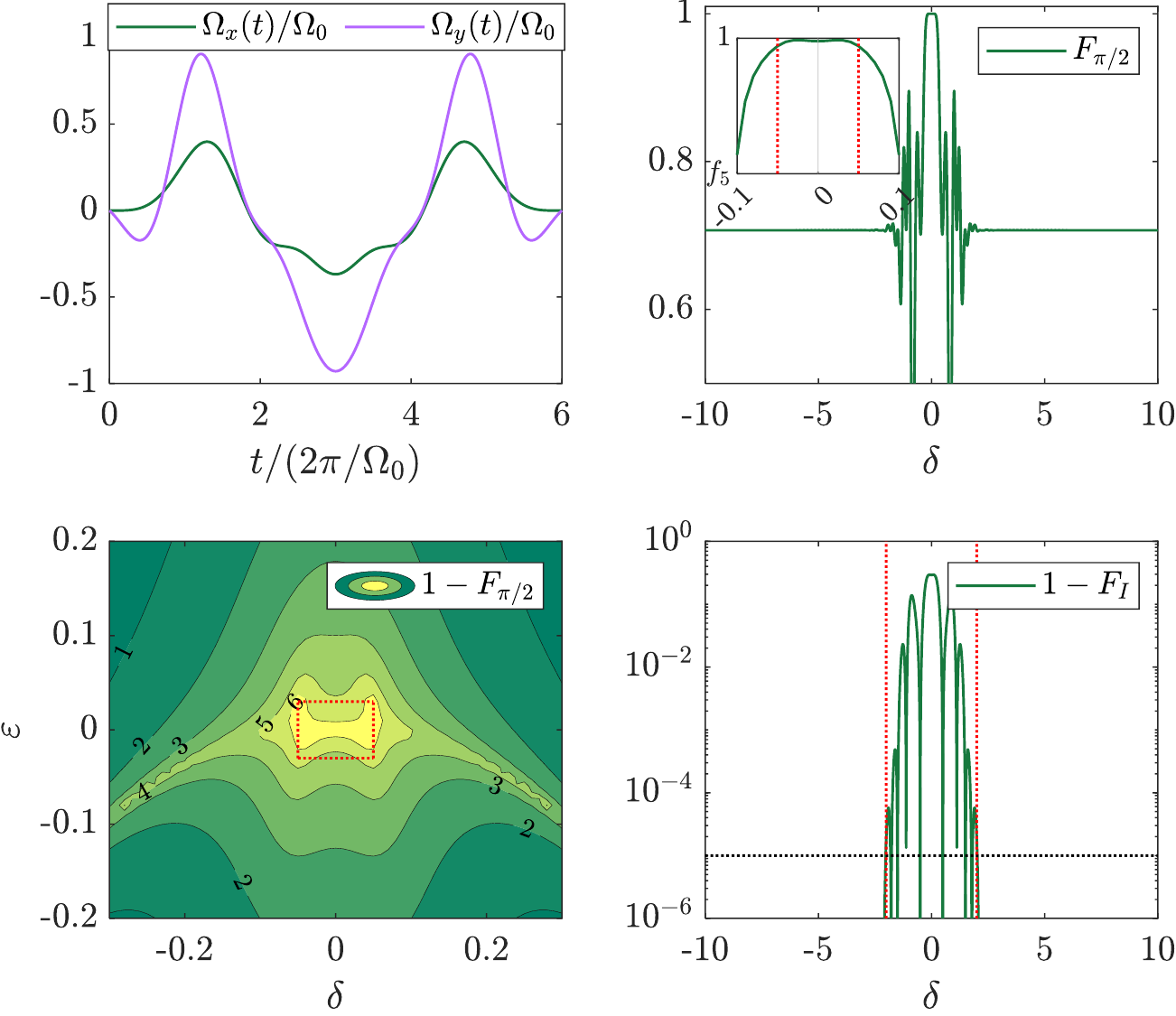}
	\caption{The optimal composite $(\pi/2)_x$ pulse. The notations are identical to those used in Fig.~\ref{fig_select_regime}. }
	\label{fig_piover2_selective}
\end{figure}

In the development of selective pulses, the unitary of the basic segment $(\alpha/2)_x$ is governed by the Hamiltonian 
\begin{align}
	H_s(\delta,\varepsilon,t) = \frac{\delta\Omega_0}{2}\sigma_z+\frac{(1+\varepsilon)}{2}\left[{\Omega_x(t)}\sigma_x + {\Omega_y(t)}\sigma_y\right].
\end{align}
In our simulation, the corresponding evolution is 
\begin{align}
	U_s(\delta,\varepsilon,T_s/2) = \mathcal{T} e^{-i\int_{0}^{T_s/2}d\tau H(\delta,\varepsilon,\tau)}
\end{align}
and the total evolution of the composite pulse is
\begin{align}
	U(\delta,\varepsilon,T_s) = \pi_x^\dagger U_s(\delta,\varepsilon,T_s/2) \pi_xU_s(\delta,\varepsilon,T_s/2).
\end{align} 
Specifically, the cost function~(\ref{eq_cost_selective}) is defined as
\begin{align}
	\label{eq_cost_amplitude_phase}
	G = c_1 \bar{I}_{I}(\vec{\delta}_l) + c_2\bar{I}_{\alpha}(\vec{\delta}_s)+ c_3 \left.\bar{D}(\vec{\delta}'_s)\right|_{\varepsilon=0}, 
\end{align}
with $\vec{\delta}_l$ and $\vec{\delta}_s$ denote the discrete values sampled within DR and RR and the weights are $\vec{c}$ = [1,~1,~$10^{-4}$]. 
The average derivative with respect to $\varepsilon$ is
\begin{align}
	\bar{D}(\vec{\delta}'_s) = \frac{1}{\tilde{n}_s}\sum_{k=1}^{\tilde{n}_s} \left\| \left. \frac{\partial U_s(\vec{\delta}'_s(k),\varepsilon,T_s/2)}{\partial\varepsilon}\right|_{\epsilon=0}\right\|_F
\end{align}
where $\vec{\delta}'_s$=[0, 0.04] represents another discrete sampling in RR, different from $\vec{\delta}_s$. The duration $T_s = 12(2\pi/g_1)$ and $\sigma = T_s/10$. In the numerical simulation of the evolution in a single $(\alpha/2)_x$ process, the number of slices is $N_T=1500$. The rest parameters of the pulses $\alpha_x$ are listed in Table.~\ref{table_parameter_gate}.
\begin{table}[h]
    \centering
	\begin{tabular}{c|c}
		\hline
		$\alpha$& Parameters\\
		\hline
		$\pi$&\makecell[c]{$\vec{a}$ = [$-$2.5136 1.6657, $-$1.1799, $-$0.1023, 0.5361 0.3596]\\
			$\vec{b}$ = [$-$2.8761 $-$3.1094 4.1194 9.4484 1.9703 2.0822]\\
			$\vec{\delta}_s$=[0, 0.02, 0.04, 0.06], $\vec{\delta}_l$=[1, 2, $\cdots$, 20]}\\
		\hline
		$\pi/2$&\makecell[c]{$\vec{a}$ = [0.5937~2.3491~0.2494~$-$0.9176~$-$0.4898~0.1127]\\
			$\vec{b}$ = [0.8659~3.9191~$-$1.3086~$-$3.6099~$-$3.1087~$-$0.8589]\\		
			$\vec{\delta}_s=[0,0.02,\cdots,1]$,~$\vec{\delta}_l=[1,2,\cdots,20]$}\\
		\hline
	\end{tabular}
	\caption{The parameters to implement selective $\pi_x$ and $(\pi/2)_x$ pulses.}
	\label{table_parameter_gate}
\end{table}
In Fig.~\ref{fig_piover2_selective} we show the composite selective $(\pi/2)_x$ pulse obtained by the optimization as we described in the main text.

\section{The error resources}   
\label{app_sec_erorr}
In this section, we briefly review the additional sources of error in the preparation protocol. We identify three main sources of concern: deviations in the selective pulses due to interactions between NV centers, the impact of the far-detuned third levels of the NV centers, and the influence of local environmental factors such as ${}^{13}$C nuclei or $P_1$ centers. The sources of error affect two main aspects of the preparation process, including the control pulses and the free evolution process. 
To demonstrate the robustness of our protocol to these errors, we numerically evaluate the corresponding preparation fidelity within the range of experimentally feasible parameters. 
\subsection{The impact of NV-NV interaction}
\begin{figure}[t]
	\centering
	\includegraphics[width=0.485\textwidth]{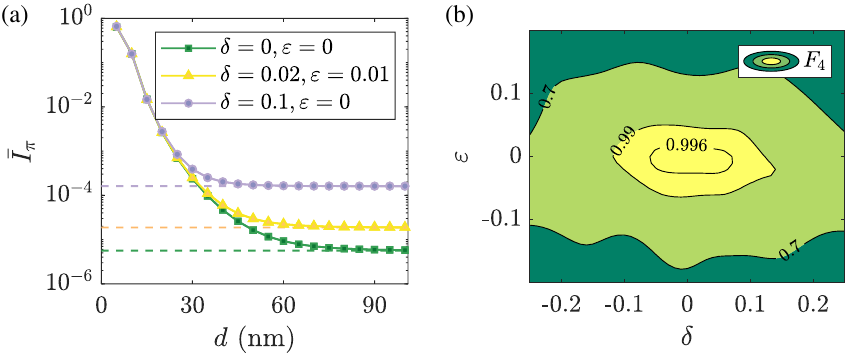}
	\caption{(a) The simulated average infidelity $\bar{I}_{\pi}$ with respect to four selective $\pi_x$ pulses. In this configuration, all four are located at the vertices of a square with sides of length $d$. The green squares, yellow triangles, and purple circles represent the average infidelities $\bar{I}_{\pi}$ with the presence of driving imperfections as $(\delta,~\varepsilon)=(0,~0),~(0.02,~0.01),~(0.1,~0)$, respectively. And the dashing lines indicate the infidelities in the absence of interactions. (b) The fidelity of cluster state preparation in the four-spin system without optimizing the length of time intervals $\vec{\tau}$ (i.e.,~Eq.~\eqref{eq:tau}). All parameters are identical to those presented in Fig.~\ref{fig_4_qubit}(b) in the main text.}
	\label{fig_distance}
\end{figure}

In the development of selective pulses, we ignored the weak interactions between NV centers. Although Gaussian-shaped pulses enhance the frequency-domain selectivity, they also reduce the ability to suppress the interactions due to their relatively longer duration. On a microsecond time scale, the coupling strengths at the kHz level still have an impact on the extremely demanding fidelities that we are aiming to achieve here. To quantify this influence, we simulate the effect of varying the nearest neighbor distance $d$ in an ideal four-spin square lattice system. Fig.~\ref{fig_distance}(a) depicts the average infidelities $\bar{I}_{\pi}$ of the selective pulses $U_{s}(j)$ we developed for four spins with $j=1,2,3,4$ compared to the ideal pulses $\pi_j^x$.
In the presence of different driving imperfections, the infidelities are approximately $10^{-4}$ when $d=30$~nm, and become negligible compared to other sources of imperfections when $d>30$~nm. For the sake of simplicity, we define the infidelity as
\begin{align}
	\bar{I}_{\pi} =1- \frac{1}{4}\sum_{j=1}^{4}\frac{1}{2^4}\operatorname{Tr}  \left[\pi_j^x U_{s}^\dagger(j)\right].
\end{align}

In the presence of interactions, the Hamiltonian to implement the desired selective pulse is
\begin{align}
	H(t) = H_{\text{s}}(t) + H_{\text{int}},
\end{align}
where $H_{\text{s}}(t)$ presents the Gaussian-shaped pulse applied on all spins. According to time-dependent perturbation theory, the weak interaction term $H_{\text{int}}$ could be regarded as a perturbation term. Therefore, to implement a selective pulse on spin $j$, the corresponding unitary operator $U_{s}(T_s) = \mathcal{T}\exp{\left[-i\int_0^{T_s}H(\tau)d\tau\right]}$ is expanded up to first order as
\begin{align}
	\label{Eq.UTs}
	U(T_s)  \simeq U_{s}(T_s)(1-i \tilde{H}_{\text{int}} T_s)\simeq U_{s}(T_s) \exp\left(-i\tilde{H}_{\text{int}} T_s\right),
\end{align}
where the unitary $U_{s}(T_s)= \mathcal{T}\exp{\left[-i\int_0^{T_s}H_s(\tau)d\tau\right]}$ represents the high fidelity selective pulse $\pi_j^x$ we developed in the absence of interactions. Consequently, the interaction can be expressed approximately as $ \tilde{H}_{\text{int}} \simeq \left(\pi_j^x\right)^\dagger H_{zz}\pi_j^x = H_{zz} - 2H_{zz}^j$. This arises because the $zz$ couplings are the crucial components of the interaction Hamiltonian (see Eq.~\eqref{eq:HI} in the main text). Here, $H_{zz}^j$ represents a subset of $H_{zz}$, which is the sum of $zz$ couplings involving spin $j$.   

In our preparation pulses, a selective pulse is usually followed by an evolution $U_{zz}(\tau)$ (i.e.,~Eq.~\eqref{eq:Uzz_2t} in the main text) governed by $H_{zz}$. The bias induced by $\tilde{H}_{\text{int}}$ in Eq.~(\ref{Eq.UTs}) typically affects the subsequent evolution, despite that $H_{zz}$ and $\tilde{H}_{\text{int}}$ are different in some certain components. Thus, it is possible and necessary to adjust the evolution time $\tau$ to compensate for the bias to some extent and thereby achieve higher fidelity. This is the reason the optimization (i.e., Eq.~\eqref{eq_op_last} in the main text) over $\vec{\tau}$ is effective. To provide a point of comparison, we present the fidelity of cluster state preparation associated with the original time length $\vec{\tau}$, as defined in Eq.~\eqref{eq:tau} in the main text, which has a maximum value of 0.9967 (see Fig.~\ref{fig_distance}(b)).

\subsection{The impact of the third NV-level}
The NV center contains ground-state triplet $( {}^3A)$, in which the degeneracy of the states $\ket{m_s = \pm 1}$ is lifted by applying an external magnetic field $B_z$ along the $z$ axis. For the sake of simplicity, we denote the three states as $\ket{0}$ and $\ket{\pm 1}$ here. With a relatively strong $B_z$, all the driving fields employed in our proposal are almost resonant with the transitions between $\ket{0}$ and $\ket{+1}$, while they are far-detuned with those between $\ket{0}$ and $\ket{-1}$. Consequently, all the transitions to $\ket{-1}$ will be suppressed by the large detuning. In addition, there are other ways to remove the effects of the third level. At first, it is helpful to take the third level $\ket{-1}$ as an additional source of error in the development of selective and broadband pulses. Second, it is also worthwhile noting that using the circularly polarized microwave fields will help to reduce further the impact of the off-resonant third level in the NV centers.

\begin{figure}[t]
	\centering
	\includegraphics[width=0.485\textwidth]{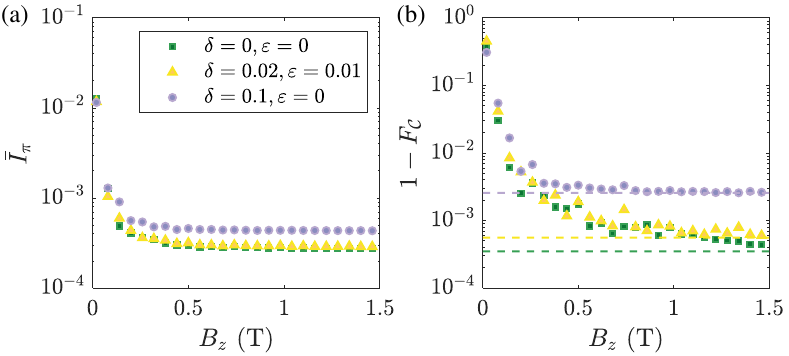}
	\caption{Simulation in the triplet states. (a) The simulated average infidelity $\bar{I}_{\pi}$ with respect to all selective $\pi_x$ pulses in the four-spin system as a function of $B_z$. (b) The infidelities of the cluster state preparation within the triplet states. The dashed lines represent the infidelities in two-level systems.}
	\label{fig_three_level}
\end{figure}

The Hamiltonian of the system is
\begin{align}
	H(t) = \sum_{i}H_i^0 + \sum_{ij} H_{ij}^{\text{dd}}+H_c(t).
\end{align}
Now, we derive the effective Hamiltonian $\tilde{H}(t) = U_0^\dagger(t)(H-H_0)U_0(t)$ in the interaction picture with respect to the unitary $U_0(t) = \exp(-iH_0t)$ with 
\begin{align}
	H_0 &= \sum_{i} \omega_c \ket{+1}_i {}_i\bra{+1} + \omega_c \ket{-1}_i {}_i\bra{-1}.
\end{align}
The two identical frequencies $\omega_c$ are signed to ensure a time-independent interaction. 
This transforms the operators into
\begin{align}
	U_0^\dagger(t) \ket{\pm1}_i {}_i\bra{0} U_0(t) &= e^{i \omega_c t} \ket{\pm1}_i {}_i\bra{0},
\end{align}
which makes spin operators $S_i^x$ and $S_i^y$ are rapidly oscillating in time.

The first term of local Hamiltonian becomes
\begin{align}
	\tilde{H}_i^0 = \Delta_i \ket{+1}_i {}_i\bra{+1}  + \Delta_i^{(-1)} \ket{-1}_i {}_i\bra{-1},
\end{align}
where the detunings $\Delta_i = D+\delta_i^D-\gamma_e B_z -\delta_i^z-\omega_c$ and $\Delta_i^{(-1)}=2 \gamma_e B_z + 2\delta_i^z + \Delta_i$. The Zeeman splitting term $\omega_e = 2\gamma_e B_z$ dominates, thus a large detuning $\Delta_i^{(-1)}$ suppresses the transitions to state $\ket{-1}$. The control field becomes
\begin{align}
	\tilde{H}_c(t) =\sum_{i=1}^N \frac{1}{\sqrt{2}} \Omega(t)\left[ \cos(\phi(t)) S_i^x + \sin(\phi(t)) S_i^y \right].
\end{align}
The magnetic dipole-dipole interaction between spin-$i$ and spin-$j$ is  
\begin{align}
	H_{ij}^{\text{dd}} = \vec{S}_i\cdot \vec{A}_{ij}\cdot\vec{S}_j,
\end{align}
and among the nine couplings in $\tilde{H}_{ij}^{\text{dd}}=U_0^\dagger(t)H_{ij}^{\text{dd}}U_0(t)$, the fast oscillating terms $S_i^xS_i^z$, $S_i^yS_i^z$, $S_i^zS_i^x$ and $S_i^zS_i^y$ are ignored due to RWA. Thus we derive the interaction Hamiltonian as
\begin{align}
	\tilde{H}_{\text{nn}} & = \sum_{i\neq j}^N A_{ij}^{xx}\left(c_{i,+}c_{j,+}^\dagger+\text{h.c.}\right) + A_{ij}^{yy}\left(c_{i,-}c_{j,-}^\dagger+\text{h.c.}\right)\nonumber\\
	& + A_{ij}^{xy} \left( ic_{i,-1}c_{j,+1}^\dagger - i c_{i,+1}c_{j,-1}^\dagger +\text{h.c.} \right) +A_{ij}^{zz} S_i^zS_j^z.
\end{align}
Here, the operators are $c_{i,\pm} = \ket{\pm}_i{}_i\bra{0}$ and $c_{i,\pm 1} = \ket{\pm 1}_i{}_i\bra{0}$ with the states defined as $\ket{\pm}_i = \left(\ket{+1}\pm\ket{-1}\right)_i/\sqrt{2}$. $A_{ij}^{\alpha\beta}$ is the element of the coupling tensor $\vec{A}_{ij}$, and the coupling strength in main text is $g_{ij} = A_{ij}^{zz}$. The Hamiltonian (Eq.~\eqref{eq:Hz_NV} in the main text) could be naturally obtained with a projection $\mathcal{P} = \otimes_i P_i = \otimes_i \left(\ket{+1}_i{}_i\bra{+1}+\ket{0}_i{}_i\bra{0}\right)$
\begin{align}
	H_I = \mathcal{P} \tilde{H}(t)\mathcal{P}.
\end{align}

In Fig.~\ref{fig_three_level}(a), we present the average infidelities $\bar{I}_{\pi}$ with respect to four selective $\pi_x$ pulses as a function of the magnetic field $B_z$. The infidelity is defined as $1-\Tr [\pi_i^x \mathcal{P} U_{\text{s}}^\dagger(j)\mathcal{P}]/2^4$, where $U_{\text{s}}^\dagger(j)$ is the selective pulse flipping spin $j$ within the triplet states. In Fig.~\ref{fig_three_level}(b), we show the cluster state preparation with different driving imperfections (the labels are the same as those in Fig.~\ref{fig_distance}(a)). The three dashed lines indicate the infidelities of the preparation in the two-level system extracted from Fig.~\ref{fig_simulation}(b) in the main text. 

\subsection{The effect of nearby nuclear spins}

\begin{figure}[t]
	\centering
	\includegraphics[width=0.5\textwidth]{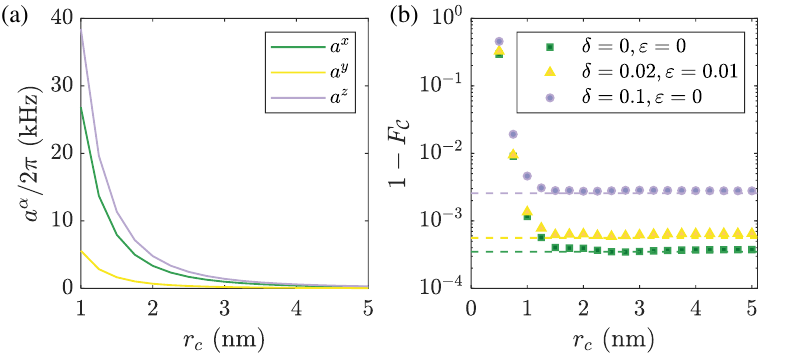}
	\caption{Simulation in the four-spin system with each NV center having a nearby $^{13}$C nuclear spin. (a) The maximum coupling strength $a^\alpha$. (b) The simulated infidelity of the system including the nuclear spins as a function of the distances between the NV centers and the nuclear spins. The imperfections of the driving field are also taken into account. For distances exceeding 1.5~nm, the influence of the nuclei becomes negligible compared to other sources of imperfection.} 		
	\label{fig_nuclear_spins}
\end{figure}

In this section, we aim to discuss the effect of the local environment on all NV centers. The main factor contributing to dephasing is the interaction with other spins, particularly the P1 centers and the $^{13}$C nuclear spin bath. To achieve long coherence times, our proposal emphasizes using highly pure samples to ensure that only a minimal number of nuclear spins interact with the NV center system. For simplicity, we assume that each NV center is coupled to a nearby nuclear spin at a distance of a few nanometers. In this case, the total Hamiltonian is~\cite{Casanova2015}
\begin{align}
	H_{\text{tot}} = H_{\text{NV}} + H_{\text{C}} + H_{\text{hyp}}.
\end{align}
Here, $H_{\text{NV}}$ is the $H_z$~(Eq.~\eqref{eq:Hz_NV} in the main text) of the NV centers. The nuclear spin Hamiltonian is $H_{\text{C}}=-\gamma_{c}\sum_i B_z \vec{I}_i$ where $\vec{I}_i= [I_i^x,~I_i^y,~I_i^z]$ are the spin-1/2 operators of the nuclear spin $i$. The hyperfine interaction is $H_{\text{hyp}} = \sum_i S_i^z \vec{a}_i\cdot \vec{I}_i$ where $\vec{a}_i$ is the coupling strength vector. Moving into the subspace $S_i^z\rightarrow(\sigma_i^z+\mathbbm{1})/2$, the total Hamiltonian is 
\begin{align}
	\label{eq:H_tot}
	{H}_{\text{tot}} = H_z + \sum_i \left(-\gamma_{c} B_z {I}_i^z + \frac{1}{2} (\sigma_i^z+\mathbbm{1}) \vec{a}_i\cdot \vec{I}_i\right).
\end{align}
Moving to the interaction picture with respect to $-\omega_z {I}_i^z$ with $\omega_z = \gamma_{c} B_z$, we could eliminate the fast oscillating terms $I_i^x\rightarrow \cos(\omega_z t) I_i^x + \sin(\omega_z t) I_i^y$ and $I_i^y\rightarrow -\sin(\omega_z t) I_i^x +\cos(\omega_z t) I_i^y$ due to RWA. We then get the effective Hamiltonian, which is 
\begin{align}
	\tilde{H}_{\text{tot}} = H_z + \sum_{i}\left(\frac{a_i^z}{2} \sigma_i^z I_i^z + \frac{a_i^z}{2} I_i^z\right).
\end{align}
These three terms mutually commute, and the $\sigma_i^z I_i^z$ interaction could be averaged together with the detuning error in Eq.~\eqref{eq:Uzz_2t} in the main text. The above derivation is based on the fact that the NV centers interact weakly with the nuclear spins. 

In the numerical simulation, we choose the Hamiltonian (\ref{eq:H_tot}) with $B_z=1.5$ T to describe the system. To evaluate the influence of the couplings between the NV centers and the nuclear spins, we assign the vector of the nuclear spin $i$ connecting the corresponding NV center $i$ as $\vec{r}_c = r_c \hat{e}_i$. The unit vectors are randomly chosen as 
\begin{align}
	\hat{e}_1 = [0.1817,~0.6198,~-0.7634]^T,\nonumber\\
	\hat{e}_2 = [0.5394,~0.1994,~-0.8181]^T,\nonumber\\
	\hat{e}_3 = [-0.1197,~0.0946,~0.9883]^T,\nonumber\\
	\hat{e}_4 = [0.6404,~-0.3121,~0.7018]^T.
\end{align}

In Fig.~\ref{fig_nuclear_spins}(a) we show the coupling strengths $a^\alpha$ between the NV centers and the corresponding nuclear spins. Here $a^\alpha = \max\left( a^\alpha_i\right)~(i=1,~2,~3,~4)$ is the maximum value among four coupling strengths with $\alpha = {x,~y,~z}$. In Fig.~\ref{fig_nuclear_spins}(b), we show the preparation infidelities with different distances $r_c$. Since a new error source, the nuclear spins, is included, we optimize the time intervals as Eq.~\eqref{eq_op_last} in the main text to get the result in Fig.~\ref{fig_nuclear_spins}(b). With a large magnetic field $B_z$, the effects of nuclear spins are highly suppressed. At short distances, nuclear spins still lead to significant infidelities in the preparation due to their large coupling strengths, which are tens of times larger than $g_1$, and the long preparation time. A reported enriched ${}^{12}$C (99.998\%) sample~\cite{Herbschleb2019} indicates that the probability
of finding a single ${}^{13}$C within 1.5~nm of the NV center is about 5\%. This low value ensures that the preparation process is free from the impact of nuclear spins. 

\begin{thebibliography}{113}%
	\makeatletter
	\providecommand \@ifxundefined [1]{%
		\@ifx{#1\undefined}
	}%
	\providecommand \@ifnum [1]{%
		\ifnum #1\expandafter \@firstoftwo
		\else \expandafter \@secondoftwo
		\fi
	}%
	\providecommand \@ifx [1]{%
		\ifx #1\expandafter \@firstoftwo
		\else \expandafter \@secondoftwo
		\fi
	}%
	\providecommand \natexlab [1]{#1}%
	\providecommand \enquote  [1]{``#1''}%
	\providecommand \bibnamefont  [1]{#1}%
	\providecommand \bibfnamefont [1]{#1}%
	\providecommand \citenamefont [1]{#1}%
	\providecommand \href@noop [0]{\@secondoftwo}%
	\providecommand \href [0]{\begingroup \@sanitize@url \@href}%
	\providecommand \@href[1]{\@@startlink{#1}\@@href}%
	\providecommand \@@href[1]{\endgroup#1\@@endlink}%
	\providecommand \@sanitize@url [0]{\catcode `\\12\catcode `\$12\catcode
		`\&12\catcode `\#12\catcode `\^12\catcode `\_12\catcode `\%12\relax}%
	\providecommand \@@startlink[1]{}%
	\providecommand \@@endlink[0]{}%
	\providecommand \url  [0]{\begingroup\@sanitize@url \@url }%
	\providecommand \@url [1]{\endgroup\@href {#1}{\urlprefix }}%
	\providecommand \urlprefix  [0]{URL }%
	\providecommand \Eprint [0]{\href }%
	\providecommand \doibase [0]{https://doi.org/}%
	\providecommand \selectlanguage [0]{\@gobble}%
	\providecommand \bibinfo  [0]{\@secondoftwo}%
	\providecommand \bibfield  [0]{\@secondoftwo}%
	\providecommand \translation [1]{[#1]}%
	\providecommand \BibitemOpen [0]{}%
	\providecommand \bibitemStop [0]{}%
	\providecommand \bibitemNoStop [0]{.\EOS\space}%
	\providecommand \EOS [0]{\spacefactor3000\relax}%
	\providecommand \BibitemShut  [1]{\csname bibitem#1\endcsname}%
	\let\auto@bib@innerbib\@empty
	\bibitem [{\citenamefont {Deutsch}(1985)}]{Deutsch1985}%
	\BibitemOpen
	\bibfield  {author} {\bibinfo {author} {\bibfnamefont {D.}~\bibnamefont
			{Deutsch}},\ }\bibfield  {title} {\bibinfo {title} {{Quantum theory, the
				Church–Turing principle and the universal quantum computer}},\ }\href
	{https://doi.org/10.1098/rspa.1985.0070} {\bibfield  {journal} {\bibinfo
			{journal} {Proc. Roy. Soc. A}\ }\textbf {\bibinfo {volume} {400}},\ \bibinfo
		{pages} {97} (\bibinfo {year} {1985})}\BibitemShut {NoStop}%
	\bibitem [{\citenamefont {Vedral}\ and\ \citenamefont
		{Plenio}(1998)}]{vedral1998basics}%
	\BibitemOpen
	\bibfield  {author} {\bibinfo {author} {\bibfnamefont {V.}~\bibnamefont
			{Vedral}}\ and\ \bibinfo {author} {\bibfnamefont {M.~B.}\ \bibnamefont
			{Plenio}},\ }\bibfield  {title} {\bibinfo {title} {Basics of quantum
			computation},\ }\href {https://doi.org/10.1016/S0079-6727(98)00004-4}
	{\bibfield  {journal} {\bibinfo  {journal} {Prog. Quantum Electron}\ }\textbf
		{\bibinfo {volume} {22}},\ \bibinfo {pages} {1} (\bibinfo {year}
		{1998})}\BibitemShut {NoStop}%
	\bibitem [{\citenamefont {Gottesman}\ and\ \citenamefont
		{Chuang}(1999)}]{GottesmannC1999}%
	\BibitemOpen
	\bibfield  {author} {\bibinfo {author} {\bibfnamefont {D.}~\bibnamefont
			{Gottesman}}\ and\ \bibinfo {author} {\bibfnamefont {I.}~\bibnamefont
			{Chuang}},\ }\bibfield  {title} {\bibinfo {title} {Demonstrating the
			viability of universal quantum computation using teleportation and
			single-qubit operations},\ }\href {https://doi.org/10.1038/46503} {\bibfield
		{journal} {\bibinfo  {journal} {Nature}\ }\textbf {\bibinfo {volume} {402}},\
		\bibinfo {pages} {390} (\bibinfo {year} {1999})}\BibitemShut {NoStop}%
	\bibitem [{\citenamefont {Eisert}\ \emph {et~al.}(2000)\citenamefont {Eisert},
		\citenamefont {Jacobs}, \citenamefont {Papadopoulos},\ and\ \citenamefont
		{Plenio}}]{eisert2000optimal}%
	\BibitemOpen
	\bibfield  {author} {\bibinfo {author} {\bibfnamefont {J.}~\bibnamefont
			{Eisert}}, \bibinfo {author} {\bibfnamefont {K.}~\bibnamefont {Jacobs}},
		\bibinfo {author} {\bibfnamefont {P.}~\bibnamefont {Papadopoulos}},\ and\
		\bibinfo {author} {\bibfnamefont {M.~B.}\ \bibnamefont {Plenio}},\ }\bibfield
	{title} {\bibinfo {title} {Optimal local implementation of nonlocal quantum
			gates},\ }\href {https://doi.org/10.1103/PhysRevA.62.052317} {\bibfield
		{journal} {\bibinfo  {journal} {Phys. Rev. A}\ }\textbf {\bibinfo {volume}
			{62}},\ \bibinfo {pages} {052317} (\bibinfo {year} {2000})}\BibitemShut
	{NoStop}%
	\bibitem [{\citenamefont {Raussendorf}\ and\ \citenamefont
		{Briegel}(2001)}]{Raussendorf2001}%
	\BibitemOpen
	\bibfield  {author} {\bibinfo {author} {\bibfnamefont {R.}~\bibnamefont
			{Raussendorf}}\ and\ \bibinfo {author} {\bibfnamefont {H.~J.}\ \bibnamefont
			{Briegel}},\ }\bibfield  {title} {\bibinfo {title} {A {One}-{Way} {Quantum}
			{Computer}},\ }\href {https://doi.org/10.1103/PhysRevLett.86.5188} {\bibfield
		{journal} {\bibinfo  {journal} {Phys. Rev. Lett.}\ }\textbf {\bibinfo
			{volume} {86}},\ \bibinfo {pages} {5188} (\bibinfo {year}
		{2001})}\BibitemShut {NoStop}%
	\bibitem [{\citenamefont {Nielsen}(2004)}]{Nielsen2004}%
	\BibitemOpen
	\bibfield  {author} {\bibinfo {author} {\bibfnamefont {M.~A.}\ \bibnamefont
			{Nielsen}},\ }\bibfield  {title} {\bibinfo {title} {Optical {Quantum}
			{Computation} {Using} {Cluster} {States}},\ }\href
	{https://doi.org/10.1103/PhysRevLett.93.040503} {\bibfield  {journal}
		{\bibinfo  {journal} {Phys. Rev. Lett.}\ }\textbf {\bibinfo {volume} {93}},\
		\bibinfo {pages} {040503} (\bibinfo {year} {2004})}\BibitemShut {NoStop}%
	\bibitem [{\citenamefont {Briegel}\ \emph {et~al.}(2009)\citenamefont
		{Briegel}, \citenamefont {Browne}, \citenamefont {Dür}, \citenamefont
		{Raussendorf},\ and\ \citenamefont {Van~den Nest}}]{Briegel2009}%
	\BibitemOpen
	\bibfield  {author} {\bibinfo {author} {\bibfnamefont {H.~J.}\ \bibnamefont
			{Briegel}}, \bibinfo {author} {\bibfnamefont {D.~E.}\ \bibnamefont {Browne}},
		\bibinfo {author} {\bibfnamefont {W.}~\bibnamefont {Dür}}, \bibinfo {author}
		{\bibfnamefont {R.}~\bibnamefont {Raussendorf}},\ and\ \bibinfo {author}
		{\bibfnamefont {M.}~\bibnamefont {Van~den Nest}},\ }\bibfield  {title}
	{\bibinfo {title} {Measurement-based quantum computation},\ }\href
	{https://doi.org/10.1038/nphys1157} {\bibfield  {journal} {\bibinfo
			{journal} {Nat. Phys.}\ }\textbf {\bibinfo {volume} {5}},\ \bibinfo {pages}
		{19} (\bibinfo {year} {2009})}\BibitemShut {NoStop}%
	\bibitem [{\citenamefont {Raussendorf}\ \emph {et~al.}(2003)\citenamefont
		{Raussendorf}, \citenamefont {Browne},\ and\ \citenamefont
		{Briegel}}]{Raussendorf2003}%
	\BibitemOpen
	\bibfield  {author} {\bibinfo {author} {\bibfnamefont {R.}~\bibnamefont
			{Raussendorf}}, \bibinfo {author} {\bibfnamefont {D.~E.}\ \bibnamefont
			{Browne}},\ and\ \bibinfo {author} {\bibfnamefont {H.~J.}\ \bibnamefont
			{Briegel}},\ }\bibfield  {title} {\bibinfo {title} {Measurement-based quantum
			computation on cluster states},\ }\href
	{https://doi.org/10.1103/PhysRevA.68.022312} {\bibfield  {journal} {\bibinfo
			{journal} {Phys. Rev. A}\ }\textbf {\bibinfo {volume} {68}},\ \bibinfo
		{pages} {022312} (\bibinfo {year} {2003})}\BibitemShut {NoStop}%
	\bibitem [{\citenamefont {Nielsen}(2006)}]{Nielsen2006}%
	\BibitemOpen
	\bibfield  {author} {\bibinfo {author} {\bibfnamefont {M.~A.}\ \bibnamefont
			{Nielsen}},\ }\bibfield  {title} {\bibinfo {title} {Cluster-state quantum
			computation},\ }\href {https://doi.org/10.1016/S0034-4877(06)80014-5}
	{\bibfield  {journal} {\bibinfo  {journal} {Rep. Math. Phys.}\ }\textbf
		{\bibinfo {volume} {57}},\ \bibinfo {pages} {147} (\bibinfo {year}
		{2006})}\BibitemShut {NoStop}%
	\bibitem [{\citenamefont {Knill}\ \emph {et~al.}(2001)\citenamefont {Knill},
		\citenamefont {Laflamme},\ and\ \citenamefont {Milburn}}]{knill2001scheme}%
	\BibitemOpen
	\bibfield  {author} {\bibinfo {author} {\bibfnamefont {E.}~\bibnamefont
			{Knill}}, \bibinfo {author} {\bibfnamefont {R.}~\bibnamefont {Laflamme}},\
		and\ \bibinfo {author} {\bibfnamefont {G.~J.}\ \bibnamefont {Milburn}},\
	}\bibfield  {title} {\bibinfo {title} {A scheme for efficient quantum
			computation with linear optics},\ }\href {https://doi.org/10.1038/35051009}
	{\bibfield  {journal} {\bibinfo  {journal} {Nature}\ }\textbf {\bibinfo
			{volume} {409}},\ \bibinfo {pages} {46} (\bibinfo {year} {2001})}\BibitemShut
	{NoStop}%
	\bibitem [{\citenamefont {Browne}\ and\ \citenamefont
		{Rudolph}(2005)}]{Browne2005}%
	\BibitemOpen
	\bibfield  {author} {\bibinfo {author} {\bibfnamefont {D.~E.}\ \bibnamefont
			{Browne}}\ and\ \bibinfo {author} {\bibfnamefont {T.}~\bibnamefont
			{Rudolph}},\ }\bibfield  {title} {\bibinfo {title} {Resource-{Efficient}
			{Linear} {Optical} {Quantum} {Computation}},\ }\href
	{https://doi.org/10.1103/PhysRevLett.95.010501} {\bibfield  {journal}
		{\bibinfo  {journal} {Phys. Rev. Lett.}\ }\textbf {\bibinfo {volume} {95}},\
		\bibinfo {pages} {010501} (\bibinfo {year} {2005})}\BibitemShut {NoStop}%
	\bibitem [{\citenamefont {Nielsen}\ and\ \citenamefont
		{Dawson}(2005)}]{nielsen2005fault}%
	\BibitemOpen
	\bibfield  {author} {\bibinfo {author} {\bibfnamefont {M.~A.}\ \bibnamefont
			{Nielsen}}\ and\ \bibinfo {author} {\bibfnamefont {C.~M.}\ \bibnamefont
			{Dawson}},\ }\bibfield  {title} {\bibinfo {title} {Fault-tolerant quantum
			computation with cluster states},\ }\href
	{https://doi.org/10.1103/PhysRevA.71.042323} {\bibfield  {journal} {\bibinfo
			{journal} {Phys. Rev. A}\ }\textbf {\bibinfo {volume} {71}},\ \bibinfo
		{pages} {042323} (\bibinfo {year} {2005})}\BibitemShut {NoStop}%
	\bibitem [{\citenamefont {Bombin}\ \emph {et~al.}(2021)\citenamefont {Bombin},
		\citenamefont {Kim}, \citenamefont {Litinski}, \citenamefont {Nickerson},
		\citenamefont {Pant}, \citenamefont {Pastawski}, \citenamefont {Roberts},\
		and\ \citenamefont {Rudolph}}]{Bombin2021}%
	\BibitemOpen
	\bibfield  {author} {\bibinfo {author} {\bibfnamefont {H.}~\bibnamefont
			{Bombin}}, \bibinfo {author} {\bibfnamefont {I.~H.}\ \bibnamefont {Kim}},
		\bibinfo {author} {\bibfnamefont {D.}~\bibnamefont {Litinski}}, \bibinfo
		{author} {\bibfnamefont {N.}~\bibnamefont {Nickerson}}, \bibinfo {author}
		{\bibfnamefont {M.}~\bibnamefont {Pant}}, \bibinfo {author} {\bibfnamefont
			{F.}~\bibnamefont {Pastawski}}, \bibinfo {author} {\bibfnamefont
			{S.}~\bibnamefont {Roberts}},\ and\ \bibinfo {author} {\bibfnamefont
			{T.}~\bibnamefont {Rudolph}},\ }\href
	{https://doi.org/10.48550/arXiv.2103.08612} {\bibinfo {title} {Interleaving:
			Modular architectures for fault-tolerant photonic quantum computing}}
	(\bibinfo {year} {2021}),\ \bibinfo {note}
	{\href{https://arxiv.org/abs/2103.08612}{arXiv:2103.08612
			[quant-ph]}}\BibitemShut {NoStop}%
	\bibitem [{\citenamefont {Bartolucci}\ \emph {et~al.}(2023)\citenamefont
		{Bartolucci}, \citenamefont {Birchall}, \citenamefont {Bombín},
		\citenamefont {Cable}, \citenamefont {Dawson}, \citenamefont
		{Gimeno-Segovia}, \citenamefont {Johnston}, \citenamefont {Kieling},
		\citenamefont {Nickerson}, \citenamefont {Pant}, \citenamefont {Pastawski},
		\citenamefont {Rudolph},\ and\ \citenamefont {Sparrow}}]{Bartolucci2023}%
	\BibitemOpen
	\bibfield  {author} {\bibinfo {author} {\bibfnamefont {S.}~\bibnamefont
			{Bartolucci}}, \bibinfo {author} {\bibfnamefont {P.}~\bibnamefont
			{Birchall}}, \bibinfo {author} {\bibfnamefont {H.}~\bibnamefont {Bombín}},
		\bibinfo {author} {\bibfnamefont {H.}~\bibnamefont {Cable}}, \bibinfo
		{author} {\bibfnamefont {C.}~\bibnamefont {Dawson}}, \bibinfo {author}
		{\bibfnamefont {M.}~\bibnamefont {Gimeno-Segovia}}, \bibinfo {author}
		{\bibfnamefont {E.}~\bibnamefont {Johnston}}, \bibinfo {author}
		{\bibfnamefont {K.}~\bibnamefont {Kieling}}, \bibinfo {author} {\bibfnamefont
			{N.}~\bibnamefont {Nickerson}}, \bibinfo {author} {\bibfnamefont
			{M.}~\bibnamefont {Pant}}, \bibinfo {author} {\bibfnamefont {F.}~\bibnamefont
			{Pastawski}}, \bibinfo {author} {\bibfnamefont {T.}~\bibnamefont {Rudolph}},\
		and\ \bibinfo {author} {\bibfnamefont {C.}~\bibnamefont {Sparrow}},\
	}\bibfield  {title} {\bibinfo {title} {Fusion-based quantum computation},\
	}\href {https://doi.org/10.1038/s41467-023-36493-1} {\bibfield  {journal}
		{\bibinfo  {journal} {Nat. Commun.}\ }\textbf {\bibinfo {volume} {14}},\
		\bibinfo {pages} {912} (\bibinfo {year} {2023})}\BibitemShut {NoStop}%
	\bibitem [{\citenamefont {Bose}\ \emph {et~al.}(1999)\citenamefont {Bose},
		\citenamefont {Knight}, \citenamefont {Plenio},\ and\ \citenamefont
		{Vedral}}]{Bose1999}%
	\BibitemOpen
	\bibfield  {author} {\bibinfo {author} {\bibfnamefont {S.}~\bibnamefont
			{Bose}}, \bibinfo {author} {\bibfnamefont {P.~L.}\ \bibnamefont {Knight}},
		\bibinfo {author} {\bibfnamefont {M.~B.}\ \bibnamefont {Plenio}},\ and\
		\bibinfo {author} {\bibfnamefont {V.}~\bibnamefont {Vedral}},\ }\bibfield
	{title} {\bibinfo {title} {Proposal for {Teleportation} of an {Atomic}
			{State} via {Cavity} {Decay}},\ }\href
	{https://doi.org/10.1103/PhysRevLett.83.5158} {\bibfield  {journal} {\bibinfo
			{journal} {Phys. Rev. Lett.}\ }\textbf {\bibinfo {volume} {83}},\ \bibinfo
		{pages} {5158} (\bibinfo {year} {1999})}\BibitemShut {NoStop}%
	\bibitem [{\citenamefont {team}(2025)}]{Alexander2024}%
	\BibitemOpen
	\bibfield  {author} {\bibinfo {author} {\bibfnamefont {P.}~\bibnamefont
			{team}},\ }\bibfield  {title} {\bibinfo {title} {A manufacturable platform
			for photonic quantum computing},\ }\href
	{https://doi.org/10.1038/s41586-025-08820-7} {\bibfield  {journal} {\bibinfo
			{journal} {Nature}\ }\textbf {\bibinfo {volume} {641}},\ \bibinfo {pages}
		{876} (\bibinfo {year} {2025})}\BibitemShut {NoStop}%
	\bibitem [{\citenamefont {Schwartz}\ \emph {et~al.}(2016)\citenamefont
		{Schwartz}, \citenamefont {Cogan}, \citenamefont {Schmidgall}, \citenamefont
		{Don}, \citenamefont {Gantz}, \citenamefont {Kenneth}, \citenamefont
		{Lindner},\ and\ \citenamefont {Gershoni}}]{Schwartz2016}%
	\BibitemOpen
	\bibfield  {author} {\bibinfo {author} {\bibfnamefont {I.}~\bibnamefont
			{Schwartz}}, \bibinfo {author} {\bibfnamefont {D.}~\bibnamefont {Cogan}},
		\bibinfo {author} {\bibfnamefont {E.~R.}\ \bibnamefont {Schmidgall}},
		\bibinfo {author} {\bibfnamefont {Y.}~\bibnamefont {Don}}, \bibinfo {author}
		{\bibfnamefont {L.}~\bibnamefont {Gantz}}, \bibinfo {author} {\bibfnamefont
			{O.}~\bibnamefont {Kenneth}}, \bibinfo {author} {\bibfnamefont {N.~H.}\
			\bibnamefont {Lindner}},\ and\ \bibinfo {author} {\bibfnamefont
			{D.}~\bibnamefont {Gershoni}},\ }\bibfield  {title} {\bibinfo {title}
		{Deterministic generation of a cluster state of entangled photons},\ }\href
	{https://doi.org/10.1126/science.aah4758} {\bibfield  {journal} {\bibinfo
			{journal} {Science}\ }\textbf {\bibinfo {volume} {354}},\ \bibinfo {pages}
		{434} (\bibinfo {year} {2016})}\BibitemShut {NoStop}%
	\bibitem [{\citenamefont {Larsen}\ \emph {et~al.}(2019)\citenamefont {Larsen},
		\citenamefont {Guo}, \citenamefont {Breum}, \citenamefont
		{Neergaard-Nielsen},\ and\ \citenamefont {Andersen}}]{Larsen2019}%
	\BibitemOpen
	\bibfield  {author} {\bibinfo {author} {\bibfnamefont {M.~V.}\ \bibnamefont
			{Larsen}}, \bibinfo {author} {\bibfnamefont {X.}~\bibnamefont {Guo}},
		\bibinfo {author} {\bibfnamefont {C.~R.}\ \bibnamefont {Breum}}, \bibinfo
		{author} {\bibfnamefont {J.~S.}\ \bibnamefont {Neergaard-Nielsen}},\ and\
		\bibinfo {author} {\bibfnamefont {U.~L.}\ \bibnamefont {Andersen}},\
	}\bibfield  {title} {\bibinfo {title} {Deterministic generation of a
			two-dimensional cluster state},\ }\href
	{https://doi.org/10.1126/science.aay4354} {\bibfield  {journal} {\bibinfo
			{journal} {Science}\ }\textbf {\bibinfo {volume} {366}},\ \bibinfo {pages}
		{369} (\bibinfo {year} {2019})}\BibitemShut {NoStop}%
	\bibitem [{\citenamefont {Istrati}\ \emph {et~al.}(2020)\citenamefont
		{Istrati}, \citenamefont {Pilnyak}, \citenamefont {Loredo}, \citenamefont
		{Antón}, \citenamefont {Somaschi}, \citenamefont {Hilaire}, \citenamefont
		{Ollivier}, \citenamefont {Esmann}, \citenamefont {Cohen}, \citenamefont
		{Vidro}, \citenamefont {Millet}, \citenamefont {Lemaître}, \citenamefont
		{Sagnes}, \citenamefont {Harouri}, \citenamefont {Lanco}, \citenamefont
		{Senellart},\ and\ \citenamefont {Eisenberg}}]{Istrati2020}%
	\BibitemOpen
	\bibfield  {author} {\bibinfo {author} {\bibfnamefont {D.}~\bibnamefont
			{Istrati}}, \bibinfo {author} {\bibfnamefont {Y.}~\bibnamefont {Pilnyak}},
		\bibinfo {author} {\bibfnamefont {J.~C.}\ \bibnamefont {Loredo}}, \bibinfo
		{author} {\bibfnamefont {C.}~\bibnamefont {Antón}}, \bibinfo {author}
		{\bibfnamefont {N.}~\bibnamefont {Somaschi}}, \bibinfo {author}
		{\bibfnamefont {P.}~\bibnamefont {Hilaire}}, \bibinfo {author} {\bibfnamefont
			{H.}~\bibnamefont {Ollivier}}, \bibinfo {author} {\bibfnamefont
			{M.}~\bibnamefont {Esmann}}, \bibinfo {author} {\bibfnamefont
			{L.}~\bibnamefont {Cohen}}, \bibinfo {author} {\bibfnamefont
			{L.}~\bibnamefont {Vidro}}, \bibinfo {author} {\bibfnamefont
			{C.}~\bibnamefont {Millet}}, \bibinfo {author} {\bibfnamefont
			{A.}~\bibnamefont {Lemaître}}, \bibinfo {author} {\bibfnamefont
			{I.}~\bibnamefont {Sagnes}}, \bibinfo {author} {\bibfnamefont
			{A.}~\bibnamefont {Harouri}}, \bibinfo {author} {\bibfnamefont
			{L.}~\bibnamefont {Lanco}}, \bibinfo {author} {\bibfnamefont
			{P.}~\bibnamefont {Senellart}},\ and\ \bibinfo {author} {\bibfnamefont
			{H.~S.}\ \bibnamefont {Eisenberg}},\ }\bibfield  {title} {\bibinfo {title}
		{Sequential generation of linear cluster states from a single photon
			emitter},\ }\href {https://doi.org/10.1038/s41467-020-19341-4} {\bibfield
		{journal} {\bibinfo  {journal} {Nat. Commun.}\ }\textbf {\bibinfo {volume}
			{11}},\ \bibinfo {pages} {5501} (\bibinfo {year} {2020})}\BibitemShut
	{NoStop}%
	\bibitem [{\citenamefont {Ferreira}\ \emph {et~al.}(2024)\citenamefont
		{Ferreira}, \citenamefont {Kim}, \citenamefont {Butler}, \citenamefont
		{Pichler},\ and\ \citenamefont {Painter}}]{Ferreira2024}%
	\BibitemOpen
	\bibfield  {author} {\bibinfo {author} {\bibfnamefont {V.~S.}\ \bibnamefont
			{Ferreira}}, \bibinfo {author} {\bibfnamefont {G.}~\bibnamefont {Kim}},
		\bibinfo {author} {\bibfnamefont {A.}~\bibnamefont {Butler}}, \bibinfo
		{author} {\bibfnamefont {H.}~\bibnamefont {Pichler}},\ and\ \bibinfo {author}
		{\bibfnamefont {O.}~\bibnamefont {Painter}},\ }\bibfield  {title} {\bibinfo
		{title} {Deterministic generation of multidimensional photonic cluster states
			with a single quantum emitter},\ }\href
	{https://doi.org/10.1038/s41567-024-02408-0} {\bibfield  {journal} {\bibinfo
			{journal} {Nat. Phys.}\ }\textbf {\bibinfo {volume} {20}},\ \bibinfo {pages}
		{865} (\bibinfo {year} {2024})}\BibitemShut {NoStop}%
	\bibitem [{\citenamefont {Gliniasty}\ \emph {et~al.}(2024)\citenamefont
		{Gliniasty}, \citenamefont {Hilaire}, \citenamefont {Emeriau}, \citenamefont
		{Wein}, \citenamefont {Salavrakos},\ and\ \citenamefont
		{Mansfield}}]{deGliniasty2024}%
	\BibitemOpen
	\bibfield  {author} {\bibinfo {author} {\bibfnamefont {G.~d.}\ \bibnamefont
			{Gliniasty}}, \bibinfo {author} {\bibfnamefont {P.}~\bibnamefont {Hilaire}},
		\bibinfo {author} {\bibfnamefont {P.-E.}\ \bibnamefont {Emeriau}}, \bibinfo
		{author} {\bibfnamefont {S.~C.}\ \bibnamefont {Wein}}, \bibinfo {author}
		{\bibfnamefont {A.}~\bibnamefont {Salavrakos}},\ and\ \bibinfo {author}
		{\bibfnamefont {S.}~\bibnamefont {Mansfield}},\ }\bibfield  {title} {\bibinfo
		{title} {A {Spin}-{Optical} {Quantum} {Computing} {Architecture}},\ }\href
	{https://doi.org/10.22331/q-2024-07-24-1423} {\bibfield  {journal} {\bibinfo
			{journal} {Quantum}\ }\textbf {\bibinfo {volume} {8}},\ \bibinfo {pages}
		{1423} (\bibinfo {year} {2024})}\BibitemShut {NoStop}%
	\bibitem [{\citenamefont {Nemoto}\ \emph {et~al.}(2014)\citenamefont {Nemoto},
		\citenamefont {Trupke}, \citenamefont {Devitt}, \citenamefont {Stephens},
		\citenamefont {Scharfenberger}, \citenamefont {Buczak}, \citenamefont
		{Nöbauer}, \citenamefont {Everitt}, \citenamefont {Schmiedmayer},\ and\
		\citenamefont {Munro}}]{Nemoto2014}%
	\BibitemOpen
	\bibfield  {author} {\bibinfo {author} {\bibfnamefont {K.}~\bibnamefont
			{Nemoto}}, \bibinfo {author} {\bibfnamefont {M.}~\bibnamefont {Trupke}},
		\bibinfo {author} {\bibfnamefont {S.~J.}\ \bibnamefont {Devitt}}, \bibinfo
		{author} {\bibfnamefont {A.~M.}\ \bibnamefont {Stephens}}, \bibinfo {author}
		{\bibfnamefont {B.}~\bibnamefont {Scharfenberger}}, \bibinfo {author}
		{\bibfnamefont {K.}~\bibnamefont {Buczak}}, \bibinfo {author} {\bibfnamefont
			{T.}~\bibnamefont {Nöbauer}}, \bibinfo {author} {\bibfnamefont {M.~S.}\
			\bibnamefont {Everitt}}, \bibinfo {author} {\bibfnamefont {J.}~\bibnamefont
			{Schmiedmayer}},\ and\ \bibinfo {author} {\bibfnamefont {W.~J.}\ \bibnamefont
			{Munro}},\ }\bibfield  {title} {\bibinfo {title} {Photonic {Architecture} for
			{Scalable} {Quantum} {Information} {Processing} in {Diamond}},\ }\href
	{https://doi.org/10.1103/PhysRevX.4.031022} {\bibfield  {journal} {\bibinfo
			{journal} {Phys. Rev. X}\ }\textbf {\bibinfo {volume} {4}},\ \bibinfo {pages}
		{031022} (\bibinfo {year} {2014})}\BibitemShut {NoStop}%
	\bibitem [{\citenamefont {Zhang}\ \emph {et~al.}(2011)\citenamefont {Zhang},
		\citenamefont {Zhou},\ and\ \citenamefont {Guo}}]{Zhang2011}%
	\BibitemOpen
	\bibfield  {author} {\bibinfo {author} {\bibfnamefont {J.-Y.}\ \bibnamefont
			{Zhang}}, \bibinfo {author} {\bibfnamefont {Z.-W.}\ \bibnamefont {Zhou}},\
		and\ \bibinfo {author} {\bibfnamefont {G.-C.}\ \bibnamefont {Guo}},\
	}\bibfield  {title} {\bibinfo {title} {Eliminating
			{Next}-{Nearest}-{Neighbor} {Interactions} in the {Preparation} of {Cluster}
			{State}},\ }\href {https://doi.org/10.1088/0256-307X/28/5/050301} {\bibfield
		{journal} {\bibinfo  {journal} {Chin. Phys. Lett.}\ }\textbf {\bibinfo
			{volume} {28}},\ \bibinfo {pages} {050301} (\bibinfo {year}
		{2011})}\BibitemShut {NoStop}%
	\bibitem [{\citenamefont {Doherty}\ \emph {et~al.}(2013)\citenamefont
		{Doherty}, \citenamefont {Manson}, \citenamefont {Delaney}, \citenamefont
		{Jelezko}, \citenamefont {Wrachtrup},\ and\ \citenamefont
		{Hollenberg}}]{Doherty2013}%
	\BibitemOpen
	\bibfield  {author} {\bibinfo {author} {\bibfnamefont {M.~W.}\ \bibnamefont
			{Doherty}}, \bibinfo {author} {\bibfnamefont {N.~B.}\ \bibnamefont {Manson}},
		\bibinfo {author} {\bibfnamefont {P.}~\bibnamefont {Delaney}}, \bibinfo
		{author} {\bibfnamefont {F.}~\bibnamefont {Jelezko}}, \bibinfo {author}
		{\bibfnamefont {J.}~\bibnamefont {Wrachtrup}},\ and\ \bibinfo {author}
		{\bibfnamefont {L.~C.}\ \bibnamefont {Hollenberg}},\ }\bibfield  {title}
	{\bibinfo {title} {The nitrogen-vacancy colour centre in diamond},\ }\href
	{https://doi.org/https://doi.org/10.1016/j.physrep.2013.02.001} {\bibfield
		{journal} {\bibinfo  {journal} {Phys. Rep.}\ }\textbf {\bibinfo {volume}
			{528}},\ \bibinfo {pages} {1} (\bibinfo {year} {2013})}\BibitemShut {NoStop}%
	\bibitem [{\citenamefont {Wu}\ \emph {et~al.}(2016)\citenamefont {Wu},
		\citenamefont {Jelezko}, \citenamefont {Plenio},\ and\ \citenamefont
		{Weil}}]{Wu2016}%
	\BibitemOpen
	\bibfield  {author} {\bibinfo {author} {\bibfnamefont {Y.}~\bibnamefont
			{Wu}}, \bibinfo {author} {\bibfnamefont {F.}~\bibnamefont {Jelezko}},
		\bibinfo {author} {\bibfnamefont {M.~B.}\ \bibnamefont {Plenio}},\ and\
		\bibinfo {author} {\bibfnamefont {T.}~\bibnamefont {Weil}},\ }\bibfield
	{title} {\bibinfo {title} {Diamond {Quantum} {Devices} in {Biology}},\ }\href
	{https://doi.org/10.1002/anie.201506556} {\bibfield  {journal} {\bibinfo
			{journal} {Angew. Chem. Int. Ed.}\ }\textbf {\bibinfo {volume} {55}},\
		\bibinfo {pages} {6586} (\bibinfo {year} {2016})}\BibitemShut {NoStop}%
	\bibitem [{\citenamefont {Dolde}\ \emph {et~al.}(2014)\citenamefont {Dolde},
		\citenamefont {Bergholm}, \citenamefont {Wang}, \citenamefont {Jakobi},
		\citenamefont {Naydenov}, \citenamefont {Pezzagna}, \citenamefont {Meijer},
		\citenamefont {Jelezko}, \citenamefont {Neumann}, \citenamefont
		{Schulte-Herbrüggen}, \citenamefont {Biamonte},\ and\ \citenamefont
		{Wrachtrup}}]{Dolde2014}%
	\BibitemOpen
	\bibfield  {author} {\bibinfo {author} {\bibfnamefont {F.}~\bibnamefont
			{Dolde}}, \bibinfo {author} {\bibfnamefont {V.}~\bibnamefont {Bergholm}},
		\bibinfo {author} {\bibfnamefont {Y.}~\bibnamefont {Wang}}, \bibinfo {author}
		{\bibfnamefont {I.}~\bibnamefont {Jakobi}}, \bibinfo {author} {\bibfnamefont
			{B.}~\bibnamefont {Naydenov}}, \bibinfo {author} {\bibfnamefont
			{S.}~\bibnamefont {Pezzagna}}, \bibinfo {author} {\bibfnamefont
			{J.}~\bibnamefont {Meijer}}, \bibinfo {author} {\bibfnamefont
			{F.}~\bibnamefont {Jelezko}}, \bibinfo {author} {\bibfnamefont
			{P.}~\bibnamefont {Neumann}}, \bibinfo {author} {\bibfnamefont
			{T.}~\bibnamefont {Schulte-Herbrüggen}}, \bibinfo {author} {\bibfnamefont
			{J.}~\bibnamefont {Biamonte}},\ and\ \bibinfo {author} {\bibfnamefont
			{J.}~\bibnamefont {Wrachtrup}},\ }\bibfield  {title} {\bibinfo {title}
		{High-fidelity spin entanglement using optimal control},\ }\href
	{https://doi.org/10.1038/ncomms4371} {\bibfield  {journal} {\bibinfo
			{journal} {Nat. Commun.}\ }\textbf {\bibinfo {volume} {5}},\ \bibinfo {pages}
		{3371} (\bibinfo {year} {2014})}\BibitemShut {NoStop}%
	\bibitem [{\citenamefont {Bradley}\ \emph {et~al.}(2019)\citenamefont
		{Bradley}, \citenamefont {Randall}, \citenamefont {Abobeih}, \citenamefont
		{Berrevoets}, \citenamefont {Degen}, \citenamefont {Bakker}, \citenamefont
		{Markham}, \citenamefont {Twitchen},\ and\ \citenamefont
		{Taminiau}}]{Bradley2019}%
	\BibitemOpen
	\bibfield  {author} {\bibinfo {author} {\bibfnamefont {C.~E.}\ \bibnamefont
			{Bradley}}, \bibinfo {author} {\bibfnamefont {J.}~\bibnamefont {Randall}},
		\bibinfo {author} {\bibfnamefont {M.~H.}\ \bibnamefont {Abobeih}}, \bibinfo
		{author} {\bibfnamefont {R.~C.}\ \bibnamefont {Berrevoets}}, \bibinfo
		{author} {\bibfnamefont {M.~J.}\ \bibnamefont {Degen}}, \bibinfo {author}
		{\bibfnamefont {M.~A.}\ \bibnamefont {Bakker}}, \bibinfo {author}
		{\bibfnamefont {M.}~\bibnamefont {Markham}}, \bibinfo {author} {\bibfnamefont
			{D.~J.}\ \bibnamefont {Twitchen}},\ and\ \bibinfo {author} {\bibfnamefont
			{T.~H.}\ \bibnamefont {Taminiau}},\ }\bibfield  {title} {\bibinfo {title} {A
			{Ten}-{Qubit} {Solid}-{State} {Spin} {Register} with {Quantum} {Memory} up to
			{One} {Minute}},\ }\href {https://doi.org/10.1103/PhysRevX.9.031045}
	{\bibfield  {journal} {\bibinfo  {journal} {Phys. Rev. X}\ }\textbf {\bibinfo
			{volume} {9}},\ \bibinfo {pages} {031045} (\bibinfo {year}
		{2019})}\BibitemShut {NoStop}%
	\bibitem [{\citenamefont {Xie}\ \emph {et~al.}(2023)\citenamefont {Xie},
		\citenamefont {Zhao}, \citenamefont {Xu}, \citenamefont {Kong}, \citenamefont
		{Yang}, \citenamefont {Wang}, \citenamefont {Wang}, \citenamefont {Shi},\
		and\ \citenamefont {Du}}]{Xie2023}%
	\BibitemOpen
	\bibfield  {author} {\bibinfo {author} {\bibfnamefont {T.}~\bibnamefont
			{Xie}}, \bibinfo {author} {\bibfnamefont {Z.}~\bibnamefont {Zhao}}, \bibinfo
		{author} {\bibfnamefont {S.}~\bibnamefont {Xu}}, \bibinfo {author}
		{\bibfnamefont {X.}~\bibnamefont {Kong}}, \bibinfo {author} {\bibfnamefont
			{Z.}~\bibnamefont {Yang}}, \bibinfo {author} {\bibfnamefont {M.}~\bibnamefont
			{Wang}}, \bibinfo {author} {\bibfnamefont {Y.}~\bibnamefont {Wang}}, \bibinfo
		{author} {\bibfnamefont {F.}~\bibnamefont {Shi}},\ and\ \bibinfo {author}
		{\bibfnamefont {J.}~\bibnamefont {Du}},\ }\bibfield  {title} {\bibinfo
		{title} {99.92\%-{Fidelity} cnot {Gates} in {Solids} by {Noise}
			{Filtering}},\ }\href {https://doi.org/10.1103/PhysRevLett.130.030601}
	{\bibfield  {journal} {\bibinfo  {journal} {Phys. Rev. Lett.}\ }\textbf
		{\bibinfo {volume} {130}},\ \bibinfo {pages} {030601} (\bibinfo {year}
		{2023})}\BibitemShut {NoStop}%
	\bibitem [{\citenamefont {Bartling}\ \emph {et~al.}(2025)\citenamefont
		{Bartling}, \citenamefont {Yun}, \citenamefont {Schymik}, \citenamefont {van
			Riggelen}, \citenamefont {Enthoven}, \citenamefont {van Ommen}, \citenamefont
		{Babaie}, \citenamefont {Sebastiano}, \citenamefont {Markham}, \citenamefont
		{Twitchen},\ and\ \citenamefont {Taminiau}}]{Bartling2024}%
	\BibitemOpen
	\bibfield  {author} {\bibinfo {author} {\bibfnamefont {H.}~\bibnamefont
			{Bartling}}, \bibinfo {author} {\bibfnamefont {J.}~\bibnamefont {Yun}},
		\bibinfo {author} {\bibfnamefont {K.}~\bibnamefont {Schymik}}, \bibinfo
		{author} {\bibfnamefont {M.}~\bibnamefont {van Riggelen}}, \bibinfo {author}
		{\bibfnamefont {L.}~\bibnamefont {Enthoven}}, \bibinfo {author}
		{\bibfnamefont {H.}~\bibnamefont {van Ommen}}, \bibinfo {author}
		{\bibfnamefont {M.}~\bibnamefont {Babaie}}, \bibinfo {author} {\bibfnamefont
			{F.}~\bibnamefont {Sebastiano}}, \bibinfo {author} {\bibfnamefont
			{M.}~\bibnamefont {Markham}}, \bibinfo {author} {\bibfnamefont
			{D.}~\bibnamefont {Twitchen}},\ and\ \bibinfo {author} {\bibfnamefont
			{T.}~\bibnamefont {Taminiau}},\ }\bibfield  {title} {\bibinfo {title}
		{Universal high-fidelity quantum gates for spin qubits in diamond},\ }\href
	{https://doi.org/10.1103/PhysRevApplied.23.034052} {\bibfield  {journal}
		{\bibinfo  {journal} {Phys. Rev. Appl.}\ }\textbf {\bibinfo {volume} {23}},\
		\bibinfo {pages} {034052} (\bibinfo {year} {2025})}\BibitemShut {NoStop}%
	\bibitem [{\citenamefont {Joas}\ \emph {et~al.}(2025)\citenamefont {Joas},
		\citenamefont {Ferlemann}, \citenamefont {Sailer}, \citenamefont {Vetter},
		\citenamefont {Zhang}, \citenamefont {Said}, \citenamefont {Teraji},
		\citenamefont {Onoda}, \citenamefont {Calarco}, \citenamefont {Genov},
		\citenamefont {M\"uller},\ and\ \citenamefont {Jelezko}}]{Joas2025}%
	\BibitemOpen
	\bibfield  {author} {\bibinfo {author} {\bibfnamefont {T.}~\bibnamefont
			{Joas}}, \bibinfo {author} {\bibfnamefont {F.}~\bibnamefont {Ferlemann}},
		\bibinfo {author} {\bibfnamefont {R.}~\bibnamefont {Sailer}}, \bibinfo
		{author} {\bibfnamefont {P.~J.}\ \bibnamefont {Vetter}}, \bibinfo {author}
		{\bibfnamefont {J.}~\bibnamefont {Zhang}}, \bibinfo {author} {\bibfnamefont
			{R.~S.}\ \bibnamefont {Said}}, \bibinfo {author} {\bibfnamefont
			{T.}~\bibnamefont {Teraji}}, \bibinfo {author} {\bibfnamefont
			{S.}~\bibnamefont {Onoda}}, \bibinfo {author} {\bibfnamefont
			{T.}~\bibnamefont {Calarco}}, \bibinfo {author} {\bibfnamefont
			{G.}~\bibnamefont {Genov}}, \bibinfo {author} {\bibfnamefont {M.~M.}\
			\bibnamefont {M\"uller}},\ and\ \bibinfo {author} {\bibfnamefont
			{F.}~\bibnamefont {Jelezko}},\ }\bibfield  {title} {\bibinfo {title}
		{High-fidelity electron spin gates for scaling diamond quantum registers},\
	}\href {https://doi.org/10.1103/PhysRevX.15.021069} {\bibfield  {journal}
		{\bibinfo  {journal} {Phys. Rev. X}\ }\textbf {\bibinfo {volume} {15}},\
		\bibinfo {pages} {021069} (\bibinfo {year} {2025})}\BibitemShut {NoStop}%
	\bibitem [{\citenamefont {Jakobi}\ \emph {et~al.}(2016)\citenamefont {Jakobi},
		\citenamefont {Waldherr}, \citenamefont {Rogers}, \citenamefont {Niemeyer},
		\citenamefont {Balasubramanian}, \citenamefont {Neumann},\ and\ \citenamefont
		{Wrachtrup}}]{Jakobi2016}%
	\BibitemOpen
	\bibfield  {author} {\bibinfo {author} {\bibfnamefont {I.}~\bibnamefont
			{Jakobi}}, \bibinfo {author} {\bibfnamefont {G.}~\bibnamefont {Waldherr}},
		\bibinfo {author} {\bibfnamefont {L.~J.}\ \bibnamefont {Rogers}}, \bibinfo
		{author} {\bibfnamefont {I.}~\bibnamefont {Niemeyer}}, \bibinfo {author}
		{\bibfnamefont {G.}~\bibnamefont {Balasubramanian}}, \bibinfo {author}
		{\bibfnamefont {P.}~\bibnamefont {Neumann}},\ and\ \bibinfo {author}
		{\bibfnamefont {J.}~\bibnamefont {Wrachtrup}},\ }\bibfield  {title} {\bibinfo
		{title} {Efficient creation of dipolar coupled nitrogen-vacancy spin qubits
			in diamond},\ }\href {https://doi.org/10.1088/1742-6596/752/1/012001}
	{\bibfield  {journal} {\bibinfo  {journal} {Journal of Physics: Condensed
				Matter}\ }\textbf {\bibinfo {volume} {752}},\ \bibinfo {pages} {012001}
		(\bibinfo {year} {2016})}\BibitemShut {NoStop}%
	\bibitem [{\citenamefont {Tobalina}\ \emph {et~al.}(2022)\citenamefont
		{Tobalina}, \citenamefont {Munuera-Javaloy}, \citenamefont {Torrontegui},
		\citenamefont {Muga},\ and\ \citenamefont {Casanova}}]{Tobalina2022}%
	\BibitemOpen
	\bibfield  {author} {\bibinfo {author} {\bibfnamefont {A.}~\bibnamefont
			{Tobalina}}, \bibinfo {author} {\bibfnamefont {C.}~\bibnamefont
			{Munuera-Javaloy}}, \bibinfo {author} {\bibfnamefont {E.}~\bibnamefont
			{Torrontegui}}, \bibinfo {author} {\bibfnamefont {J.~G.}\ \bibnamefont
			{Muga}},\ and\ \bibinfo {author} {\bibfnamefont {J.}~\bibnamefont
			{Casanova}},\ }\bibfield  {title} {\bibinfo {title} {Tailored ion beam for
			precise colour centre creation},\ }\href
	{https://doi.org/10.1098/rsta.2021.0271} {\bibfield  {journal} {\bibinfo
			{journal} {Philosophical Transactions of the Royal Society A: Mathematical,
				Physical and Engineering Sciences}\ }\textbf {\bibinfo {volume} {380}},\
		\bibinfo {pages} {20210271} (\bibinfo {year} {2022})}\BibitemShut {NoStop}%
	\bibitem [{\citenamefont {Jacob}\ \emph {et~al.}(2016)\citenamefont {Jacob},
		\citenamefont {Groot-Berning}, \citenamefont {Wolf}, \citenamefont {Ulm},
		\citenamefont {Couturier}, \citenamefont {Dawkins}, \citenamefont
		{Poschinger}, \citenamefont {Schmidt-Kaler},\ and\ \citenamefont
		{Singer}}]{Jacob2016}%
	\BibitemOpen
	\bibfield  {author} {\bibinfo {author} {\bibfnamefont {G.}~\bibnamefont
			{Jacob}}, \bibinfo {author} {\bibfnamefont {K.}~\bibnamefont
			{Groot-Berning}}, \bibinfo {author} {\bibfnamefont {S.}~\bibnamefont {Wolf}},
		\bibinfo {author} {\bibfnamefont {S.}~\bibnamefont {Ulm}}, \bibinfo {author}
		{\bibfnamefont {L.}~\bibnamefont {Couturier}}, \bibinfo {author}
		{\bibfnamefont {S.~T.}\ \bibnamefont {Dawkins}}, \bibinfo {author}
		{\bibfnamefont {U.~G.}\ \bibnamefont {Poschinger}}, \bibinfo {author}
		{\bibfnamefont {F.}~\bibnamefont {Schmidt-Kaler}},\ and\ \bibinfo {author}
		{\bibfnamefont {K.}~\bibnamefont {Singer}},\ }\bibfield  {title} {\bibinfo
		{title} {Transmission microscopy with nanometer resolution using a
			deterministic single ion source},\ }\href
	{https://doi.org/10.1103/PhysRevLett.117.043001} {\bibfield  {journal}
		{\bibinfo  {journal} {Phys. Rev. Lett.}\ }\textbf {\bibinfo {volume} {117}},\
		\bibinfo {pages} {043001} (\bibinfo {year} {2016})}\BibitemShut {NoStop}%
	\bibitem [{\citenamefont {Groot-Berning}\ \emph {et~al.}(2019)\citenamefont
		{Groot-Berning}, \citenamefont {Kornher}, \citenamefont {Jacob},
		\citenamefont {Stopp}, \citenamefont {Dawkins}, \citenamefont {Kolesov},
		\citenamefont {{Wrachtrup}}, \citenamefont {Singer},\ and\ \citenamefont
		{Schmidt-Kaler}}]{Groot-Berning2019}%
	\BibitemOpen
	\bibfield  {author} {\bibinfo {author} {\bibfnamefont {K.}~\bibnamefont
			{Groot-Berning}}, \bibinfo {author} {\bibfnamefont {T.}~\bibnamefont
			{Kornher}}, \bibinfo {author} {\bibfnamefont {G.}~\bibnamefont {Jacob}},
		\bibinfo {author} {\bibfnamefont {F.}~\bibnamefont {Stopp}}, \bibinfo
		{author} {\bibfnamefont {S.~T.}\ \bibnamefont {Dawkins}}, \bibinfo {author}
		{\bibfnamefont {R.}~\bibnamefont {Kolesov}}, \bibinfo {author} {\bibfnamefont
			{J.}~\bibnamefont {{Wrachtrup}}}, \bibinfo {author} {\bibfnamefont
			{K.}~\bibnamefont {Singer}},\ and\ \bibinfo {author} {\bibfnamefont
			{F.}~\bibnamefont {Schmidt-Kaler}},\ }\bibfield  {title} {\bibinfo {title}
		{Deterministic single-ion implantation of rare-earth ions for
			nanometer-resolution color-center generation},\ }\href
	{https://doi.org/10.1103/PhysRevLett.123.106802} {\bibfield  {journal}
		{\bibinfo  {journal} {Phys. Rev. Lett.}\ }\textbf {\bibinfo {volume} {123}},\
		\bibinfo {pages} {106802} (\bibinfo {year} {2019})}\BibitemShut {NoStop}%
	\bibitem [{\citenamefont {Haruyama}\ \emph {et~al.}(2019)\citenamefont
		{Haruyama}, \citenamefont {Onoda}, \citenamefont {Higuchi}, \citenamefont
		{Kada}, \citenamefont {Chiba}, \citenamefont {Hirano}, \citenamefont
		{Teraji}, \citenamefont {Igarashi}, \citenamefont {Kawai}, \citenamefont
		{Kawarada}, \citenamefont {Ishii}, \citenamefont {Fukuda}, \citenamefont
		{Tanii}, \citenamefont {Isoya}, \citenamefont {Ohshima},\ and\ \citenamefont
		{Hanaizumi}}]{Haruyama2019}%
	\BibitemOpen
	\bibfield  {author} {\bibinfo {author} {\bibfnamefont {M.}~\bibnamefont
			{Haruyama}}, \bibinfo {author} {\bibfnamefont {S.}~\bibnamefont {Onoda}},
		\bibinfo {author} {\bibfnamefont {T.}~\bibnamefont {Higuchi}}, \bibinfo
		{author} {\bibfnamefont {W.}~\bibnamefont {Kada}}, \bibinfo {author}
		{\bibfnamefont {A.}~\bibnamefont {Chiba}}, \bibinfo {author} {\bibfnamefont
			{Y.}~\bibnamefont {Hirano}}, \bibinfo {author} {\bibfnamefont
			{T.}~\bibnamefont {Teraji}}, \bibinfo {author} {\bibfnamefont
			{R.}~\bibnamefont {Igarashi}}, \bibinfo {author} {\bibfnamefont
			{S.}~\bibnamefont {Kawai}}, \bibinfo {author} {\bibfnamefont
			{H.}~\bibnamefont {Kawarada}}, \bibinfo {author} {\bibfnamefont
			{Y.}~\bibnamefont {Ishii}}, \bibinfo {author} {\bibfnamefont
			{R.}~\bibnamefont {Fukuda}}, \bibinfo {author} {\bibfnamefont
			{T.}~\bibnamefont {Tanii}}, \bibinfo {author} {\bibfnamefont
			{J.}~\bibnamefont {Isoya}}, \bibinfo {author} {\bibfnamefont
			{T.}~\bibnamefont {Ohshima}},\ and\ \bibinfo {author} {\bibfnamefont
			{O.}~\bibnamefont {Hanaizumi}},\ }\bibfield  {title} {\bibinfo {title}
		{Triple nitrogen-vacancy centre fabrication by {C5N4Hn} ion implantation},\
	}\href {https://doi.org/10.1038/s41467-019-10529-x} {\bibfield  {journal}
		{\bibinfo  {journal} {Nat. Commun.}\ }\textbf {\bibinfo {volume} {10}},\
		\bibinfo {pages} {2664} (\bibinfo {year} {2019})}\BibitemShut {NoStop}%
	\bibitem [{\citenamefont {Kimura}\ \emph {et~al.}(2022)\citenamefont {Kimura},
		\citenamefont {Onoda}, \citenamefont {Yamada}, \citenamefont {Kada},
		\citenamefont {Teraji}, \citenamefont {Isoya}, \citenamefont {Hanaizumi},\
		and\ \citenamefont {Ohshima}}]{Kimura2022}%
	\BibitemOpen
	\bibfield  {author} {\bibinfo {author} {\bibfnamefont {K.}~\bibnamefont
			{Kimura}}, \bibinfo {author} {\bibfnamefont {S.}~\bibnamefont {Onoda}},
		\bibinfo {author} {\bibfnamefont {K.}~\bibnamefont {Yamada}}, \bibinfo
		{author} {\bibfnamefont {W.}~\bibnamefont {Kada}}, \bibinfo {author}
		{\bibfnamefont {T.}~\bibnamefont {Teraji}}, \bibinfo {author} {\bibfnamefont
			{J.}~\bibnamefont {Isoya}}, \bibinfo {author} {\bibfnamefont
			{O.}~\bibnamefont {Hanaizumi}},\ and\ \bibinfo {author} {\bibfnamefont
			{T.}~\bibnamefont {Ohshima}},\ }\bibfield  {title} {\bibinfo {title}
		{Creation of multiple {NV} centers by phthalocyanine ion implantation},\
	}\href {https://doi.org/10.35848/1882-0786/ac7030} {\bibfield  {journal}
		{\bibinfo  {journal} {Appl. Phys. Express}\ }\textbf {\bibinfo {volume}
			{15}},\ \bibinfo {pages} {066501} (\bibinfo {year} {2022})}\BibitemShut
	{NoStop}%
	\bibitem [{\citenamefont {Reinhardt}\ \emph {et~al.}(2024)\citenamefont
		{Reinhardt}, \citenamefont {Lühmann}, \citenamefont {Räcke}, \citenamefont
		{Heupel}, \citenamefont {Kieschnick}, \citenamefont {Mändl}, \citenamefont
		{Popov}, \citenamefont {Meijer},\ and\ \citenamefont
		{Wunderlich}}]{Reinhardt2024}%
	\BibitemOpen
	\bibfield  {author} {\bibinfo {author} {\bibfnamefont {D.}~\bibnamefont
			{Reinhardt}}, \bibinfo {author} {\bibfnamefont {T.}~\bibnamefont {Lühmann}},
		\bibinfo {author} {\bibfnamefont {P.}~\bibnamefont {Räcke}}, \bibinfo
		{author} {\bibfnamefont {J.}~\bibnamefont {Heupel}}, \bibinfo {author}
		{\bibfnamefont {M.}~\bibnamefont {Kieschnick}}, \bibinfo {author}
		{\bibfnamefont {S.}~\bibnamefont {Mändl}}, \bibinfo {author} {\bibfnamefont
			{C.}~\bibnamefont {Popov}}, \bibinfo {author} {\bibfnamefont {J.~B.}\
			\bibnamefont {Meijer}},\ and\ \bibinfo {author} {\bibfnamefont
			{R.}~\bibnamefont {Wunderlich}},\ }\bibfield  {title} {\bibinfo {title}
		{Subdiffraction distance measurement of dipolar emitting qubit pairs},\
	}\href {https://doi.org/10.1021/acsphotonics.4c00257} {\bibfield  {journal}
		{\bibinfo  {journal} {ACS Photonics}\ }\textbf {\bibinfo {volume} {11}},\
		\bibinfo {pages} {1382} (\bibinfo {year} {2024})}\BibitemShut {NoStop}%
	\bibitem [{\citenamefont {Grotz}\ \emph {et~al.}(2011)\citenamefont {Grotz},
		\citenamefont {Beck}, \citenamefont {Neumann}, \citenamefont {Naydenov},
		\citenamefont {Reuter}, \citenamefont {Reinhard}, \citenamefont {Jelezko},
		\citenamefont {Wrachtrup}, \citenamefont {Schweinfurth}, \citenamefont
		{Sarkar},\ and\ \citenamefont {Hemmer}}]{Grotz2011}%
	\BibitemOpen
	\bibfield  {author} {\bibinfo {author} {\bibfnamefont {B.}~\bibnamefont
			{Grotz}}, \bibinfo {author} {\bibfnamefont {J.}~\bibnamefont {Beck}},
		\bibinfo {author} {\bibfnamefont {P.}~\bibnamefont {Neumann}}, \bibinfo
		{author} {\bibfnamefont {B.}~\bibnamefont {Naydenov}}, \bibinfo {author}
		{\bibfnamefont {R.}~\bibnamefont {Reuter}}, \bibinfo {author} {\bibfnamefont
			{F.}~\bibnamefont {Reinhard}}, \bibinfo {author} {\bibfnamefont
			{F.}~\bibnamefont {Jelezko}}, \bibinfo {author} {\bibfnamefont
			{J.}~\bibnamefont {Wrachtrup}}, \bibinfo {author} {\bibfnamefont
			{D.}~\bibnamefont {Schweinfurth}}, \bibinfo {author} {\bibfnamefont
			{B.}~\bibnamefont {Sarkar}},\ and\ \bibinfo {author} {\bibfnamefont
			{P.}~\bibnamefont {Hemmer}},\ }\bibfield  {title} {\bibinfo {title} {Sensing
			external spins with nitrogen-vacancy diamond},\ }\href
	{https://doi.org/10.1088/1367-2630/13/5/055004} {\bibfield  {journal}
		{\bibinfo  {journal} {New J. Phys.}\ }\textbf {\bibinfo {volume} {13}},\
		\bibinfo {pages} {055004} (\bibinfo {year} {2011})}\BibitemShut {NoStop}%
	\bibitem [{\citenamefont {Du}\ \emph {et~al.}(2024)\citenamefont {Du},
		\citenamefont {Shi}, \citenamefont {Kong}, \citenamefont {Jelezko},\ and\
		\citenamefont {Wrachtrup}}]{Du2024}%
	\BibitemOpen
	\bibfield  {author} {\bibinfo {author} {\bibfnamefont {J.}~\bibnamefont
			{Du}}, \bibinfo {author} {\bibfnamefont {F.}~\bibnamefont {Shi}}, \bibinfo
		{author} {\bibfnamefont {X.}~\bibnamefont {Kong}}, \bibinfo {author}
		{\bibfnamefont {F.}~\bibnamefont {Jelezko}},\ and\ \bibinfo {author}
		{\bibfnamefont {J.}~\bibnamefont {Wrachtrup}},\ }\bibfield  {title} {\bibinfo
		{title} {Single-molecule scale magnetic resonance spectroscopy using quantum
			diamond sensors},\ }\href {https://doi.org/10.1103/RevModPhys.96.025001}
	{\bibfield  {journal} {\bibinfo  {journal} {Rev. Mod. Phys.}\ }\textbf
		{\bibinfo {volume} {96}},\ \bibinfo {pages} {025001} (\bibinfo {year}
		{2024})}\BibitemShut {NoStop}%
	\bibitem [{\citenamefont {Choi}\ \emph {et~al.}(2017)\citenamefont {Choi},
		\citenamefont {Choi}, \citenamefont {Landig}, \citenamefont {Kucsko},
		\citenamefont {Zhou}, \citenamefont {Isoya}, \citenamefont {Jelezko},
		\citenamefont {Onoda}, \citenamefont {Sumiya}, \citenamefont {Khemani},
		\citenamefont {von Keyserlingk}, \citenamefont {Yao}, \citenamefont
		{Demler},\ and\ \citenamefont {Lukin}}]{Choi2017}%
	\BibitemOpen
	\bibfield  {author} {\bibinfo {author} {\bibfnamefont {S.}~\bibnamefont
			{Choi}}, \bibinfo {author} {\bibfnamefont {J.}~\bibnamefont {Choi}}, \bibinfo
		{author} {\bibfnamefont {R.}~\bibnamefont {Landig}}, \bibinfo {author}
		{\bibfnamefont {G.}~\bibnamefont {Kucsko}}, \bibinfo {author} {\bibfnamefont
			{H.}~\bibnamefont {Zhou}}, \bibinfo {author} {\bibfnamefont {J.}~\bibnamefont
			{Isoya}}, \bibinfo {author} {\bibfnamefont {F.}~\bibnamefont {Jelezko}},
		\bibinfo {author} {\bibfnamefont {S.}~\bibnamefont {Onoda}}, \bibinfo
		{author} {\bibfnamefont {H.}~\bibnamefont {Sumiya}}, \bibinfo {author}
		{\bibfnamefont {V.}~\bibnamefont {Khemani}}, \bibinfo {author} {\bibfnamefont
			{C.}~\bibnamefont {von Keyserlingk}}, \bibinfo {author} {\bibfnamefont
			{N.~Y.}\ \bibnamefont {Yao}}, \bibinfo {author} {\bibfnamefont
			{E.}~\bibnamefont {Demler}},\ and\ \bibinfo {author} {\bibfnamefont {M.~D.}\
			\bibnamefont {Lukin}},\ }\bibfield  {title} {\bibinfo {title} {Observation of
			discrete time-crystalline order in a disordered dipolar many-body system},\
	}\href {https://doi.org/10.1038/nature21426} {\bibfield  {journal} {\bibinfo
			{journal} {Nature}\ }\textbf {\bibinfo {volume} {543}},\ \bibinfo {pages}
		{221} (\bibinfo {year} {2017})}\BibitemShut {NoStop}%
	\bibitem [{\citenamefont {Kucsko}\ \emph {et~al.}(2018)\citenamefont {Kucsko},
		\citenamefont {Choi}, \citenamefont {Choi}, \citenamefont {Maurer},
		\citenamefont {Zhou}, \citenamefont {Landig}, \citenamefont {Sumiya},
		\citenamefont {Onoda}, \citenamefont {Isoya}, \citenamefont {Jelezko},
		\citenamefont {Demler}, \citenamefont {Yao},\ and\ \citenamefont
		{Lukin}}]{Kucsko2018}%
	\BibitemOpen
	\bibfield  {author} {\bibinfo {author} {\bibfnamefont {G.}~\bibnamefont
			{Kucsko}}, \bibinfo {author} {\bibfnamefont {S.}~\bibnamefont {Choi}},
		\bibinfo {author} {\bibfnamefont {J.}~\bibnamefont {Choi}}, \bibinfo {author}
		{\bibfnamefont {P.~C.}\ \bibnamefont {Maurer}}, \bibinfo {author}
		{\bibfnamefont {H.}~\bibnamefont {Zhou}}, \bibinfo {author} {\bibfnamefont
			{R.}~\bibnamefont {Landig}}, \bibinfo {author} {\bibfnamefont
			{H.}~\bibnamefont {Sumiya}}, \bibinfo {author} {\bibfnamefont
			{S.}~\bibnamefont {Onoda}}, \bibinfo {author} {\bibfnamefont
			{J.}~\bibnamefont {Isoya}}, \bibinfo {author} {\bibfnamefont
			{F.}~\bibnamefont {Jelezko}}, \bibinfo {author} {\bibfnamefont
			{E.}~\bibnamefont {Demler}}, \bibinfo {author} {\bibfnamefont {N.~Y.}\
			\bibnamefont {Yao}},\ and\ \bibinfo {author} {\bibfnamefont {M.~D.}\
			\bibnamefont {Lukin}},\ }\bibfield  {title} {\bibinfo {title} {Critical
			{Thermalization} of a {Disordered} {Dipolar} {Spin} {System} in {Diamond}},\
	}\href {https://doi.org/10.1103/PhysRevLett.121.023601} {\bibfield  {journal}
		{\bibinfo  {journal} {Phys. Rev. Lett.}\ }\textbf {\bibinfo {volume} {121}},\
		\bibinfo {pages} {023601} (\bibinfo {year} {2018})}\BibitemShut {NoStop}%
	\bibitem [{\citenamefont {Cai}\ \emph {et~al.}(2013)\citenamefont {Cai},
		\citenamefont {Retzker}, \citenamefont {Jelezko},\ and\ \citenamefont
		{Plenio}}]{Cai2013}%
	\BibitemOpen
	\bibfield  {author} {\bibinfo {author} {\bibfnamefont {J.}~\bibnamefont
			{Cai}}, \bibinfo {author} {\bibfnamefont {A.}~\bibnamefont {Retzker}},
		\bibinfo {author} {\bibfnamefont {F.}~\bibnamefont {Jelezko}},\ and\ \bibinfo
		{author} {\bibfnamefont {M.~B.}\ \bibnamefont {Plenio}},\ }\bibfield  {title}
	{\bibinfo {title} {A large-scale quantum simulator on a diamond surface at
			room temperature},\ }\href {https://doi.org/10.1038/nphys2519} {\bibfield
		{journal} {\bibinfo  {journal} {Nat. Phys.}\ }\textbf {\bibinfo {volume}
			{9}},\ \bibinfo {pages} {168} (\bibinfo {year} {2013})}\BibitemShut {NoStop}%
	\bibitem [{\citenamefont {de~Leon}\ \emph {et~al.}(2021)\citenamefont
		{de~Leon}, \citenamefont {Itoh}, \citenamefont {Kim}, \citenamefont {Mehta},
		\citenamefont {Northup}, \citenamefont {Paik}, \citenamefont {Palmer},
		\citenamefont {Samarth}, \citenamefont {Sangtawesin},\ and\ \citenamefont
		{Steuerman}}]{de_leon_2021}%
	\BibitemOpen
	\bibfield  {author} {\bibinfo {author} {\bibfnamefont {N.~P.}\ \bibnamefont
			{de~Leon}}, \bibinfo {author} {\bibfnamefont {K.~M.}\ \bibnamefont {Itoh}},
		\bibinfo {author} {\bibfnamefont {D.}~\bibnamefont {Kim}}, \bibinfo {author}
		{\bibfnamefont {K.~K.}\ \bibnamefont {Mehta}}, \bibinfo {author}
		{\bibfnamefont {T.~E.}\ \bibnamefont {Northup}}, \bibinfo {author}
		{\bibfnamefont {H.}~\bibnamefont {Paik}}, \bibinfo {author} {\bibfnamefont
			{B.~S.}\ \bibnamefont {Palmer}}, \bibinfo {author} {\bibfnamefont
			{N.}~\bibnamefont {Samarth}}, \bibinfo {author} {\bibfnamefont
			{S.}~\bibnamefont {Sangtawesin}},\ and\ \bibinfo {author} {\bibfnamefont
			{D.~W.}\ \bibnamefont {Steuerman}},\ }\bibfield  {title} {\bibinfo {title}
		{Materials challenges and opportunities for quantum computing hardware},\
	}\href {https://doi.org/10.1126/science.abb2823} {\bibfield  {journal}
		{\bibinfo  {journal} {Science}\ }\textbf {\bibinfo {volume} {372}},\ \bibinfo
		{pages} {eabb2823} (\bibinfo {year} {2021})}\BibitemShut {NoStop}%
	\bibitem [{\citenamefont {Zhang}\ \emph {et~al.}(2017)\citenamefont {Zhang},
		\citenamefont {Arai}, \citenamefont {Belthangady}, \citenamefont {Jaskula},\
		and\ \citenamefont {Walsworth}}]{Zhang2017}%
	\BibitemOpen
	\bibfield  {author} {\bibinfo {author} {\bibfnamefont {H.}~\bibnamefont
			{Zhang}}, \bibinfo {author} {\bibfnamefont {K.}~\bibnamefont {Arai}},
		\bibinfo {author} {\bibfnamefont {C.}~\bibnamefont {Belthangady}}, \bibinfo
		{author} {\bibfnamefont {J.-C.}\ \bibnamefont {Jaskula}},\ and\ \bibinfo
		{author} {\bibfnamefont {R.~L.}\ \bibnamefont {Walsworth}},\ }\bibfield
	{title} {\bibinfo {title} {Selective addressing of solid-state spins at the
			nanoscale via magnetic resonance frequency encoding},\ }\href
	{https://doi.org/10.1038/s41534-017-0033-3} {\bibfield  {journal} {\bibinfo
			{journal} {npj Quantum Inf.}\ }\textbf {\bibinfo {volume} {3}},\ \bibinfo
		{pages} {1} (\bibinfo {year} {2017})}\BibitemShut {NoStop}%
	\bibitem [{\citenamefont {Bodenstedt}\ \emph {et~al.}(2018)\citenamefont
		{Bodenstedt}, \citenamefont {Jakobi}, \citenamefont {Michl}, \citenamefont
		{Gerhardt}, \citenamefont {Neumann},\ and\ \citenamefont
		{Wrachtrup}}]{Bodenstedt2018}%
	\BibitemOpen
	\bibfield  {author} {\bibinfo {author} {\bibfnamefont {S.}~\bibnamefont
			{Bodenstedt}}, \bibinfo {author} {\bibfnamefont {I.}~\bibnamefont {Jakobi}},
		\bibinfo {author} {\bibfnamefont {J.}~\bibnamefont {Michl}}, \bibinfo
		{author} {\bibfnamefont {I.}~\bibnamefont {Gerhardt}}, \bibinfo {author}
		{\bibfnamefont {P.}~\bibnamefont {Neumann}},\ and\ \bibinfo {author}
		{\bibfnamefont {J.}~\bibnamefont {Wrachtrup}},\ }\bibfield  {title} {\bibinfo
		{title} {Nanoscale {Spin} {Manipulation} with {Pulsed} {Magnetic} {Gradient}
			{Fields} from a {Hard} {Disc} {Drive} {Writer}},\ }\href
	{https://doi.org/10.1021/acs.nanolett.8b01387} {\bibfield  {journal}
		{\bibinfo  {journal} {Nano Lett.}\ }\textbf {\bibinfo {volume} {18}},\
		\bibinfo {pages} {5389} (\bibinfo {year} {2018})}\BibitemShut {NoStop}%
	\bibitem [{\citenamefont {Bar-Gill}\ \emph {et~al.}(2012)\citenamefont
		{Bar-Gill}, \citenamefont {Pham}, \citenamefont {Belthangady}, \citenamefont
		{Le~Sage}, \citenamefont {Cappellaro}, \citenamefont {Maze}, \citenamefont
		{Lukin}, \citenamefont {Yacoby},\ and\ \citenamefont
		{Walsworth}}]{Bar-Gill2012}%
	\BibitemOpen
	\bibfield  {author} {\bibinfo {author} {\bibfnamefont {N.}~\bibnamefont
			{Bar-Gill}}, \bibinfo {author} {\bibfnamefont {L.~M.}\ \bibnamefont {Pham}},
		\bibinfo {author} {\bibfnamefont {C.}~\bibnamefont {Belthangady}}, \bibinfo
		{author} {\bibfnamefont {D.}~\bibnamefont {Le~Sage}}, \bibinfo {author}
		{\bibfnamefont {P.}~\bibnamefont {Cappellaro}}, \bibinfo {author}
		{\bibfnamefont {J.~R.}\ \bibnamefont {Maze}}, \bibinfo {author}
		{\bibfnamefont {M.~D.}\ \bibnamefont {Lukin}}, \bibinfo {author}
		{\bibfnamefont {A.}~\bibnamefont {Yacoby}},\ and\ \bibinfo {author}
		{\bibfnamefont {R.}~\bibnamefont {Walsworth}},\ }\bibfield  {title} {\bibinfo
		{title} {Suppression of spin-bath dynamics for improved coherence of
			multi-spin-qubit systems},\ }\href {https://doi.org/10.1038/ncomms1856}
	{\bibfield  {journal} {\bibinfo  {journal} {Nat. Commun.}\ }\textbf {\bibinfo
			{volume} {3}},\ \bibinfo {pages} {858} (\bibinfo {year} {2012})}\BibitemShut
	{NoStop}%
	\bibitem [{\citenamefont {Carr}\ and\ \citenamefont
		{Purcell}(1954)}]{Carr1954}%
	\BibitemOpen
	\bibfield  {author} {\bibinfo {author} {\bibfnamefont {H.~Y.}\ \bibnamefont
			{Carr}}\ and\ \bibinfo {author} {\bibfnamefont {E.~M.}\ \bibnamefont
			{Purcell}},\ }\bibfield  {title} {\bibinfo {title} {{Effects} of {Diffusion}
			on {Free} {Precession} in {Nuclear} {Magnetic} {Resonance} {Experiments}},\
	}\href {https://doi.org/10.1103/PhysRev.94.630} {\bibfield  {journal}
		{\bibinfo  {journal} {Phys. Rev.}\ }\textbf {\bibinfo {volume} {94}},\
		\bibinfo {pages} {630} (\bibinfo {year} {1954})}\BibitemShut {NoStop}%
	\bibitem [{\citenamefont {Meiboom}\ and\ \citenamefont
		{Gill}(1958)}]{Meiboom1958}%
	\BibitemOpen
	\bibfield  {author} {\bibinfo {author} {\bibfnamefont {S.}~\bibnamefont
			{Meiboom}}\ and\ \bibinfo {author} {\bibfnamefont {D.}~\bibnamefont {Gill}},\
	}\bibfield  {title} {\bibinfo {title} {Modified {Spin}‐{Echo} {Method} for
			{Measuring} {Nuclear} {Relaxation} {Times}},\ }\href
	{https://doi.org/10.1063/1.1716296} {\bibfield  {journal} {\bibinfo
			{journal} {Rev. Sci. Instrum.}\ }\textbf {\bibinfo {volume} {29}},\ \bibinfo
		{pages} {688} (\bibinfo {year} {1958})}\BibitemShut {NoStop}%
	\bibitem [{\citenamefont {Ryan}\ \emph {et~al.}(2010)\citenamefont {Ryan},
		\citenamefont {Hodges},\ and\ \citenamefont {Cory}}]{Ryan2010}%
	\BibitemOpen
	\bibfield  {author} {\bibinfo {author} {\bibfnamefont {C.~A.}\ \bibnamefont
			{Ryan}}, \bibinfo {author} {\bibfnamefont {J.~S.}\ \bibnamefont {Hodges}},\
		and\ \bibinfo {author} {\bibfnamefont {D.~G.}\ \bibnamefont {Cory}},\
	}\bibfield  {title} {\bibinfo {title} {{Robust} {Decoupling} {Techniques} to
			{Extend} {Quantum} {Coherence} in {Diamond}},\ }\href
	{https://doi.org/10.1103/PhysRevLett.105.200402} {\bibfield  {journal}
		{\bibinfo  {journal} {Phys. Rev. Lett.}\ }\textbf {\bibinfo {volume} {105}},\
		\bibinfo {pages} {200402} (\bibinfo {year} {2010})}\BibitemShut {NoStop}%
	\bibitem [{\citenamefont {Souza}\ \emph {et~al.}(2011)\citenamefont {Souza},
		\citenamefont {\'Alvarez},\ and\ \citenamefont {Suter}}]{Souza2011}%
	\BibitemOpen
	\bibfield  {author} {\bibinfo {author} {\bibfnamefont {A.~M.}\ \bibnamefont
			{Souza}}, \bibinfo {author} {\bibfnamefont {G.~A.}\ \bibnamefont
			{\'Alvarez}},\ and\ \bibinfo {author} {\bibfnamefont {D.}~\bibnamefont
			{Suter}},\ }\bibfield  {title} {\bibinfo {title} {{Robust} {Dynamical}
			{Decoupling} for {Quantum} {Computing} and {Quantum} {Memory}},\ }\href
	{https://doi.org/10.1103/PhysRevLett.106.240501} {\bibfield  {journal}
		{\bibinfo  {journal} {Phys. Rev. Lett.}\ }\textbf {\bibinfo {volume} {106}},\
		\bibinfo {pages} {240501} (\bibinfo {year} {2011})}\BibitemShut {NoStop}%
	\bibitem [{\citenamefont {Ali~Ahmed}\ \emph {et~al.}(2013)\citenamefont
		{Ali~Ahmed}, \citenamefont {\'Alvarez},\ and\ \citenamefont
		{Suter}}]{Ali2013}%
	\BibitemOpen
	\bibfield  {author} {\bibinfo {author} {\bibfnamefont {M.~A.}\ \bibnamefont
			{Ali~Ahmed}}, \bibinfo {author} {\bibfnamefont {G.~A.}\ \bibnamefont
			{\'Alvarez}},\ and\ \bibinfo {author} {\bibfnamefont {D.}~\bibnamefont
			{Suter}},\ }\bibfield  {title} {\bibinfo {title} {{Robustness} of dynamical
			decoupling sequences},\ }\href {https://doi.org/10.1103/PhysRevA.87.042309}
	{\bibfield  {journal} {\bibinfo  {journal} {Phys. Rev. A}\ }\textbf {\bibinfo
			{volume} {87}},\ \bibinfo {pages} {042309} (\bibinfo {year}
		{2013})}\BibitemShut {NoStop}%
	\bibitem [{\citenamefont {Abobeih}\ \emph {et~al.}(2018)\citenamefont
		{Abobeih}, \citenamefont {Cramer}, \citenamefont {Bakker}, \citenamefont
		{Kalb}, \citenamefont {Markham}, \citenamefont {Twitchen},\ and\
		\citenamefont {Taminiau}}]{Abobeih2018}%
	\BibitemOpen
	\bibfield  {author} {\bibinfo {author} {\bibfnamefont {M.~H.}\ \bibnamefont
			{Abobeih}}, \bibinfo {author} {\bibfnamefont {J.}~\bibnamefont {Cramer}},
		\bibinfo {author} {\bibfnamefont {M.~A.}\ \bibnamefont {Bakker}}, \bibinfo
		{author} {\bibfnamefont {N.}~\bibnamefont {Kalb}}, \bibinfo {author}
		{\bibfnamefont {M.}~\bibnamefont {Markham}}, \bibinfo {author} {\bibfnamefont
			{D.~J.}\ \bibnamefont {Twitchen}},\ and\ \bibinfo {author} {\bibfnamefont
			{T.~H.}\ \bibnamefont {Taminiau}},\ }\bibfield  {title} {\bibinfo {title}
		{One-second coherence for a single electron spin coupled to a multi-qubit
			nuclear-spin environment},\ }\href
	{https://doi.org/10.1038/s41467-018-04916-z} {\bibfield  {journal} {\bibinfo
			{journal} {Nat. Commun.}\ }\textbf {\bibinfo {volume} {9}},\ \bibinfo {pages}
		{2552} (\bibinfo {year} {2018})}\BibitemShut {NoStop}%
	\bibitem [{\citenamefont {Rong}\ \emph {et~al.}(2015)\citenamefont {Rong},
		\citenamefont {Geng}, \citenamefont {Shi}, \citenamefont {Liu}, \citenamefont
		{Xu}, \citenamefont {Ma}, \citenamefont {Kong}, \citenamefont {Jiang},
		\citenamefont {Wu},\ and\ \citenamefont {Du}}]{Rong2015}%
	\BibitemOpen
	\bibfield  {author} {\bibinfo {author} {\bibfnamefont {X.}~\bibnamefont
			{Rong}}, \bibinfo {author} {\bibfnamefont {J.}~\bibnamefont {Geng}}, \bibinfo
		{author} {\bibfnamefont {F.}~\bibnamefont {Shi}}, \bibinfo {author}
		{\bibfnamefont {Y.}~\bibnamefont {Liu}}, \bibinfo {author} {\bibfnamefont
			{K.}~\bibnamefont {Xu}}, \bibinfo {author} {\bibfnamefont {W.}~\bibnamefont
			{Ma}}, \bibinfo {author} {\bibfnamefont {F.}~\bibnamefont {Kong}}, \bibinfo
		{author} {\bibfnamefont {Z.}~\bibnamefont {Jiang}}, \bibinfo {author}
		{\bibfnamefont {Y.}~\bibnamefont {Wu}},\ and\ \bibinfo {author}
		{\bibfnamefont {J.}~\bibnamefont {Du}},\ }\bibfield  {title} {\bibinfo
		{title} {Experimental fault-tolerant universal quantum gates with solid-state
			spins under ambient conditions},\ }\href {https://doi.org/10.1038/ncomms9748}
	{\bibfield  {journal} {\bibinfo  {journal} {Nat. Commun.}\ }\textbf {\bibinfo
			{volume} {6}},\ \bibinfo {pages} {8748} (\bibinfo {year} {2015})}\BibitemShut
	{NoStop}%
	\bibitem [{\citenamefont {Torosov}\ \emph {et~al.}(2021)\citenamefont
		{Torosov}, \citenamefont {Shore},\ and\ \citenamefont
		{Vitanov}}]{Torosov2021}%
	\BibitemOpen
	\bibfield  {author} {\bibinfo {author} {\bibfnamefont {B.~T.}\ \bibnamefont
			{Torosov}}, \bibinfo {author} {\bibfnamefont {B.~W.}\ \bibnamefont {Shore}},\
		and\ \bibinfo {author} {\bibfnamefont {N.~V.}\ \bibnamefont {Vitanov}},\
	}\bibfield  {title} {\bibinfo {title} {Coherent control techniques for
			two-state quantum systems: A comparative study},\ }\href
	{https://doi.org/10.1103/PhysRevA.103.033110} {\bibfield  {journal} {\bibinfo
			{journal} {Phys. Rev. A}\ }\textbf {\bibinfo {volume} {103}},\ \bibinfo
		{pages} {033110} (\bibinfo {year} {2021})}\BibitemShut {NoStop}%
	\bibitem [{\citenamefont {Levitt}(1986)}]{Levitt1986}%
	\BibitemOpen
	\bibfield  {author} {\bibinfo {author} {\bibfnamefont {M.~H.}\ \bibnamefont
			{Levitt}},\ }\bibfield  {title} {\bibinfo {title} {Composite pulses},\ }\href
	{https://doi.org/10.1016/0079-6565(86)80005-X} {\bibfield  {journal}
		{\bibinfo  {journal} {Prog. Nucl. Magn. Reson. Spectrosc.}\ }\textbf
		{\bibinfo {volume} {18}},\ \bibinfo {pages} {61} (\bibinfo {year}
		{1986})}\BibitemShut {NoStop}%
	\bibitem [{\citenamefont {Tycko}(1983)}]{Tycko1983}%
	\BibitemOpen
	\bibfield  {author} {\bibinfo {author} {\bibfnamefont {R.}~\bibnamefont
			{Tycko}},\ }\bibfield  {title} {\bibinfo {title} {Broadband {Population}
			{Inversion}},\ }\href {https://doi.org/10.1103/PhysRevLett.51.775} {\bibfield
		{journal} {\bibinfo  {journal} {Phys. Rev. Lett.}\ }\textbf {\bibinfo
			{volume} {51}},\ \bibinfo {pages} {775} (\bibinfo {year} {1983})}\BibitemShut
	{NoStop}%
	\bibitem [{\citenamefont {Jones}(2009)}]{Jones2009}%
	\BibitemOpen
	\bibfield  {author} {\bibinfo {author} {\bibfnamefont {J.~A.}\ \bibnamefont
			{Jones}},\ }\bibfield  {title} {\bibinfo {title} {Composite pulses in {NMR}
			quantum computation},\ }\href
	{https://doi.org/https://journal.iisc.ac.in/index.php/iisc/article/view/106}
	{\bibfield  {journal} {\bibinfo  {journal} {J. Indian Inst. Sci.}\ }\textbf
		{\bibinfo {volume} {89}},\ \bibinfo {pages} {303} (\bibinfo {year}
		{2009})}\BibitemShut {NoStop}%
	\bibitem [{\citenamefont {Jones}(2013)}]{Jones2013}%
	\BibitemOpen
	\bibfield  {author} {\bibinfo {author} {\bibfnamefont {J.~A.}\ \bibnamefont
			{Jones}},\ }\bibfield  {title} {\bibinfo {title} {Designing short robust
			{NOT} gates for quantum computation},\ }\href
	{https://doi.org/10.1103/PhysRevA.87.052317} {\bibfield  {journal} {\bibinfo
			{journal} {Phys. Rev. A}\ }\textbf {\bibinfo {volume} {87}},\ \bibinfo
		{pages} {052317} (\bibinfo {year} {2013})}\BibitemShut {NoStop}%
	\bibitem [{\citenamefont {Gevorgyan}\ and\ \citenamefont
		{Vitanov}(2021)}]{Gevorgyan2021}%
	\BibitemOpen
	\bibfield  {author} {\bibinfo {author} {\bibfnamefont {H.~L.}\ \bibnamefont
			{Gevorgyan}}\ and\ \bibinfo {author} {\bibfnamefont {N.~V.}\ \bibnamefont
			{Vitanov}},\ }\bibfield  {title} {\bibinfo {title} {Ultrahigh-fidelity
			composite rotational quantum gates},\ }\href
	{https://doi.org/10.1103/PhysRevA.104.012609} {\bibfield  {journal} {\bibinfo
			{journal} {Phys. Rev. A}\ }\textbf {\bibinfo {volume} {104}},\ \bibinfo
		{pages} {012609} (\bibinfo {year} {2021})}\BibitemShut {NoStop}%
	\bibitem [{\citenamefont {Maudsley}(1986)}]{Maudsley1986}%
	\BibitemOpen
	\bibfield  {author} {\bibinfo {author} {\bibfnamefont {A.~A.}\ \bibnamefont
			{Maudsley}},\ }\bibfield  {title} {\bibinfo {title} {Modified
			{Carr}-{Purcell}-{Meiboom}-{Gill} sequence for {NMR} fourier imaging
			applications},\ }\href {https://doi.org/10.1016/0022-2364(86)90160-5}
	{\bibfield  {journal} {\bibinfo  {journal} {J. Magn. Reson.}\ }\textbf
		{\bibinfo {volume} {69}},\ \bibinfo {pages} {488} (\bibinfo {year}
		{1986})}\BibitemShut {NoStop}%
	\bibitem [{\citenamefont {Gullion}\ \emph {et~al.}(1990)\citenamefont
		{Gullion}, \citenamefont {Baker},\ and\ \citenamefont
		{Conradi}}]{Gullion1990}%
	\BibitemOpen
	\bibfield  {author} {\bibinfo {author} {\bibfnamefont {T.}~\bibnamefont
			{Gullion}}, \bibinfo {author} {\bibfnamefont {D.~B.}\ \bibnamefont {Baker}},\
		and\ \bibinfo {author} {\bibfnamefont {M.~S.}\ \bibnamefont {Conradi}},\
	}\bibfield  {title} {\bibinfo {title} {New, compensated {Carr}-{Purcell}
			sequences},\ }\href {https://doi.org/10.1016/0022-2364(90)90331-3} {\bibfield
		{journal} {\bibinfo  {journal} {J. Magn. Reson.}\ }\textbf {\bibinfo
			{volume} {89}},\ \bibinfo {pages} {479} (\bibinfo {year} {1990})}\BibitemShut
	{NoStop}%
	\bibitem [{\citenamefont {Casanova}\ \emph {et~al.}(2015)\citenamefont
		{Casanova}, \citenamefont {Wang}, \citenamefont {Haase},\ and\ \citenamefont
		{Plenio}}]{Casanova2015}%
	\BibitemOpen
	\bibfield  {author} {\bibinfo {author} {\bibfnamefont {J.}~\bibnamefont
			{Casanova}}, \bibinfo {author} {\bibfnamefont {Z.-Y.}\ \bibnamefont {Wang}},
		\bibinfo {author} {\bibfnamefont {J.~F.}\ \bibnamefont {Haase}},\ and\
		\bibinfo {author} {\bibfnamefont {M.~B.}\ \bibnamefont {Plenio}},\ }\bibfield
	{title} {\bibinfo {title} {Robust {dynamical} {decoupling} {sequences} for
			{individual}-nuclear-spin {addressing}},\ }\href
	{https://doi.org/10.1103/PhysRevA.92.042304} {\bibfield  {journal} {\bibinfo
			{journal} {Phys. Rev. A}\ }\textbf {\bibinfo {volume} {92}},\ \bibinfo
		{pages} {042304} (\bibinfo {year} {2015})}\BibitemShut {NoStop}%
	\bibitem [{\citenamefont {Steffen}\ and\ \citenamefont
		{Koch}(2007)}]{Steffen2007}%
	\BibitemOpen
	\bibfield  {author} {\bibinfo {author} {\bibfnamefont {M.}~\bibnamefont
			{Steffen}}\ and\ \bibinfo {author} {\bibfnamefont {R.~H.}\ \bibnamefont
			{Koch}},\ }\bibfield  {title} {\bibinfo {title} {Shaped pulses for quantum
			computing},\ }\href {https://doi.org/10.1103/PhysRevA.75.062326} {\bibfield
		{journal} {\bibinfo  {journal} {Phys. Rev. A}\ }\textbf {\bibinfo {volume}
			{75}},\ \bibinfo {pages} {062326} (\bibinfo {year} {2007})}\BibitemShut
	{NoStop}%
	\bibitem [{\citenamefont {Vandersypen}\ and\ \citenamefont
		{Chuang}(2005)}]{Vandersypen2005}%
	\BibitemOpen
	\bibfield  {author} {\bibinfo {author} {\bibfnamefont {L.~M.~K.}\
			\bibnamefont {Vandersypen}}\ and\ \bibinfo {author} {\bibfnamefont {I.~L.}\
			\bibnamefont {Chuang}},\ }\bibfield  {title} {\bibinfo {title} {{NMR}
			techniques for quantum control and computation},\ }\href
	{https://doi.org/10.1103/RevModPhys.76.1037} {\bibfield  {journal} {\bibinfo
			{journal} {Rev. Mod. Phys.}\ }\textbf {\bibinfo {volume} {76}},\ \bibinfo
		{pages} {1037} (\bibinfo {year} {2005})}\BibitemShut {NoStop}%
	\bibitem [{\citenamefont {Haase}\ \emph {et~al.}(2018)\citenamefont {Haase},
		\citenamefont {Wang}, \citenamefont {Casanova},\ and\ \citenamefont
		{Plenio}}]{Haase2018}%
	\BibitemOpen
	\bibfield  {author} {\bibinfo {author} {\bibfnamefont {J.~F.}\ \bibnamefont
			{Haase}}, \bibinfo {author} {\bibfnamefont {Z.-Y.}\ \bibnamefont {Wang}},
		\bibinfo {author} {\bibfnamefont {J.}~\bibnamefont {Casanova}},\ and\
		\bibinfo {author} {\bibfnamefont {M.~B.}\ \bibnamefont {Plenio}},\ }\bibfield
	{title} {\bibinfo {title} {Soft {Quantum} {Control} for {Highly} {Selective}
			{Interactions} among {Joint} {Quantum} {Systems}},\ }\href
	{https://doi.org/10.1103/PhysRevLett.121.050402} {\bibfield  {journal}
		{\bibinfo  {journal} {Phys. Rev. Lett.}\ }\textbf {\bibinfo {volume} {121}},\
		\bibinfo {pages} {050402} (\bibinfo {year} {2018})}\BibitemShut {NoStop}%
	\bibitem [{\citenamefont {Johnson}\ \emph {et~al.}(2017)\citenamefont
		{Johnson}, \citenamefont {Dolan},\ and\ \citenamefont {Smith}}]{Johnson2017}%
	\BibitemOpen
	\bibfield  {author} {\bibinfo {author} {\bibfnamefont {S.}~\bibnamefont
			{Johnson}}, \bibinfo {author} {\bibfnamefont {P.~R.}\ \bibnamefont {Dolan}},\
		and\ \bibinfo {author} {\bibfnamefont {J.~M.}\ \bibnamefont {Smith}},\
	}\bibfield  {title} {\bibinfo {title} {Diamond photonics for distributed
			quantum networks},\ }\href {https://doi.org/10.1016/j.pquantelec.2017.05.003}
	{\bibfield  {journal} {\bibinfo  {journal} {Prog. Quantum Electron.}\
		}\textbf {\bibinfo {volume} {55}},\ \bibinfo {pages} {129} (\bibinfo {year}
		{2017})}\BibitemShut {NoStop}%
	\bibitem [{\citenamefont {Aharonovich}\ \emph {et~al.}(2011)\citenamefont
		{Aharonovich}, \citenamefont {Greentree},\ and\ \citenamefont
		{Prawer}}]{Aharonovich2011}%
	\BibitemOpen
	\bibfield  {author} {\bibinfo {author} {\bibfnamefont {I.}~\bibnamefont
			{Aharonovich}}, \bibinfo {author} {\bibfnamefont {A.~D.}\ \bibnamefont
			{Greentree}},\ and\ \bibinfo {author} {\bibfnamefont {S.}~\bibnamefont
			{Prawer}},\ }\bibfield  {title} {\bibinfo {title} {Diamond photonics},\
	}\href {https://doi.org/10.1038/nphoton.2011.54} {\bibfield  {journal}
		{\bibinfo  {journal} {Nat. Photonics}\ }\textbf {\bibinfo {volume} {5}},\
		\bibinfo {pages} {397} (\bibinfo {year} {2011})}\BibitemShut {NoStop}%
	\bibitem [{\citenamefont {Beukers}\ \emph {et~al.}(2024)\citenamefont
		{Beukers}, \citenamefont {Pasini}, \citenamefont {Choi}, \citenamefont
		{Englund}, \citenamefont {Hanson},\ and\ \citenamefont
		{Borregaard}}]{Beukers2024}%
	\BibitemOpen
	\bibfield  {author} {\bibinfo {author} {\bibfnamefont {H.~K.}\ \bibnamefont
			{Beukers}}, \bibinfo {author} {\bibfnamefont {M.}~\bibnamefont {Pasini}},
		\bibinfo {author} {\bibfnamefont {H.}~\bibnamefont {Choi}}, \bibinfo {author}
		{\bibfnamefont {D.}~\bibnamefont {Englund}}, \bibinfo {author} {\bibfnamefont
			{R.}~\bibnamefont {Hanson}},\ and\ \bibinfo {author} {\bibfnamefont
			{J.}~\bibnamefont {Borregaard}},\ }\bibfield  {title} {\bibinfo {title}
		{Remote-{Entanglement} {Protocols} for {Stationary} {Qubits} with {Photonic}
			{Interfaces}},\ }\href {https://doi.org/10.1103/PRXQuantum.5.010202}
	{\bibfield  {journal} {\bibinfo  {journal} {PRX Quantum}\ }\textbf {\bibinfo
			{volume} {5}},\ \bibinfo {pages} {010202} (\bibinfo {year}
		{2024})}\BibitemShut {NoStop}%
	\bibitem [{\citenamefont {Ruf}\ \emph {et~al.}(2021)\citenamefont {Ruf},
		\citenamefont {Wan}, \citenamefont {Choi}, \citenamefont {Englund},\ and\
		\citenamefont {Hanson}}]{Ruf2021a}%
	\BibitemOpen
	\bibfield  {author} {\bibinfo {author} {\bibfnamefont {M.}~\bibnamefont
			{Ruf}}, \bibinfo {author} {\bibfnamefont {N.~H.}\ \bibnamefont {Wan}},
		\bibinfo {author} {\bibfnamefont {H.}~\bibnamefont {Choi}}, \bibinfo {author}
		{\bibfnamefont {D.}~\bibnamefont {Englund}},\ and\ \bibinfo {author}
		{\bibfnamefont {R.}~\bibnamefont {Hanson}},\ }\bibfield  {title} {\bibinfo
		{title} {Quantum networks based on color centers in diamond},\ }\href
	{https://doi.org/10.1063/5.0056534} {\bibfield  {journal} {\bibinfo
			{journal} {J. Appl. Phys.}\ }\textbf {\bibinfo {volume} {130}},\ \bibinfo
		{pages} {070901} (\bibinfo {year} {2021})}\BibitemShut {NoStop}%
	\bibitem [{\citenamefont {Browne}\ \emph {et~al.}(2003)\citenamefont {Browne},
		\citenamefont {Plenio},\ and\ \citenamefont {Huelga}}]{Browne2003}%
	\BibitemOpen
	\bibfield  {author} {\bibinfo {author} {\bibfnamefont {D.~E.}\ \bibnamefont
			{Browne}}, \bibinfo {author} {\bibfnamefont {M.~B.}\ \bibnamefont {Plenio}},\
		and\ \bibinfo {author} {\bibfnamefont {S.~F.}\ \bibnamefont {Huelga}},\
	}\bibfield  {title} {\bibinfo {title} {Robust {Creation} of {Entanglement}
			between {Ions} in {Spatially} {Separate} {Cavities}},\ }\href
	{https://doi.org/10.1103/PhysRevLett.91.067901} {\bibfield  {journal}
		{\bibinfo  {journal} {Phys. Rev. Lett.}\ }\textbf {\bibinfo {volume} {91}},\
		\bibinfo {pages} {067901} (\bibinfo {year} {2003})}\BibitemShut {NoStop}%
	\bibitem [{\citenamefont {Barrett}\ and\ \citenamefont
		{Kok}(2005)}]{Barrett2005}%
	\BibitemOpen
	\bibfield  {author} {\bibinfo {author} {\bibfnamefont {S.~D.}\ \bibnamefont
			{Barrett}}\ and\ \bibinfo {author} {\bibfnamefont {P.}~\bibnamefont {Kok}},\
	}\bibfield  {title} {\bibinfo {title} {Efficient high-fidelity quantum
			computation using matter qubits and linear optics},\ }\href
	{https://doi.org/10.1103/PhysRevA.71.060310} {\bibfield  {journal} {\bibinfo
			{journal} {Phys. Rev. A}\ }\textbf {\bibinfo {volume} {71}},\ \bibinfo
		{pages} {060310(R)} (\bibinfo {year} {2005})}\BibitemShut {NoStop}%
	\bibitem [{\citenamefont {Bernien}\ \emph {et~al.}(2013)\citenamefont
		{Bernien}, \citenamefont {Hensen}, \citenamefont {Pfaff}, \citenamefont
		{Koolstra}, \citenamefont {Blok}, \citenamefont {Robledo}, \citenamefont
		{Taminiau}, \citenamefont {Markham}, \citenamefont {Twitchen}, \citenamefont
		{Childress},\ and\ \citenamefont {Hanson}}]{Bernien2013}%
	\BibitemOpen
	\bibfield  {author} {\bibinfo {author} {\bibfnamefont {H.}~\bibnamefont
			{Bernien}}, \bibinfo {author} {\bibfnamefont {B.}~\bibnamefont {Hensen}},
		\bibinfo {author} {\bibfnamefont {W.}~\bibnamefont {Pfaff}}, \bibinfo
		{author} {\bibfnamefont {G.}~\bibnamefont {Koolstra}}, \bibinfo {author}
		{\bibfnamefont {M.~S.}\ \bibnamefont {Blok}}, \bibinfo {author}
		{\bibfnamefont {L.}~\bibnamefont {Robledo}}, \bibinfo {author} {\bibfnamefont
			{T.~H.}\ \bibnamefont {Taminiau}}, \bibinfo {author} {\bibfnamefont
			{M.}~\bibnamefont {Markham}}, \bibinfo {author} {\bibfnamefont {D.~J.}\
			\bibnamefont {Twitchen}}, \bibinfo {author} {\bibfnamefont {L.}~\bibnamefont
			{Childress}},\ and\ \bibinfo {author} {\bibfnamefont {R.}~\bibnamefont
			{Hanson}},\ }\bibfield  {title} {\bibinfo {title} {Heralded entanglement
			between solid-state qubits separated by three metres},\ }\href
	{https://doi.org/10.1038/nature12016} {\bibfield  {journal} {\bibinfo
			{journal} {Nature}\ }\textbf {\bibinfo {volume} {497}},\ \bibinfo {pages}
		{86} (\bibinfo {year} {2013})}\BibitemShut {NoStop}%
	\bibitem [{\citenamefont {Hensen}\ \emph {et~al.}(2015)\citenamefont {Hensen},
		\citenamefont {Bernien}, \citenamefont {Dréau}, \citenamefont {Reiserer},
		\citenamefont {Kalb}, \citenamefont {Blok}, \citenamefont {Ruitenberg},
		\citenamefont {Vermeulen}, \citenamefont {Schouten}, \citenamefont
		{Abellán}, \citenamefont {Amaya}, \citenamefont {Pruneri}, \citenamefont
		{Mitchell}, \citenamefont {Markham}, \citenamefont {Twitchen}, \citenamefont
		{Elkouss}, \citenamefont {Wehner}, \citenamefont {Taminiau},\ and\
		\citenamefont {Hanson}}]{Hensen2015}%
	\BibitemOpen
	\bibfield  {author} {\bibinfo {author} {\bibfnamefont {B.}~\bibnamefont
			{Hensen}}, \bibinfo {author} {\bibfnamefont {H.}~\bibnamefont {Bernien}},
		\bibinfo {author} {\bibfnamefont {A.~E.}\ \bibnamefont {Dréau}}, \bibinfo
		{author} {\bibfnamefont {A.}~\bibnamefont {Reiserer}}, \bibinfo {author}
		{\bibfnamefont {N.}~\bibnamefont {Kalb}}, \bibinfo {author} {\bibfnamefont
			{M.~S.}\ \bibnamefont {Blok}}, \bibinfo {author} {\bibfnamefont
			{J.}~\bibnamefont {Ruitenberg}}, \bibinfo {author} {\bibfnamefont {R.~F.~L.}\
			\bibnamefont {Vermeulen}}, \bibinfo {author} {\bibfnamefont {R.~N.}\
			\bibnamefont {Schouten}}, \bibinfo {author} {\bibfnamefont {C.}~\bibnamefont
			{Abellán}}, \bibinfo {author} {\bibfnamefont {W.}~\bibnamefont {Amaya}},
		\bibinfo {author} {\bibfnamefont {V.}~\bibnamefont {Pruneri}}, \bibinfo
		{author} {\bibfnamefont {M.~W.}\ \bibnamefont {Mitchell}}, \bibinfo {author}
		{\bibfnamefont {M.}~\bibnamefont {Markham}}, \bibinfo {author} {\bibfnamefont
			{D.~J.}\ \bibnamefont {Twitchen}}, \bibinfo {author} {\bibfnamefont
			{D.}~\bibnamefont {Elkouss}}, \bibinfo {author} {\bibfnamefont
			{S.}~\bibnamefont {Wehner}}, \bibinfo {author} {\bibfnamefont {T.~H.}\
			\bibnamefont {Taminiau}},\ and\ \bibinfo {author} {\bibfnamefont
			{R.}~\bibnamefont {Hanson}},\ }\bibfield  {title} {\bibinfo {title}
		{Loophole-free {Bell} inequality violation using electron spins separated by
			1.3 kilometres},\ }\href {https://doi.org/10.1038/nature15759} {\bibfield
		{journal} {\bibinfo  {journal} {Nature}\ }\textbf {\bibinfo {volume} {526}},\
		\bibinfo {pages} {682} (\bibinfo {year} {2015})}\BibitemShut {NoStop}%
	\bibitem [{\citenamefont {Pompili}\ \emph {et~al.}(2021)\citenamefont
		{Pompili}, \citenamefont {Hermans}, \citenamefont {Baier}, \citenamefont
		{Beukers}, \citenamefont {Humphreys}, \citenamefont {Schouten}, \citenamefont
		{Vermeulen}, \citenamefont {Tiggelman}, \citenamefont {dos Santos~Martins},
		\citenamefont {Dirkse}, \citenamefont {Wehner},\ and\ \citenamefont
		{Hanson}}]{Pompili2021}%
	\BibitemOpen
	\bibfield  {author} {\bibinfo {author} {\bibfnamefont {M.}~\bibnamefont
			{Pompili}}, \bibinfo {author} {\bibfnamefont {S.~L.~N.}\ \bibnamefont
			{Hermans}}, \bibinfo {author} {\bibfnamefont {S.}~\bibnamefont {Baier}},
		\bibinfo {author} {\bibfnamefont {H.~K.~C.}\ \bibnamefont {Beukers}},
		\bibinfo {author} {\bibfnamefont {P.~C.}\ \bibnamefont {Humphreys}}, \bibinfo
		{author} {\bibfnamefont {R.~N.}\ \bibnamefont {Schouten}}, \bibinfo {author}
		{\bibfnamefont {R.~F.~L.}\ \bibnamefont {Vermeulen}}, \bibinfo {author}
		{\bibfnamefont {M.~J.}\ \bibnamefont {Tiggelman}}, \bibinfo {author}
		{\bibfnamefont {L.}~\bibnamefont {dos Santos~Martins}}, \bibinfo {author}
		{\bibfnamefont {B.}~\bibnamefont {Dirkse}}, \bibinfo {author} {\bibfnamefont
			{S.}~\bibnamefont {Wehner}},\ and\ \bibinfo {author} {\bibfnamefont
			{R.}~\bibnamefont {Hanson}},\ }\bibfield  {title} {\bibinfo {title}
		{Realization of a multinode quantum network of remote solid-state qubits},\
	}\href {https://doi.org/10.1126/science.abg1919} {\bibfield  {journal}
		{\bibinfo  {journal} {Science}\ }\textbf {\bibinfo {volume} {372}},\ \bibinfo
		{pages} {259} (\bibinfo {year} {2021})}\BibitemShut {NoStop}%
	\bibitem [{\citenamefont {Kok}\ \emph {et~al.}(2007)\citenamefont {Kok},
		\citenamefont {Munro}, \citenamefont {Nemoto}, \citenamefont {Ralph},
		\citenamefont {Dowling},\ and\ \citenamefont {Milburn}}]{Kok2007}%
	\BibitemOpen
	\bibfield  {author} {\bibinfo {author} {\bibfnamefont {P.}~\bibnamefont
			{Kok}}, \bibinfo {author} {\bibfnamefont {W.~J.}\ \bibnamefont {Munro}},
		\bibinfo {author} {\bibfnamefont {K.}~\bibnamefont {Nemoto}}, \bibinfo
		{author} {\bibfnamefont {T.~C.}\ \bibnamefont {Ralph}}, \bibinfo {author}
		{\bibfnamefont {J.~P.}\ \bibnamefont {Dowling}},\ and\ \bibinfo {author}
		{\bibfnamefont {G.~J.}\ \bibnamefont {Milburn}},\ }\bibfield  {title}
	{\bibinfo {title} {Linear optical quantum computing with photonic qubits},\
	}\href {https://doi.org/10.1103/RevModPhys.79.135} {\bibfield  {journal}
		{\bibinfo  {journal} {Rev. Mod. Phys.}\ }\textbf {\bibinfo {volume} {79}},\
		\bibinfo {pages} {135} (\bibinfo {year} {2007})}\BibitemShut {NoStop}%
	\bibitem [{\citenamefont {Reiserer}\ and\ \citenamefont
		{Rempe}(2015)}]{Reiserer2015}%
	\BibitemOpen
	\bibfield  {author} {\bibinfo {author} {\bibfnamefont {A.}~\bibnamefont
			{Reiserer}}\ and\ \bibinfo {author} {\bibfnamefont {G.}~\bibnamefont
			{Rempe}},\ }\bibfield  {title} {\bibinfo {title} {Cavity-based quantum
			networks with single atoms and optical photons},\ }\href
	{https://doi.org/10.1103/RevModPhys.87.1379} {\bibfield  {journal} {\bibinfo
			{journal} {Rev. Mod. Phys.}\ }\textbf {\bibinfo {volume} {87}},\ \bibinfo
		{pages} {1379} (\bibinfo {year} {2015})}\BibitemShut {NoStop}%
	\bibitem [{\citenamefont {Bhaskar}\ \emph {et~al.}(2020)\citenamefont
		{Bhaskar}, \citenamefont {Riedinger}, \citenamefont {Machielse},
		\citenamefont {Levonian}, \citenamefont {Nguyen}, \citenamefont {Knall},
		\citenamefont {Park}, \citenamefont {Englund}, \citenamefont {Lončar},
		\citenamefont {Sukachev},\ and\ \citenamefont {Lukin}}]{Bhaskar2020}%
	\BibitemOpen
	\bibfield  {author} {\bibinfo {author} {\bibfnamefont {M.~K.}\ \bibnamefont
			{Bhaskar}}, \bibinfo {author} {\bibfnamefont {R.}~\bibnamefont {Riedinger}},
		\bibinfo {author} {\bibfnamefont {B.}~\bibnamefont {Machielse}}, \bibinfo
		{author} {\bibfnamefont {D.~S.}\ \bibnamefont {Levonian}}, \bibinfo {author}
		{\bibfnamefont {C.~T.}\ \bibnamefont {Nguyen}}, \bibinfo {author}
		{\bibfnamefont {E.~N.}\ \bibnamefont {Knall}}, \bibinfo {author}
		{\bibfnamefont {H.}~\bibnamefont {Park}}, \bibinfo {author} {\bibfnamefont
			{D.}~\bibnamefont {Englund}}, \bibinfo {author} {\bibfnamefont
			{M.}~\bibnamefont {Lončar}}, \bibinfo {author} {\bibfnamefont {D.~D.}\
			\bibnamefont {Sukachev}},\ and\ \bibinfo {author} {\bibfnamefont {M.~D.}\
			\bibnamefont {Lukin}},\ }\bibfield  {title} {\bibinfo {title} {Experimental
			demonstration of memory-enhanced quantum communication},\ }\href
	{https://doi.org/10.1038/s41586-020-2103-5} {\bibfield  {journal} {\bibinfo
			{journal} {Nature}\ }\textbf {\bibinfo {volume} {580}},\ \bibinfo {pages}
		{60} (\bibinfo {year} {2020})}\BibitemShut {NoStop}%
	\bibitem [{\citenamefont {Riedel}\ \emph {et~al.}(2017)\citenamefont {Riedel},
		\citenamefont {Söllner}, \citenamefont {Shields}, \citenamefont
		{Starosielec}, \citenamefont {Appel}, \citenamefont {Neu}, \citenamefont
		{Maletinsky},\ and\ \citenamefont {Warburton}}]{Riedel2017}%
	\BibitemOpen
	\bibfield  {author} {\bibinfo {author} {\bibfnamefont {D.}~\bibnamefont
			{Riedel}}, \bibinfo {author} {\bibfnamefont {I.}~\bibnamefont {Söllner}},
		\bibinfo {author} {\bibfnamefont {B.~J.}\ \bibnamefont {Shields}}, \bibinfo
		{author} {\bibfnamefont {S.}~\bibnamefont {Starosielec}}, \bibinfo {author}
		{\bibfnamefont {P.}~\bibnamefont {Appel}}, \bibinfo {author} {\bibfnamefont
			{E.}~\bibnamefont {Neu}}, \bibinfo {author} {\bibfnamefont {P.}~\bibnamefont
			{Maletinsky}},\ and\ \bibinfo {author} {\bibfnamefont {R.~J.}\ \bibnamefont
			{Warburton}},\ }\bibfield  {title} {\bibinfo {title} {Deterministic
			{Enhancement} of {Coherent} {Photon} {Generation} from a {Nitrogen}-{Vacancy}
			{Center} in {Ultrapure} {Diamond}},\ }\href
	{https://doi.org/10.1103/PhysRevX.7.031040} {\bibfield  {journal} {\bibinfo
			{journal} {Phys. Rev. X}\ }\textbf {\bibinfo {volume} {7}},\ \bibinfo {pages}
		{031040} (\bibinfo {year} {2017})}\BibitemShut {NoStop}%
	\bibitem [{\citenamefont {Yurgens}\ \emph {et~al.}(2024)\citenamefont
		{Yurgens}, \citenamefont {Fontana}, \citenamefont {Corazza}, \citenamefont
		{Shields}, \citenamefont {Maletinsky},\ and\ \citenamefont
		{Warburton}}]{Yurgens2024}%
	\BibitemOpen
	\bibfield  {author} {\bibinfo {author} {\bibfnamefont {V.}~\bibnamefont
			{Yurgens}}, \bibinfo {author} {\bibfnamefont {Y.}~\bibnamefont {Fontana}},
		\bibinfo {author} {\bibfnamefont {A.}~\bibnamefont {Corazza}}, \bibinfo
		{author} {\bibfnamefont {B.~J.}\ \bibnamefont {Shields}}, \bibinfo {author}
		{\bibfnamefont {P.}~\bibnamefont {Maletinsky}},\ and\ \bibinfo {author}
		{\bibfnamefont {R.~J.}\ \bibnamefont {Warburton}},\ }\bibfield  {title}
	{\bibinfo {title} {Cavity-assisted resonance fluorescence from a
			nitrogen-vacancy center in diamond},\ }\href
	{https://doi.org/10.1038/s41534-024-00915-9} {\bibfield  {journal} {\bibinfo
			{journal} {npj Quantum Inf.}\ }\textbf {\bibinfo {volume} {10}},\ \bibinfo
		{pages} {112} (\bibinfo {year} {2024})}\BibitemShut {NoStop}%
	\bibitem [{\citenamefont {Omlor}\ \emph {et~al.}(2025)\citenamefont {Omlor},
		\citenamefont {Tissot},\ and\ \citenamefont {Burkard}}]{Omlor2025}%
	\BibitemOpen
	\bibfield  {author} {\bibinfo {author} {\bibfnamefont {F.}~\bibnamefont
			{Omlor}}, \bibinfo {author} {\bibfnamefont {B.}~\bibnamefont {Tissot}},\ and\
		\bibinfo {author} {\bibfnamefont {G.}~\bibnamefont {Burkard}},\ }\bibfield
	{title} {\bibinfo {title} {Entanglement generation using single-photon pulse
			reflection in realistic networks},\ }\href
	{https://doi.org/10.1103/PhysRevA.111.012612} {\bibfield  {journal} {\bibinfo
			{journal} {Phys. Rev. A}\ }\textbf {\bibinfo {volume} {111}},\ \bibinfo
		{pages} {012612} (\bibinfo {year} {2025})}\BibitemShut {NoStop}%
	\bibitem [{\citenamefont {Koshino}\ and\ \citenamefont
		{Matsuzaki}(2012)}]{Koshino2012}%
	\BibitemOpen
	\bibfield  {author} {\bibinfo {author} {\bibfnamefont {K.}~\bibnamefont
			{Koshino}}\ and\ \bibinfo {author} {\bibfnamefont {Y.}~\bibnamefont
			{Matsuzaki}},\ }\bibfield  {title} {\bibinfo {title} {Entangling
			homogeneously broadened matter qubits in the weak-coupling cavity-qed
			regime},\ }\href {https://doi.org/10.1103/PhysRevA.86.020305} {\bibfield
		{journal} {\bibinfo  {journal} {Phys. Rev. A}\ }\textbf {\bibinfo {volume}
			{86}},\ \bibinfo {pages} {020305} (\bibinfo {year} {2012})}\BibitemShut
	{NoStop}%
	\bibitem [{\citenamefont {Lukens}\ and\ \citenamefont
		{Lougovski}(2017)}]{Lukens2017}%
	\BibitemOpen
	\bibfield  {author} {\bibinfo {author} {\bibfnamefont {J.~M.}\ \bibnamefont
			{Lukens}}\ and\ \bibinfo {author} {\bibfnamefont {P.}~\bibnamefont
			{Lougovski}},\ }\bibfield  {title} {\bibinfo {title} {Frequency-encoded
			photonic qubits for scalable quantum information processing},\ }\href
	{https://doi.org/10.1364/OPTICA.4.000008} {\bibfield  {journal} {\bibinfo
			{journal} {Optica}\ }\textbf {\bibinfo {volume} {4}},\ \bibinfo {pages} {8}
		(\bibinfo {year} {2017})}\BibitemShut {NoStop}%
	\bibitem [{\citenamefont {Clementi}\ \emph {et~al.}(2023)\citenamefont
		{Clementi}, \citenamefont {Sabattoli}, \citenamefont {Borghi}, \citenamefont
		{Gianini}, \citenamefont {Tagliavacche}, \citenamefont {El~Dirani},
		\citenamefont {Youssef}, \citenamefont {Bergamasco}, \citenamefont
		{Petit-Etienne}, \citenamefont {Pargon}, \citenamefont {Sipe}, \citenamefont
		{Liscidini}, \citenamefont {Sciancalepore}, \citenamefont {Galli},\ and\
		\citenamefont {Bajoni}}]{Clementi2023}%
	\BibitemOpen
	\bibfield  {author} {\bibinfo {author} {\bibfnamefont {M.}~\bibnamefont
			{Clementi}}, \bibinfo {author} {\bibfnamefont {F.~A.}\ \bibnamefont
			{Sabattoli}}, \bibinfo {author} {\bibfnamefont {M.}~\bibnamefont {Borghi}},
		\bibinfo {author} {\bibfnamefont {L.}~\bibnamefont {Gianini}}, \bibinfo
		{author} {\bibfnamefont {N.}~\bibnamefont {Tagliavacche}}, \bibinfo {author}
		{\bibfnamefont {H.}~\bibnamefont {El~Dirani}}, \bibinfo {author}
		{\bibfnamefont {L.}~\bibnamefont {Youssef}}, \bibinfo {author} {\bibfnamefont
			{N.}~\bibnamefont {Bergamasco}}, \bibinfo {author} {\bibfnamefont
			{C.}~\bibnamefont {Petit-Etienne}}, \bibinfo {author} {\bibfnamefont
			{E.}~\bibnamefont {Pargon}}, \bibinfo {author} {\bibfnamefont {J.~E.}\
			\bibnamefont {Sipe}}, \bibinfo {author} {\bibfnamefont {M.}~\bibnamefont
			{Liscidini}}, \bibinfo {author} {\bibfnamefont {C.}~\bibnamefont
			{Sciancalepore}}, \bibinfo {author} {\bibfnamefont {M.}~\bibnamefont
			{Galli}},\ and\ \bibinfo {author} {\bibfnamefont {D.}~\bibnamefont
			{Bajoni}},\ }\bibfield  {title} {\bibinfo {title} {Programmable frequency-bin
			quantum states in a nano-engineered silicon device},\ }\href
	{https://doi.org/10.1038/s41467-022-35773-6} {\bibfield  {journal} {\bibinfo
			{journal} {Nat. Commun.}\ }\textbf {\bibinfo {volume} {14}},\ \bibinfo
		{pages} {176} (\bibinfo {year} {2023})}\BibitemShut {NoStop}%
	\bibitem [{\citenamefont {Tscherbul}\ \emph {et~al.}(2023)\citenamefont
		{Tscherbul}, \citenamefont {Ye},\ and\ \citenamefont {Rey}}]{Tscherbul2023}%
	\BibitemOpen
	\bibfield  {author} {\bibinfo {author} {\bibfnamefont {T.~V.}\ \bibnamefont
			{Tscherbul}}, \bibinfo {author} {\bibfnamefont {J.}~\bibnamefont {Ye}},\ and\
		\bibinfo {author} {\bibfnamefont {A.~M.}\ \bibnamefont {Rey}},\ }\bibfield
	{title} {\bibinfo {title} {Robust nuclear spin entanglement via dipolar
			interactions in polar molecules},\ }\href
	{https://doi.org/10.1103/PhysRevLett.130.143002} {\bibfield  {journal}
		{\bibinfo  {journal} {Phys. Rev. Lett.}\ }\textbf {\bibinfo {volume} {130}},\
		\bibinfo {pages} {143002} (\bibinfo {year} {2023})}\BibitemShut {NoStop}%
	\bibitem [{\citenamefont {Le}\ \emph {et~al.}(2023)\citenamefont {Le},
		\citenamefont {Cykiert},\ and\ \citenamefont {Ginossar}}]{Le2023}%
	\BibitemOpen
	\bibfield  {author} {\bibinfo {author} {\bibfnamefont {N.~H.}\ \bibnamefont
			{Le}}, \bibinfo {author} {\bibfnamefont {M.}~\bibnamefont {Cykiert}},\ and\
		\bibinfo {author} {\bibfnamefont {E.}~\bibnamefont {Ginossar}},\ }\bibfield
	{title} {\bibinfo {title} {Scalable and robust quantum computing on qubit
			arrays with fixed coupling},\ }\href
	{https://doi.org/10.1038/s41534-022-00668-3} {\bibfield  {journal} {\bibinfo
			{journal} {npj Quantum Inf.}\ }\textbf {\bibinfo {volume} {9}},\ \bibinfo
		{pages} {1} (\bibinfo {year} {2023})}\BibitemShut {NoStop}%
	\bibitem [{\citenamefont {Rajabi}\ \emph {et~al.}(2019)\citenamefont {Rajabi},
		\citenamefont {Motlakunta}, \citenamefont {Shih}, \citenamefont
		{Kotibhaskar}, \citenamefont {Quraishi}, \citenamefont {Ajoy},\ and\
		\citenamefont {Islam}}]{Rajabi2019}%
	\BibitemOpen
	\bibfield  {author} {\bibinfo {author} {\bibfnamefont {F.}~\bibnamefont
			{Rajabi}}, \bibinfo {author} {\bibfnamefont {S.}~\bibnamefont {Motlakunta}},
		\bibinfo {author} {\bibfnamefont {C.-Y.}\ \bibnamefont {Shih}}, \bibinfo
		{author} {\bibfnamefont {N.}~\bibnamefont {Kotibhaskar}}, \bibinfo {author}
		{\bibfnamefont {Q.}~\bibnamefont {Quraishi}}, \bibinfo {author}
		{\bibfnamefont {A.}~\bibnamefont {Ajoy}},\ and\ \bibinfo {author}
		{\bibfnamefont {R.}~\bibnamefont {Islam}},\ }\bibfield  {title} {\bibinfo
		{title} {Dynamical {Hamiltonian} engineering of {2D} rectangular lattices in
			a one-dimensional ion chain},\ }\href
	{https://doi.org/10.1038/s41534-019-0147-x} {\bibfield  {journal} {\bibinfo
			{journal} {npj Quantum Inf.}\ }\textbf {\bibinfo {volume} {5}},\ \bibinfo
		{pages} {1} (\bibinfo {year} {2019})}\BibitemShut {NoStop}%
	\bibitem [{\citenamefont {Torosov}\ and\ \citenamefont
		{Vitanov}(2014)}]{Torosov2014}%
	\BibitemOpen
	\bibfield  {author} {\bibinfo {author} {\bibfnamefont {B.~T.}\ \bibnamefont
			{Torosov}}\ and\ \bibinfo {author} {\bibfnamefont {N.~V.}\ \bibnamefont
			{Vitanov}},\ }\bibfield  {title} {\bibinfo {title} {High-fidelity
			error-resilient composite phase gates},\ }\href
	{https://doi.org/10.1103/PhysRevA.90.012341} {\bibfield  {journal} {\bibinfo
			{journal} {Phys. Rev. A}\ }\textbf {\bibinfo {volume} {90}},\ \bibinfo
		{pages} {012341} (\bibinfo {year} {2014})}\BibitemShut {NoStop}%
	\bibitem [{\citenamefont {Barenco}\ \emph {et~al.}(1995)\citenamefont
		{Barenco}, \citenamefont {Bennett}, \citenamefont {Cleve}, \citenamefont
		{DiVincenzo}, \citenamefont {Margolus}, \citenamefont {Shor}, \citenamefont
		{Sleator}, \citenamefont {Smolin},\ and\ \citenamefont
		{Weinfurter}}]{Barenco1995}%
	\BibitemOpen
	\bibfield  {author} {\bibinfo {author} {\bibfnamefont {A.}~\bibnamefont
			{Barenco}}, \bibinfo {author} {\bibfnamefont {C.~H.}\ \bibnamefont
			{Bennett}}, \bibinfo {author} {\bibfnamefont {R.}~\bibnamefont {Cleve}},
		\bibinfo {author} {\bibfnamefont {D.~P.}\ \bibnamefont {DiVincenzo}},
		\bibinfo {author} {\bibfnamefont {N.}~\bibnamefont {Margolus}}, \bibinfo
		{author} {\bibfnamefont {P.}~\bibnamefont {Shor}}, \bibinfo {author}
		{\bibfnamefont {T.}~\bibnamefont {Sleator}}, \bibinfo {author} {\bibfnamefont
			{J.~A.}\ \bibnamefont {Smolin}},\ and\ \bibinfo {author} {\bibfnamefont
			{H.}~\bibnamefont {Weinfurter}},\ }\bibfield  {title} {\bibinfo {title}
		{Elementary gates for quantum computation},\ }\href
	{https://doi.org/10.1103/PhysRevA.52.3457} {\bibfield  {journal} {\bibinfo
			{journal} {Phys. Rev. A}\ }\textbf {\bibinfo {volume} {52}},\ \bibinfo
		{pages} {3457} (\bibinfo {year} {1995})}\BibitemShut {NoStop}%
	\bibitem [{\citenamefont {Dutt}\ \emph {et~al.}(2007)\citenamefont {Dutt},
		\citenamefont {Childress}, \citenamefont {Jiang}, \citenamefont {Togan},
		\citenamefont {Maze}, \citenamefont {Jelezko}, \citenamefont {Zibrov},
		\citenamefont {Hemmer},\ and\ \citenamefont {Lukin}}]{Dutt2007}%
	\BibitemOpen
	\bibfield  {author} {\bibinfo {author} {\bibfnamefont {M.~V.~G.}\
			\bibnamefont {Dutt}}, \bibinfo {author} {\bibfnamefont {L.}~\bibnamefont
			{Childress}}, \bibinfo {author} {\bibfnamefont {L.}~\bibnamefont {Jiang}},
		\bibinfo {author} {\bibfnamefont {E.}~\bibnamefont {Togan}}, \bibinfo
		{author} {\bibfnamefont {J.}~\bibnamefont {Maze}}, \bibinfo {author}
		{\bibfnamefont {F.}~\bibnamefont {Jelezko}}, \bibinfo {author} {\bibfnamefont
			{A.~S.}\ \bibnamefont {Zibrov}}, \bibinfo {author} {\bibfnamefont {P.~R.}\
			\bibnamefont {Hemmer}},\ and\ \bibinfo {author} {\bibfnamefont {M.~D.}\
			\bibnamefont {Lukin}},\ }\bibfield  {title} {\bibinfo {title} {Quantum
			{Register} {Based} on {Individual} {Electronic} and {Nuclear} {Spin} {Qubits}
			in {Diamond}},\ }\href {https://doi.org/10.1126/science.1139831} {\bibfield
		{journal} {\bibinfo  {journal} {Science}\ }\textbf {\bibinfo {volume}
			{316}},\ \bibinfo {pages} {1312} (\bibinfo {year} {2007})}\BibitemShut
	{NoStop}%
	\bibitem [{\citenamefont {Neumann}\ \emph {et~al.}(2010)\citenamefont
		{Neumann}, \citenamefont {Kolesov}, \citenamefont {Naydenov}, \citenamefont
		{Beck}, \citenamefont {Rempp}, \citenamefont {Steiner}, \citenamefont
		{Jacques}, \citenamefont {Balasubramanian}, \citenamefont {Markham},
		\citenamefont {Twitchen}, \citenamefont {Pezzagna}, \citenamefont {Meijer},
		\citenamefont {Twamley}, \citenamefont {Jelezko},\ and\ \citenamefont
		{Wrachtrup}}]{Neumann2010}%
	\BibitemOpen
	\bibfield  {author} {\bibinfo {author} {\bibfnamefont {P.}~\bibnamefont
			{Neumann}}, \bibinfo {author} {\bibfnamefont {R.}~\bibnamefont {Kolesov}},
		\bibinfo {author} {\bibfnamefont {B.}~\bibnamefont {Naydenov}}, \bibinfo
		{author} {\bibfnamefont {J.}~\bibnamefont {Beck}}, \bibinfo {author}
		{\bibfnamefont {F.}~\bibnamefont {Rempp}}, \bibinfo {author} {\bibfnamefont
			{M.}~\bibnamefont {Steiner}}, \bibinfo {author} {\bibfnamefont
			{V.}~\bibnamefont {Jacques}}, \bibinfo {author} {\bibfnamefont
			{G.}~\bibnamefont {Balasubramanian}}, \bibinfo {author} {\bibfnamefont
			{M.~L.}\ \bibnamefont {Markham}}, \bibinfo {author} {\bibfnamefont {D.~J.}\
			\bibnamefont {Twitchen}}, \bibinfo {author} {\bibfnamefont {S.}~\bibnamefont
			{Pezzagna}}, \bibinfo {author} {\bibfnamefont {J.}~\bibnamefont {Meijer}},
		\bibinfo {author} {\bibfnamefont {J.}~\bibnamefont {Twamley}}, \bibinfo
		{author} {\bibfnamefont {F.}~\bibnamefont {Jelezko}},\ and\ \bibinfo {author}
		{\bibfnamefont {J.}~\bibnamefont {Wrachtrup}},\ }\bibfield  {title} {\bibinfo
		{title} {Quantum register based on coupled electron spins in a
			room-temperature solid},\ }\href {https://doi.org/10.1038/nphys1536}
	{\bibfield  {journal} {\bibinfo  {journal} {Nat. Phys.}\ }\textbf {\bibinfo
			{volume} {6}},\ \bibinfo {pages} {249} (\bibinfo {year} {2010})}\BibitemShut
	{NoStop}%
	\bibitem [{\citenamefont {Wu}\ \emph {et~al.}(2019)\citenamefont {Wu},
		\citenamefont {Wang}, \citenamefont {Qin}, \citenamefont {Rong},\ and\
		\citenamefont {Du}}]{Wu2019}%
	\BibitemOpen
	\bibfield  {author} {\bibinfo {author} {\bibfnamefont {Y.}~\bibnamefont
			{Wu}}, \bibinfo {author} {\bibfnamefont {Y.}~\bibnamefont {Wang}}, \bibinfo
		{author} {\bibfnamefont {X.}~\bibnamefont {Qin}}, \bibinfo {author}
		{\bibfnamefont {X.}~\bibnamefont {Rong}},\ and\ \bibinfo {author}
		{\bibfnamefont {J.}~\bibnamefont {Du}},\ }\bibfield  {title} {\bibinfo
		{title} {A programmable two-qubit solid-state quantum processor under ambient
			conditions},\ }\href {https://doi.org/10.1038/s41534-019-0129-z} {\bibfield
		{journal} {\bibinfo  {journal} {npj Quantum Inf.}\ }\textbf {\bibinfo
			{volume} {5}},\ \bibinfo {pages} {9} (\bibinfo {year} {2019})}\BibitemShut
	{NoStop}%
	\bibitem [{\citenamefont {Bayliss}\ \emph {et~al.}(2020)\citenamefont
		{Bayliss}, \citenamefont {Laorenza}, \citenamefont {Mintun}, \citenamefont
		{Kovos}, \citenamefont {Freedman},\ and\ \citenamefont
		{Awschalom}}]{Bayliss2020}%
	\BibitemOpen
	\bibfield  {author} {\bibinfo {author} {\bibfnamefont {S.~L.}\ \bibnamefont
			{Bayliss}}, \bibinfo {author} {\bibfnamefont {D.~W.}\ \bibnamefont
			{Laorenza}}, \bibinfo {author} {\bibfnamefont {P.~J.}\ \bibnamefont
			{Mintun}}, \bibinfo {author} {\bibfnamefont {B.~D.}\ \bibnamefont {Kovos}},
		\bibinfo {author} {\bibfnamefont {D.~E.}\ \bibnamefont {Freedman}},\ and\
		\bibinfo {author} {\bibfnamefont {D.~D.}\ \bibnamefont {Awschalom}},\
	}\bibfield  {title} {\bibinfo {title} {Optically addressable molecular spins
			for quantum information processing},\ }\href
	{https://doi.org/10.1126/science.abb9352} {\bibfield  {journal} {\bibinfo
			{journal} {Science}\ }\textbf {\bibinfo {volume} {370}},\ \bibinfo {pages}
		{1309} (\bibinfo {year} {2020})}\BibitemShut {NoStop}%
	\bibitem [{\citenamefont {Gaita-Ariño}\ \emph {et~al.}(2019)\citenamefont
		{Gaita-Ariño}, \citenamefont {Luis}, \citenamefont {Hill},\ and\
		\citenamefont {Coronado}}]{Gaita2019}%
	\BibitemOpen
	\bibfield  {author} {\bibinfo {author} {\bibfnamefont {A.}~\bibnamefont
			{Gaita-Ariño}}, \bibinfo {author} {\bibfnamefont {F.}~\bibnamefont {Luis}},
		\bibinfo {author} {\bibfnamefont {S.}~\bibnamefont {Hill}},\ and\ \bibinfo
		{author} {\bibfnamefont {E.}~\bibnamefont {Coronado}},\ }\bibfield  {title}
	{\bibinfo {title} {Molecular spins for quantum computation},\ }\href
	{https://doi.org/10.1038/s41557-019-0232-y} {\bibfield  {journal} {\bibinfo
			{journal} {Nat. Chem.}\ }\textbf {\bibinfo {volume} {11}},\ \bibinfo {pages}
		{301} (\bibinfo {year} {2019})}\BibitemShut {NoStop}%
	\bibitem [{\citenamefont {Bayliss}\ \emph {et~al.}(2022)\citenamefont
		{Bayliss}, \citenamefont {Deb}, \citenamefont {Laorenza}, \citenamefont
		{Onizhuk}, \citenamefont {Galli}, \citenamefont {Freedman},\ and\
		\citenamefont {Awschalom}}]{Bayliss2022}%
	\BibitemOpen
	\bibfield  {author} {\bibinfo {author} {\bibfnamefont {S.~L.}\ \bibnamefont
			{Bayliss}}, \bibinfo {author} {\bibfnamefont {P.}~\bibnamefont {Deb}},
		\bibinfo {author} {\bibfnamefont {D.~W.}\ \bibnamefont {Laorenza}}, \bibinfo
		{author} {\bibfnamefont {M.}~\bibnamefont {Onizhuk}}, \bibinfo {author}
		{\bibfnamefont {G.}~\bibnamefont {Galli}}, \bibinfo {author} {\bibfnamefont
			{D.~E.}\ \bibnamefont {Freedman}},\ and\ \bibinfo {author} {\bibfnamefont
			{D.~D.}\ \bibnamefont {Awschalom}},\ }\bibfield  {title} {\bibinfo {title}
		{Enhancing spin coherence in optically addressable molecular qubits through
			host-matrix control},\ }\href {https://doi.org/10.1103/PhysRevX.12.031028}
	{\bibfield  {journal} {\bibinfo  {journal} {Phys. Rev. X}\ }\textbf {\bibinfo
			{volume} {12}},\ \bibinfo {pages} {031028} (\bibinfo {year}
		{2022})}\BibitemShut {NoStop}%
	\bibitem [{\citenamefont {Ivanov}\ \emph {et~al.}(2008)\citenamefont {Ivanov},
		\citenamefont {Vitanov},\ and\ \citenamefont {Plenio}}]{Ivanov2008}%
	\BibitemOpen
	\bibfield  {author} {\bibinfo {author} {\bibfnamefont {P.~A.}\ \bibnamefont
			{Ivanov}}, \bibinfo {author} {\bibfnamefont {N.~V.}\ \bibnamefont
			{Vitanov}},\ and\ \bibinfo {author} {\bibfnamefont {M.~B.}\ \bibnamefont
			{Plenio}},\ }\bibfield  {title} {\bibinfo {title} {Creation of cluster states
			of trapped ions by collective addressing},\ }\href
	{https://doi.org/10.1103/PhysRevA.78.012323} {\bibfield  {journal} {\bibinfo
			{journal} {Phys. Rev. A}\ }\textbf {\bibinfo {volume} {78}},\ \bibinfo
		{pages} {012323} (\bibinfo {year} {2008})}\BibitemShut {NoStop}%
	\bibitem [{\citenamefont {Wunderlich}\ \emph {et~al.}(2009)\citenamefont
		{Wunderlich}, \citenamefont {Wunderlich}, \citenamefont {Singer},\ and\
		\citenamefont {Schmidt-Kaler}}]{Wunderlich2009}%
	\BibitemOpen
	\bibfield  {author} {\bibinfo {author} {\bibfnamefont {H.}~\bibnamefont
			{Wunderlich}}, \bibinfo {author} {\bibfnamefont {C.}~\bibnamefont
			{Wunderlich}}, \bibinfo {author} {\bibfnamefont {K.}~\bibnamefont {Singer}},\
		and\ \bibinfo {author} {\bibfnamefont {F.}~\bibnamefont {Schmidt-Kaler}},\
	}\bibfield  {title} {\bibinfo {title} {Two-dimensional cluster-state
			preparation with linear ion traps},\ }\href
	{https://doi.org/10.1103/PhysRevA.79.052324} {\bibfield  {journal} {\bibinfo
			{journal} {Phys. Rev. A}\ }\textbf {\bibinfo {volume} {79}},\ \bibinfo
		{pages} {052324} (\bibinfo {year} {2009})}\BibitemShut {NoStop}%
	\bibitem [{\citenamefont {Ichikawa}\ \emph {et~al.}(2013)\citenamefont
		{Ichikawa}, \citenamefont {G\"ung\"ord\"u}, \citenamefont {Bando},
		\citenamefont {Kondo},\ and\ \citenamefont {Nakahara}}]{Ichikawa2013}%
	\BibitemOpen
	\bibfield  {author} {\bibinfo {author} {\bibfnamefont {T.}~\bibnamefont
			{Ichikawa}}, \bibinfo {author} {\bibfnamefont {U.}~\bibnamefont
			{G\"ung\"ord\"u}}, \bibinfo {author} {\bibfnamefont {M.}~\bibnamefont
			{Bando}}, \bibinfo {author} {\bibfnamefont {Y.}~\bibnamefont {Kondo}},\ and\
		\bibinfo {author} {\bibfnamefont {M.}~\bibnamefont {Nakahara}},\ }\bibfield
	{title} {\bibinfo {title} {Minimal and robust composite two-qubit gates with
			{I}sing-type interaction},\ }\href
	{https://doi.org/10.1103/PhysRevA.87.022323} {\bibfield  {journal} {\bibinfo
			{journal} {Phys. Rev. A}\ }\textbf {\bibinfo {volume} {87}},\ \bibinfo
		{pages} {022323} (\bibinfo {year} {2013})}\BibitemShut {NoStop}%
	\bibitem [{\citenamefont {Ivanov}\ and\ \citenamefont
		{Vitanov}(2015)}]{Ivanov2015}%
	\BibitemOpen
	\bibfield  {author} {\bibinfo {author} {\bibfnamefont {S.~S.}\ \bibnamefont
			{Ivanov}}\ and\ \bibinfo {author} {\bibfnamefont {N.~V.}\ \bibnamefont
			{Vitanov}},\ }\bibfield  {title} {\bibinfo {title} {Composite two-qubit
			gates},\ }\href {https://doi.org/10.1103/PhysRevA.92.022333} {\bibfield
		{journal} {\bibinfo  {journal} {Phys. Rev. A}\ }\textbf {\bibinfo {volume}
			{92}},\ \bibinfo {pages} {022333} (\bibinfo {year} {2015})}\BibitemShut
	{NoStop}%
	\bibitem [{\citenamefont {Niskanen}\ \emph {et~al.}(2007)\citenamefont
		{Niskanen}, \citenamefont {Harrabi}, \citenamefont {Yoshihara}, \citenamefont
		{Nakamura}, \citenamefont {Lloyd},\ and\ \citenamefont
		{Tsai}}]{Niskanen2007}%
	\BibitemOpen
	\bibfield  {author} {\bibinfo {author} {\bibfnamefont {A.~O.}\ \bibnamefont
			{Niskanen}}, \bibinfo {author} {\bibfnamefont {K.}~\bibnamefont {Harrabi}},
		\bibinfo {author} {\bibfnamefont {F.}~\bibnamefont {Yoshihara}}, \bibinfo
		{author} {\bibfnamefont {Y.}~\bibnamefont {Nakamura}}, \bibinfo {author}
		{\bibfnamefont {S.}~\bibnamefont {Lloyd}},\ and\ \bibinfo {author}
		{\bibfnamefont {J.~S.}\ \bibnamefont {Tsai}},\ }\bibfield  {title} {\bibinfo
		{title} {Quantum {Coherent} {Tunable} {Coupling} of {Superconducting}
			{Qubits}},\ }\href {https://doi.org/10.1126/science.1141324} {\bibfield
		{journal} {\bibinfo  {journal} {Science}\ }\textbf {\bibinfo {volume}
			{316}},\ \bibinfo {pages} {723} (\bibinfo {year} {2007})}\BibitemShut
	{NoStop}%
	\bibitem [{\citenamefont {Sung}\ \emph {et~al.}(2021)\citenamefont {Sung},
		\citenamefont {Ding}, \citenamefont {Braumüller}, \citenamefont
		{Vepsäläinen}, \citenamefont {Kannan}, \citenamefont {Kjaergaard},
		\citenamefont {Greene}, \citenamefont {Samach}, \citenamefont {McNally},
		\citenamefont {Kim}, \citenamefont {Melville}, \citenamefont {Niedzielski},
		\citenamefont {Schwartz}, \citenamefont {Yoder}, \citenamefont {Orlando},
		\citenamefont {Gustavsson},\ and\ \citenamefont {Oliver}}]{Sung2021}%
	\BibitemOpen
	\bibfield  {author} {\bibinfo {author} {\bibfnamefont {Y.}~\bibnamefont
			{Sung}}, \bibinfo {author} {\bibfnamefont {L.}~\bibnamefont {Ding}}, \bibinfo
		{author} {\bibfnamefont {J.}~\bibnamefont {Braumüller}}, \bibinfo {author}
		{\bibfnamefont {A.}~\bibnamefont {Vepsäläinen}}, \bibinfo {author}
		{\bibfnamefont {B.}~\bibnamefont {Kannan}}, \bibinfo {author} {\bibfnamefont
			{M.}~\bibnamefont {Kjaergaard}}, \bibinfo {author} {\bibfnamefont
			{A.}~\bibnamefont {Greene}}, \bibinfo {author} {\bibfnamefont {G.~O.}\
			\bibnamefont {Samach}}, \bibinfo {author} {\bibfnamefont {C.}~\bibnamefont
			{McNally}}, \bibinfo {author} {\bibfnamefont {D.}~\bibnamefont {Kim}},
		\bibinfo {author} {\bibfnamefont {A.}~\bibnamefont {Melville}}, \bibinfo
		{author} {\bibfnamefont {B.~M.}\ \bibnamefont {Niedzielski}}, \bibinfo
		{author} {\bibfnamefont {M.~E.}\ \bibnamefont {Schwartz}}, \bibinfo {author}
		{\bibfnamefont {J.~L.}\ \bibnamefont {Yoder}}, \bibinfo {author}
		{\bibfnamefont {T.~P.}\ \bibnamefont {Orlando}}, \bibinfo {author}
		{\bibfnamefont {S.}~\bibnamefont {Gustavsson}},\ and\ \bibinfo {author}
		{\bibfnamefont {W.~D.}\ \bibnamefont {Oliver}},\ }\bibfield  {title}
	{\bibinfo {title} {Realization of {High}-{Fidelity} {CZ} and ${ZZ}$-{Free}
			{iSWAP} {Gates} with a {Tunable} {Coupler}},\ }\href
	{https://doi.org/10.1103/PhysRevX.11.021058} {\bibfield  {journal} {\bibinfo
			{journal} {Phys. Rev. X}\ }\textbf {\bibinfo {volume} {11}},\ \bibinfo
		{pages} {021058} (\bibinfo {year} {2021})}\BibitemShut {NoStop}%
	\bibitem [{\citenamefont {Li}\ \emph {et~al.}(2020)\citenamefont {Li},
		\citenamefont {Cai}, \citenamefont {Yan}, \citenamefont {Wang}, \citenamefont
		{Pan}, \citenamefont {Ma}, \citenamefont {Cai}, \citenamefont {Han},
		\citenamefont {Hua}, \citenamefont {Han}, \citenamefont {Wu}, \citenamefont
		{Zhang}, \citenamefont {Wang}, \citenamefont {Song}, \citenamefont {Duan},\
		and\ \citenamefont {Sun}}]{Li2020}%
	\BibitemOpen
	\bibfield  {author} {\bibinfo {author} {\bibfnamefont {X.}~\bibnamefont
			{Li}}, \bibinfo {author} {\bibfnamefont {T.}~\bibnamefont {Cai}}, \bibinfo
		{author} {\bibfnamefont {H.}~\bibnamefont {Yan}}, \bibinfo {author}
		{\bibfnamefont {Z.}~\bibnamefont {Wang}}, \bibinfo {author} {\bibfnamefont
			{X.}~\bibnamefont {Pan}}, \bibinfo {author} {\bibfnamefont {Y.}~\bibnamefont
			{Ma}}, \bibinfo {author} {\bibfnamefont {W.}~\bibnamefont {Cai}}, \bibinfo
		{author} {\bibfnamefont {J.}~\bibnamefont {Han}}, \bibinfo {author}
		{\bibfnamefont {Z.}~\bibnamefont {Hua}}, \bibinfo {author} {\bibfnamefont
			{X.}~\bibnamefont {Han}}, \bibinfo {author} {\bibfnamefont {Y.}~\bibnamefont
			{Wu}}, \bibinfo {author} {\bibfnamefont {H.}~\bibnamefont {Zhang}}, \bibinfo
		{author} {\bibfnamefont {H.}~\bibnamefont {Wang}}, \bibinfo {author}
		{\bibfnamefont {Y.}~\bibnamefont {Song}}, \bibinfo {author} {\bibfnamefont
			{L.}~\bibnamefont {Duan}},\ and\ \bibinfo {author} {\bibfnamefont
			{L.}~\bibnamefont {Sun}},\ }\bibfield  {title} {\bibinfo {title} {{Tunable}
			{Coupler} for {Realizing} a {Controlled}-{Phase} {Gate} with {Dynamically}
			{Decoupled} {Regime} in a {Superconducting} {Circuit}},\ }\href
	{https://doi.org/10.1103/PhysRevApplied.14.024070} {\bibfield  {journal}
		{\bibinfo  {journal} {Phys. Rev. Appl.}\ }\textbf {\bibinfo {volume} {14}},\
		\bibinfo {pages} {024070} (\bibinfo {year} {2020})}\BibitemShut {NoStop}%
	\bibitem [{\citenamefont {Zhao}\ \emph {et~al.}(2020)\citenamefont {Zhao},
		\citenamefont {Xu}, \citenamefont {Lan}, \citenamefont {Chu}, \citenamefont
		{Tan}, \citenamefont {Yu},\ and\ \citenamefont {Yu}}]{Zhao2020}%
	\BibitemOpen
	\bibfield  {author} {\bibinfo {author} {\bibfnamefont {P.}~\bibnamefont
			{Zhao}}, \bibinfo {author} {\bibfnamefont {P.}~\bibnamefont {Xu}}, \bibinfo
		{author} {\bibfnamefont {D.}~\bibnamefont {Lan}}, \bibinfo {author}
		{\bibfnamefont {J.}~\bibnamefont {Chu}}, \bibinfo {author} {\bibfnamefont
			{X.}~\bibnamefont {Tan}}, \bibinfo {author} {\bibfnamefont {H.}~\bibnamefont
			{Yu}},\ and\ \bibinfo {author} {\bibfnamefont {Y.}~\bibnamefont {Yu}},\
	}\bibfield  {title} {\bibinfo {title} {High-{Contrast} ${ZZ}$ {Interaction}
			{Using} {Superconducting} {Qubits} with {Opposite}-{Sign} {Anharmonicity}},\
	}\href {https://doi.org/10.1103/PhysRevLett.125.200503} {\bibfield  {journal}
		{\bibinfo  {journal} {Phys. Rev. Lett.}\ }\textbf {\bibinfo {volume} {125}},\
		\bibinfo {pages} {200503} (\bibinfo {year} {2020})}\BibitemShut {NoStop}%
	\bibitem [{\citenamefont {Ku}\ \emph {et~al.}(2020)\citenamefont {Ku},
		\citenamefont {Xu}, \citenamefont {Brink}, \citenamefont {McKay},
		\citenamefont {Hertzberg}, \citenamefont {Ansari},\ and\ \citenamefont
		{Plourde}}]{Ku2020}%
	\BibitemOpen
	\bibfield  {author} {\bibinfo {author} {\bibfnamefont {J.}~\bibnamefont
			{Ku}}, \bibinfo {author} {\bibfnamefont {X.}~\bibnamefont {Xu}}, \bibinfo
		{author} {\bibfnamefont {M.}~\bibnamefont {Brink}}, \bibinfo {author}
		{\bibfnamefont {D.~C.}\ \bibnamefont {McKay}}, \bibinfo {author}
		{\bibfnamefont {J.~B.}\ \bibnamefont {Hertzberg}}, \bibinfo {author}
		{\bibfnamefont {M.~H.}\ \bibnamefont {Ansari}},\ and\ \bibinfo {author}
		{\bibfnamefont {B.~L.~T.}\ \bibnamefont {Plourde}},\ }\bibfield  {title}
	{\bibinfo {title} {Suppression of {Unwanted} ${ZZ}$ {Interactions} in a
			{Hybrid} {Two}-{Qubit} {System}},\ }\href
	{https://doi.org/10.1103/PhysRevLett.125.200504} {\bibfield  {journal}
		{\bibinfo  {journal} {Phys. Rev. Lett.}\ }\textbf {\bibinfo {volume} {125}},\
		\bibinfo {pages} {200504} (\bibinfo {year} {2020})}\BibitemShut {NoStop}%
	\bibitem [{\citenamefont {Liang}\ \emph {et~al.}(2024)\citenamefont {Liang},
		\citenamefont {Liang}, \citenamefont {Li}, \citenamefont {Wang},\ and\
		\citenamefont {Xue}}]{Liang2024}%
	\BibitemOpen
	\bibfield  {author} {\bibinfo {author} {\bibfnamefont {Y.}~\bibnamefont
			{Liang}}, \bibinfo {author} {\bibfnamefont {M.-J.}\ \bibnamefont {Liang}},
		\bibinfo {author} {\bibfnamefont {S.}~\bibnamefont {Li}}, \bibinfo {author}
		{\bibfnamefont {Z.~D.}\ \bibnamefont {Wang}},\ and\ \bibinfo {author}
		{\bibfnamefont {Z.-Y.}\ \bibnamefont {Xue}},\ }\bibfield  {title} {\bibinfo
		{title} {Scalable protocol to mitigate $zz$ crosstalk in universal quantum
			gates},\ }\href {https://doi.org/10.1103/PhysRevApplied.21.024016} {\bibfield
		{journal} {\bibinfo  {journal} {Phys. Rev. Appl.}\ }\textbf {\bibinfo
			{volume} {21}},\ \bibinfo {pages} {024016} (\bibinfo {year}
		{2024})}\BibitemShut {NoStop}%
	\bibitem [{\citenamefont {Xie}\ \emph {et~al.}(2022)\citenamefont {Xie},
		\citenamefont {Zhai}, \citenamefont {Zhang}, \citenamefont {Allcock},
		\citenamefont {Zhang},\ and\ \citenamefont {Zheng}}]{xie2022}%
	\BibitemOpen
	\bibfield  {author} {\bibinfo {author} {\bibfnamefont {L.}~\bibnamefont
			{Xie}}, \bibinfo {author} {\bibfnamefont {J.}~\bibnamefont {Zhai}}, \bibinfo
		{author} {\bibfnamefont {Z.}~\bibnamefont {Zhang}}, \bibinfo {author}
		{\bibfnamefont {J.}~\bibnamefont {Allcock}}, \bibinfo {author} {\bibfnamefont
			{S.}~\bibnamefont {Zhang}},\ and\ \bibinfo {author} {\bibfnamefont {Y.-C.}\
			\bibnamefont {Zheng}},\ }\bibfield  {title} {\bibinfo {title} {Suppressing
			{ZZ} crosstalk of {Quantum} {Computers} through {Pulse} and {Scheduling}
			{Co-Optimization}},\ }in\ \href {https://doi.org/10.1145/3503222.3507761}
	{\emph {\bibinfo {booktitle}
			{\href{https://doi.org/10.1145/3503222.3507761}{ASPLOS ’22: Proceedings of
					the 27th {ACM} International Conference on Architectural Support for
					Programming Languages and Operating Systems. {Pages} 499-513}}}}\ (\bibinfo
	{address} {New York, NY, USA},\ \bibinfo {year} {2022})\BibitemShut {NoStop}%
	\bibitem [{\citenamefont {Tripathi}\ \emph {et~al.}(2022)\citenamefont
		{Tripathi}, \citenamefont {Chen}, \citenamefont {Khezri}, \citenamefont
		{Yip}, \citenamefont {Levenson-Falk},\ and\ \citenamefont
		{Lidar}}]{Tripathi2022}%
	\BibitemOpen
	\bibfield  {author} {\bibinfo {author} {\bibfnamefont {V.}~\bibnamefont
			{Tripathi}}, \bibinfo {author} {\bibfnamefont {H.}~\bibnamefont {Chen}},
		\bibinfo {author} {\bibfnamefont {M.}~\bibnamefont {Khezri}}, \bibinfo
		{author} {\bibfnamefont {K.-W.}\ \bibnamefont {Yip}}, \bibinfo {author}
		{\bibfnamefont {E.~M.}\ \bibnamefont {Levenson-Falk}},\ and\ \bibinfo
		{author} {\bibfnamefont {D.~A.}\ \bibnamefont {Lidar}},\ }\bibfield  {title}
	{\bibinfo {title} {Suppression of {Crosstalk} in {Superconducting} {Qubits}
			{Using} {Dynamical} {Decoupling}},\ }\href
	{https://doi.org/10.1103/PhysRevApplied.18.024068} {\bibfield  {journal}
		{\bibinfo  {journal} {Phys. Rev. Appl.}\ }\textbf {\bibinfo {volume} {18}},\
		\bibinfo {pages} {024068} (\bibinfo {year} {2022})}\BibitemShut {NoStop}%
	\bibitem [{Opt(023a)}]{Optimization2023}%
	\BibitemOpen
	\href {https://de.mathworks.com/products/optimization.html} {\bibinfo {title}
		{Matlab optimization toolbox}} (\bibinfo {year} {r2023a}),\ \bibinfo {note}
	{the MathWorks, Natick, MA, USA}\BibitemShut {NoStop}%
	\bibitem [{\citenamefont {Van~Loan}(1978)}]{VanLoan1978}%
	\BibitemOpen
	\bibfield  {author} {\bibinfo {author} {\bibfnamefont {C.}~\bibnamefont
			{Van~Loan}},\ }\bibfield  {title} {\bibinfo {title} {Computing integrals
			involving the matrix exponential},\ }\href
	{https://doi.org/10.1109/TAC.1978.1101743} {\bibfield  {journal} {\bibinfo
			{journal} {IEEE Trans. Autom. Control}\ }\textbf {\bibinfo {volume} {23}},\
		\bibinfo {pages} {395} (\bibinfo {year} {1978})}\BibitemShut {NoStop}%
	\bibitem [{\citenamefont {Najfeld}\ and\ \citenamefont
		{Havel}(1995)}]{Najfeld1995}%
	\BibitemOpen
	\bibfield  {author} {\bibinfo {author} {\bibfnamefont {I.}~\bibnamefont
			{Najfeld}}\ and\ \bibinfo {author} {\bibfnamefont {T.~F.}\ \bibnamefont
			{Havel}},\ }\bibfield  {title} {\bibinfo {title} {Derivatives of the {Matrix}
			{Exponential} and {Their} {Computation}},\ }\href
	{https://doi.org/10.1006/aama.1995.1017} {\bibfield  {journal} {\bibinfo
			{journal} {Adv. Appl. Math.}\ }\textbf {\bibinfo {volume} {16}},\ \bibinfo
		{pages} {321} (\bibinfo {year} {1995})}\BibitemShut {NoStop}%
	\bibitem [{\citenamefont {Carbonell}\ \emph {et~al.}(2008)\citenamefont
		{Carbonell}, \citenamefont {Jímenez},\ and\ \citenamefont
		{Pedroso}}]{Carbonell2008}%
	\BibitemOpen
	\bibfield  {author} {\bibinfo {author} {\bibfnamefont {F.}~\bibnamefont
			{Carbonell}}, \bibinfo {author} {\bibfnamefont {J.~C.}\ \bibnamefont
			{Jímenez}},\ and\ \bibinfo {author} {\bibfnamefont {L.~M.}\ \bibnamefont
			{Pedroso}},\ }\bibfield  {title} {\bibinfo {title} {Computing multiple
			integrals involving matrix exponentials},\ }\href
	{https://doi.org/10.1016/j.cam.2007.01.007} {\bibfield  {journal} {\bibinfo
			{journal} {J. Comput. Appl. Math.}\ }\textbf {\bibinfo {volume} {213}},\
		\bibinfo {pages} {300} (\bibinfo {year} {2008})}\BibitemShut {NoStop}%
	\bibitem [{\citenamefont {Goodwin}\ and\ \citenamefont
		{Kuprov}(2015)}]{Goodwin2015}%
	\BibitemOpen
	\bibfield  {author} {\bibinfo {author} {\bibfnamefont {D.~L.}\ \bibnamefont
			{Goodwin}}\ and\ \bibinfo {author} {\bibfnamefont {I.}~\bibnamefont
			{Kuprov}},\ }\bibfield  {title} {\bibinfo {title} {Auxiliary matrix formalism
			for interaction representation transformations, optimal control, and spin
			relaxation theories},\ }\href {https://doi.org/10.1063/1.4928978} {\bibfield
		{journal} {\bibinfo  {journal} {J. Chem. Phys.}\ }\textbf {\bibinfo {volume}
			{143}},\ \bibinfo {pages} {084113} (\bibinfo {year} {2015})}\BibitemShut
	{NoStop}%
	\bibitem [{\citenamefont {Haas}\ \emph {et~al.}(2019)\citenamefont {Haas},
		\citenamefont {Puzzuoli}, \citenamefont {Zhang},\ and\ \citenamefont
		{Cory}}]{Haas2019}%
	\BibitemOpen
	\bibfield  {author} {\bibinfo {author} {\bibfnamefont {H.}~\bibnamefont
			{Haas}}, \bibinfo {author} {\bibfnamefont {D.}~\bibnamefont {Puzzuoli}},
		\bibinfo {author} {\bibfnamefont {F.}~\bibnamefont {Zhang}},\ and\ \bibinfo
		{author} {\bibfnamefont {D.~G.}\ \bibnamefont {Cory}},\ }\bibfield  {title}
	{\bibinfo {title} {Engineering effective {Hamiltonians}},\ }\href
	{https://doi.org/10.1088/1367-2630/ab4525} {\bibfield  {journal} {\bibinfo
			{journal} {New J. Phys.}\ }\textbf {\bibinfo {volume} {21}},\ \bibinfo
		{pages} {103011} (\bibinfo {year} {2019})}\BibitemShut {NoStop}%
	\bibitem [{\citenamefont {Herbschleb}\ \emph {et~al.}(2019)\citenamefont
		{Herbschleb}, \citenamefont {Kato}, \citenamefont {Maruyama}, \citenamefont
		{Danjo}, \citenamefont {Makino}, \citenamefont {Yamasaki}, \citenamefont
		{Ohki}, \citenamefont {Hayashi}, \citenamefont {Morishita}, \citenamefont
		{Fujiwara},\ and\ \citenamefont {Mizuochi}}]{Herbschleb2019}%
	\BibitemOpen
	\bibfield  {author} {\bibinfo {author} {\bibfnamefont {E.~D.}\ \bibnamefont
			{Herbschleb}}, \bibinfo {author} {\bibfnamefont {H.}~\bibnamefont {Kato}},
		\bibinfo {author} {\bibfnamefont {Y.}~\bibnamefont {Maruyama}}, \bibinfo
		{author} {\bibfnamefont {T.}~\bibnamefont {Danjo}}, \bibinfo {author}
		{\bibfnamefont {T.}~\bibnamefont {Makino}}, \bibinfo {author} {\bibfnamefont
			{S.}~\bibnamefont {Yamasaki}}, \bibinfo {author} {\bibfnamefont
			{I.}~\bibnamefont {Ohki}}, \bibinfo {author} {\bibfnamefont {K.}~\bibnamefont
			{Hayashi}}, \bibinfo {author} {\bibfnamefont {H.}~\bibnamefont {Morishita}},
		\bibinfo {author} {\bibfnamefont {M.}~\bibnamefont {Fujiwara}},\ and\
		\bibinfo {author} {\bibfnamefont {N.}~\bibnamefont {Mizuochi}},\ }\bibfield
	{title} {\bibinfo {title} {Ultra-long coherence times amongst
			room-temperature solid-state spins},\ }\href
	{https://doi.org/10.1038/s41467-019-11776-8} {\bibfield  {journal} {\bibinfo
			{journal} {Nat. Commun.}\ }\textbf {\bibinfo {volume} {10}},\ \bibinfo
		{pages} {3766} (\bibinfo {year} {2019})}\BibitemShut {NoStop}%
\end{thebibliography}
%

\end{document}